\pdfoutput=1 
\documentclass[11pt,a4paper]{article}
\usepackage{jheppub} 
\usepackage[T1]{fontenc} 
\usepackage{preprintcover}  
\usepackage{graphicx}
\usepackage{subfigure}
\usepackage{amsmath}
\usepackage{lineno}

\PreprintCoverPaperTitle{}  

\PreprintIdNumber{CERN-PH-EP-2013-023}  

\PreprintCoverPaperTitle{Measurement of the production cross section of jets in association with a   $Z$ boson  in pp collisions at
  $\sqrt{s} = 7 \rm TeV$ with the ATLAS detector
}  

\PreprintCoverAbstract{Measurements of  the production of jets of particles in association with a $Z$ boson  in \pp\  collisions at $\sqrt{s} = 7 \tev $ are presented, 
using data corresponding to an integrated luminosity of 4.6~$\ifb$ collected by the ATLAS experiment at the Large Hadron Collider. 
Inclusive and differential  jet cross sections in $Z$  events, with
$Z$  decaying into electron or muon pairs, are measured for jets with
transverse momentum $\pt > 30 \gev$ and rapidity $|y| < 4.4$. The
results are compared to next-to-leading-order perturbative QCD
calculations, and to predictions from different Monte Carlo generators
based on leading-order and next-to-leading-order matrix elements supplemented by parton showers.}  
\PreprintJournalName{JHEP}  

\title{Measurement of the production cross section of
  jets in association with a   $Z$ boson  in pp collisions at
  $\sqrt{s} = 7 \rm TeV$ with the ATLAS detector}

\def\alp{ALPGEN}
\def\her{HERWIG}
\def\pyt{PYTHIA}
\def\she{SHERPA}

\def\bhs{{\sc BlackHat}+SHERPA}
\def\mca{MC@NLO}
\def\jim{JIMMY}
\def\pho{PHOTOS}
\def\ace{AcerMC}
\def\pow{POWHEG}
\def\per{PERUGIA2011C}
\def\ctt{CT10}
\def\cts{CTEQ61L}
\def\aue{AUET2}
\def\few{FEWZ}
\def\gea{GEANT4}

\def\ifb{\ensuremath{\rm fb^{-1}}}

\def\pt{\ensuremath{p_{\rm T}}}
\def\as{\ensuremath{\alpha_{\rm s}}}
\def\epem{\ensuremath{e^+e^-}}
\def\mpmm{\ensuremath{\mu^+\mu^-}}

\def\dr{\ensuremath{\Delta R}}
\def\rts{\ensuremath{\sqrt{s}}}
\def\pp{\ensuremath{pp}}
\def\tev{\ifmmode {\mathrm{\ Te\kern -0.1em V}}\else
                   \textrm{Te\kern -0.1em V}\fi}%
\def\gev{\ifmmode {\mathrm{\ Ge\kern -0.1em V}}\else
                   \textrm{Ge\kern -0.1em V}\fi}%
\def\el{\ensuremath{e}}
\def\mgeq{\ensuremath{\geq}}

\def\htj{\ensuremath{H_{\rm T}}}
\def\stj{\ensuremath{S_{\rm T}}}

\def\ptll{\ensuremath{p^{\ell\ell}_{\rm T}}}

\def\ptj{\ensuremath{p^{\rm jet}_{\rm T}}}
\def\ayj{\ensuremath{|y^{\rm jet}|}}
\def\yj{\ensuremath{y^{\rm jet}}}
\def\mjj{\ensuremath{m^{jj}}}
\def\mll{\ensuremath{m^{\ell\ell}}}
\def\mee{\ensuremath{m^{ee}}}
\def\mmm{\ensuremath{m^{\mu\mu}}}
\def\mz{\ensuremath{m_{\rm Z }}}
\def\Rmulti{\ensuremath{R_{(n+1)/n}}}
\def\Rinmulti{\ensuremath{R_{\geq (n+1)/\geq n}}}
\def\Rintwoone{\ensuremath{R_{\geq 2/\geq 1}}}

\def\Rthreetwo{\ensuremath{R_{3/2}}}
\def\Rtwoone{\ensuremath{R_{2/1}}}
 
\def\nj{\ensuremath{N_{\rm jet}}}
\def\drjj{\ensuremath{\Delta R^{jj}}}
\def\drlj{\ensuremath{\Delta R^{\ell j}}}
\def\drll{\ensuremath{\Delta R^{\ell\ell}}}
\def\dpjj{\ensuremath{\Delta\phi^{jj}}}
\def\dyjj{\ensuremath{|\Delta y^{jj}}|}

\def\Zmm{\ensuremath{Z\to\mu\mu}}
\def\Zee{\ensuremath{Z\to\el\el}}

\def\Zjets{\ensuremath{\Zg\,\texttt{+}\,\mathrm{jets}}}
\def\Zmmjets{\ensuremath{\Zgmm\,\texttt{+}\,\mathrm{jets}}}
\def\Zeejets{\ensuremath{\Zgee\,\texttt{+}\,\mathrm{jets}}}
\def\Zlljets{\ensuremath{\Zgll\,\texttt{+}\,\mathrm{jets}}}
\def\Zttjets{\ensuremath{\Zgtau\,\texttt{+}\,\mathrm{jets}}}
\def\Wjets{\ensuremath{W\,\texttt{+}\,\mathrm{jets}}}
\def\antibar#1{\ensuremath{#1\bar{#1}}}
\def\ttbar{\antibar{t}}
\def\db{diboson}
\def\Wln{\ensuremath{W\to\ell\nu}}
\def\Wen{\ensuremath{W\to\el\nu}}
\def\Wmn{\ensuremath{W\to\mu\nu}}
\def\Zg{\ensuremath{Z}}
\def\Zzero{\ensuremath{Z}}
\def\Zgll{\ensuremath{\Zg\,(\to\ell \ell)}}
\def\Zgee{\ensuremath{\Zg\,(\to\el\el)}}
\def\Zgmm{\ensuremath{\Zg\,(\to\mu\mu)}}
\def\Zgtau{\ensuremath{\Zg\,(\to\tau\tau)}}

\def\Zmj{\ensuremath{\Zgmm\,\texttt{+}\mgeq\mathrm{1~jet}}}

\def\Zmjjjjjj{\ensuremath{\Zgmm\,\texttt{+}\mgeq\mathrm{6~jets}}}

\def\Zej{\ensuremath{\Zgee\,\texttt{+}\mgeq\mathrm{1~jet}}}

\def\Zejjjjjj{\ensuremath{\Zgee\,\texttt{+}\mgeq\mathrm{6~jets}}}

\def\Zlnj{\ensuremath{\Zgll\,\texttt{+}\mgeq n\,\mathrm{jets}}}

\def\Znp{\ensuremath{\Zg\,\texttt{+}\mgeq n\,\mathrm{partons}}}
\def\Zlj{\ensuremath{\Zgll\,\texttt{+}\mgeq\mathrm{1~jet}}}
\def\Zljj{\ensuremath{\Zgll\,\texttt{+}\mgeq\mathrm{2~jets}}}

\def\Zj{\ensuremath{\Zg\,(\texttt{+}\mgeq\mathrm{1~jet})}}

\def\Zjm{\ensuremath{\Zg\,(\texttt{+}\leq\mathrm{1~jet})}}

\def\Zpp{\ensuremath{\Zg\,(\texttt{+}\mgeq\mathrm{2~partons})}}

\def\Zjjjj{\ensuremath{\Zg\,(\texttt{+}\mgeq\mathrm{4~jets})}}

\def\ZjZjjjj{\ensuremath{\Zg\,(\texttt{+}\,(1-4)\mathrm{~jets})}}

\def\zj{\ensuremath{\mgeq\mathrm{0~jets}}}
\def\j{\ensuremath{\mgeq\mathrm{1~jet}}}
\def\jj{\ensuremath{\mgeq\mathrm{2~jets}}}
\def\jjj{\ensuremath{\mgeq\mathrm{3~jets}}}
\def\jjjj{\ensuremath{\mgeq\mathrm{4~jets}}}
\def\jjjjj{\ensuremath{\mgeq\mathrm{5~jets}}}
\def\jjjjjj{\ensuremath{\mgeq\mathrm{6~jets}}}
\def\jjjjjjj{\ensuremath{\mgeq\mathrm{7~jets}}}
\def\Zzjex{\ensuremath{\Zg\,(\texttt{+}\mathrm{0~jets})}}
\def\Zjex{\ensuremath{\Zg\,(\texttt{+}\mathrm{1~jet})}}
\def\Zpex{\ensuremath{\Zg\,(\texttt{+}\mathrm{1~parton})}}

\def\Zljex{\ensuremath{\Zgll\,\texttt{+}\mathrm{1~jet}}}

\def\Zzjex{\ensuremath{\Zzero\,(\texttt{+}\mathrm{1~jet})}}

\def\Zzjj{\ensuremath{\Zzero\,(\texttt{+}\mgeq\mathrm{2~jets})}}

\newcommand{\antikt}{anti-$k_{t}$}

\abstract{Measurements of  the production of jets of particles in association with a $Z$ boson  in \pp\  collisions at $\sqrt{s} = 7 \tev $ are presented, 
using data corresponding to an integrated luminosity of 4.6~$\ifb$ collected by the ATLAS experiment at the Large Hadron Collider. 
Inclusive and differential  jet cross sections in $Z$  events, with
$Z$  decaying into electron or muon pairs, are measured for jets with
transverse momentum $\pt > 30 \gev$ and rapidity $|y| < 4.4$. The
results are compared to next-to-leading-order perturbative QCD
calculations, and to predictions from different Monte Carlo generators
based on leading-order and next-to-leading-order matrix elements supplemented by parton showers.}

\begin{document}

\maketitle
\flushbottom

\section{Introduction \label{sec:Intro}}

The production of jets of particles in association with a \Zzero\ boson\footnote{The notation \Zzero\ refers to the complete $Z/\gamma^*$ interference.} at hadron colliders provides an important test of perturbative quantum 
chromodynamics (pQCD).  Such events also constitute a non-negligible  background for studies of the Higgs boson candidate~\cite{higgsobsatl,higgsobscms} and
searches  for new phenomena. 
In these searches, the multiplicity and kinematics of jets  in  \Zjets\ events are exploited  to achieve a separation  of signal from background.  This procedure  often 
introduces  scales larger than the mass of  the \Zzero\ boson, resulting in large logarithmic contributions in the calculation of higher-order QCD corrections 
to the predicted \Zjets\ cross section~\cite{scaling12,gkfactors}.
The  measured \Zjets\ cross section can be compared directly to fixed-order predictions at   next-to-leading-order (NLO) in pQCD~\cite{blackhat,bhz3jets,bhz4jets} 
and to Monte Carlo (MC) generators based on  next-to-leading-order or leading-order (LO) matrix elements supplemented by parton showers~\cite{alpgen,sherpa,sherpaMenlops}. 
The simulations based on LO matrix elements  are affected by  large uncertainties in the factorization and renormalization scales and need to be tuned and validated using data. 

Measurements of the \Zjets\ cross section have been reported for lower jet energies  and lower jet multiplicities in proton--antiproton collisions 
at a center-of-mass energy of \linebreak  $\rts = 1.96 \tev$~\cite{zjetscdf,zjetsd0,zjetsd0b} and in proton--proton collisions based on a  data set of  0.036~\ifb\ collected at $\rts = 7 \tev$~\cite{zjetsatlas, vjetscms}. 
This article extends these measurements, using  4.6~\ifb\ of proton--proton collision data collected by the ATLAS experiment in 2011 at $\rts = 7 \tev$.
The large data set  allows cross sections to be measured for the production of up to seven jets 
 in association with a \Zzero\ boson. Differential jet cross sections are accessible for large jet multiplicities and for 
energy regimes up to 1~\tev , which allows the modelling of the
\Zjets\ process to be probed for typical phase-space regimes expected
from new phenomena and from  Higgs boson production, for example via  vector-boson-fusion (VBF).

Selected events contain  a \Zzero\ boson decaying into a pair of
electrons or muons. Associated jets are identified in a rapidity (\yj ) range
of $\ayj  < 4.4$ and with transverse momentum (\ptj ) of    $\ptj > 30 \gev$.
The measurements comprise inclusive and exclusive jet multiplicities for different phase-space constraints and differential jet cross sections 
as a function of the  transverse momentum  and   the rapidity   of the  four jets with the largest transverse momentum  (`leading jets').
 Cross sections for events with at least two jets in the final state are measured as a function of the 
invariant mass (\mjj )  and the angular separation of the two leading  jets.  Differential cross sections  in events with at least one jet are
 measured as a function of the scalar \pt\ sum of the jets (\stj ), of the scalar \pt\ sum of the leptons and jets (\htj ), and the transverse momentum of the \Zzero\   boson 
candidate  (\ptll ).
The results of the measurements are unfolded  for detector effects and quoted at the particle (hadron) level, where they are  compared to predictions from fixed-order NLO pQCD programs and  from  several MC generators.

The paper is organized as follows. The detector  and the data sample are  described in the
next section. Section 3 provides details of the simulations used in the measurements, while section 4 describes the lepton and jet  reconstruction and the event selection. 
The estimation of background contributions is described in section 5 and selected uncorrected distributions are presented in section 6.
The procedures used to unfold the measurements for detector effects and to combine  electron and muon channels are detailed in section 7.
Systematic uncertainties are discussed in section 8. The NLO pQCD predictions are described in section 9. 
Measured cross sections are presented  in section 10 and compared to generator and NLO pQCD predictions. Finally, section 11 provides a summary.

\section{Experimental setup \label{sec:Detector}}

The ATLAS detector~\cite{atlasdet} at the LHC covers nearly the entire solid angle around the collision point. It consists of an inner tracking detector surrounded by a thin superconducting solenoid, followed by electromagnetic and hadronic calorimeters  and a muon spectrometer incorporating three large superconducting toroid magnets (each with eight coils).
The inner detector (ID) is immersed in a 2~T axial magnetic field and provides charged-particle tracking in the pseudorapidity\footnote{ATLAS uses a 
right-handed coordinate system with its origin at the nominal interaction point (IP) in the centre of the detector and the $z$-axis along the beam pipe. The $x$-axis 
points from the IP to the centre of the LHC ring, and the $y$-axis points upward. Cylindrical coordinates $(r,\phi)$ are used in the transverse plane, $\phi$ being the 
azimuthal angle around the beam pipe. The pseudorapidity is defined in terms of the polar angle $\theta$ as $\eta=-\ln\tan(\theta/2)$.} range $|\eta| < 2.5$.  The high-granularity silicon pixel detector covers the vertex region and typically provides three measurements per track, the first hit being normally in the innermost layer. It is followed by the silicon microstrip tracker, which provides typically eight measurements (four space-points)  per track. These silicon detectors are complemented by the transition radiation tracker, which covers a region up to $|\eta| = 2.0$. 
The transition radiation tracker also provides electron identification information based on the fraction of hits 
above a high energy-deposit threshold corresponding to transition radiation.
The calorimeter system covers the pseudorapidity range $|\eta|< 4.9$. Within the region \linebreak $|\eta|< 3.2$, electromagnetic calorimetry is provided by barrel 
and endcap high-granularity lead/liquid-argon (LAr)  calorimeters.  An additional thin LAr presampler
covers $|\eta| < 1.8$  to correct for energy loss in material upstream of the calorimeters. 
Hadronic calorimetry is provided by a  steel/scintillating-tile calorimeter, segmented radially into three barrel structures within $|\eta| < 1.7$, and two copper/LAr hadronic endcap calorimeters,
that cover the region $1.5 < |\eta| < 3.2$.
The solid angle coverage is completed in the region of $3.1 < |\eta| < 4.9$ with forward copper/LAr and tungsten/LAr calorimeter modules optimized for electromagnetic and hadronic 
measurements respectively. 
The muon spectrometer (MS) comprises separate trigger and high-precision tracking chambers measuring the deflection of muons in a magnetic field generated by superconducting air-core toroids. The precision chamber system covers the region $|\eta| < 2.7$ with three layers of monitored drift tubes, complemented by cathode strip chambers in the forward region, where the background is highest. The muon trigger system covers the range $|\eta| < 2.4$ with resistive plate chambers in the barrel, and thin gap chambers in the endcap regions.
A three-level trigger system is used to select interesting events. The Level-1 trigger is implemented in hardware and uses a subset of detector information to reduce the event rate to a design value of at most 75~kHz. This is followed by two software-based trigger levels which together reduce the event rate to about 400~Hz.

The analysis is based on a sample of 
proton--proton collisions at  $\sqrt{s} \,=\, 7~\tev$, collected in 2011 during periods of stable beam operation.
Di-electron final states are selected with a trigger requiring at least two electrons of $\pt >12~\gev$, using an electron identification similar to the one 
used in offline selection. 
Di-muon final states are selected with a trigger requiring at least one muon of $\pt >18~\gev$, using a higher-level trigger  algorithm similar to the one 
used in the offline selection.
The integrated luminosity used in both channels is 4.64$\pm$0.08~\ifb  ~\cite{lumiatlas}.

\section{Monte Carlo simulation \label{sec:MonteCarlo}}

Monte Carlo event samples are used to determine background contributions, correct the measurements for detector effects, 
correct the theory calculations for non-perturbative effects, calculate acceptance corrections, and estimate systematic uncertainties on the final results.

Signal events (\Zmmjets\ and \Zeejets)  are generated using \alp  \linebreak  v2.13~\cite{alpgen}
interfaced to   \her ~v6.520~\cite{herwig} for parton shower and fragmentation and to  \jim ~v4.31~\cite{jimmy} 
for modelling  interactions of the proton remnants,  referred to as `underlying event' in the following, 
using the  \aue -\cts\ tune~\cite{auet2lo}.  In the following sections,  the expression `ALPGEN'
refers to this version unless stated otherwise.
Similar samples are produced with \alp ~v2.14 interfaced to \pyt\ v6.425~\cite{pythia} 
using the \per ~\cite{perugia2011} tune.  For both  \alp\ samples,   \cts ~\cite{cteq6} parton distribution functions (PDFs)  are
employed.
Signal samples are also generated with \she ~v1.4.1 using the ME{\sc nlo}PS approach~\cite{sherpaMenlops} and with \mca  ~v4.01~\cite{mcatnlo}, interfaced to \her  , 
both using  the \ctt ~\cite{cteq10} PDF set.
The program \pho ~\cite{photos} is used to simulate QED final state radiation (FSR) in the \alp\ samples. 
 QED-FSR simulation in \she\  is based on  the YFS method~\cite{yfsm}.  \alp\ and \she\  matrix elements are generated for up to five partons.
The signal samples do not include \Zjets\ events produced via VBF.
Based on generator-level studies, the expected contribution of these events to the  measured cross sections 
is at the per-mille to per-cent  level for the selections and kinematic ranges explored in this paper and always significantly below the 
statistical and systematic precision of the measurement.

Background samples from \Wjets\ and \Zttjets\ final states 
are generated similarly to the signal samples, using \alp\ interfaced to \her  .
The \Wjets\ and \Zjets\ samples  are normalized globally to next-to-next-to-leading-order (NNLO) pQCD inclusive Drell--Yan predictions as 
determined by the \few ~\cite{nnlo2} program using the \linebreak MSTW2008NNLO PDF set~\cite{mstw2008}.  The uncertainties of about 5\%  are taken from an envelope of predictions using different PDF
 sets and factorization and renormalization scales, as described in ref.~\cite{wzatlas}.
  Single-top-quark events are produced with \ace ~\cite{acer}, interfaced to
  \pyt  , using  \cts\  PDFs. Diboson processes ($WW$, $WZ$ and $ZZ$) are simulated with \her\  using the \aue -LO* tune~\cite{auet2lo}. 
Reference cross sections  for single-top-quark and \db\ processes  are calculated using the  \mca\  generator  with the MSTW2008 PDF set~\cite{mstw2008}.
The \ttbar\ samples used for the relative normalization of final states in top-quark pair-production are  generated with  {\mca  } interfaced to \her\  
and with \pow ~\cite{powheg1,powheg2} interfaced to  \pyt  , both using  the \ctt\ PDF set.

All samples are processed through the \gea -based simulation~\cite{geant4, atlsim} of the ATLAS detector. The simulation includes the modelling of
additional \pp\  interactions in the same and neighbouring bunch crossings (pile-up), with an average of nine interactions per crossing, that matches 
the distribution of  interactions per crossing  measured in data.

\section{Event selection \label{sec:Selection}}

Table~\ref{t:selection} summarizes the kinematic regions in which \Zzero\ bosons and jets are selected.
They are defined to provide  a good experimental coverage for the reconstruction of 
electrons, muons and jets in the event.  Events with less than three tracks associated to the hard 
scattering vertex, defined as the vertex with the highest \pt\ sum of its associated tracks, are discarded.

Electrons are reconstructed from clusters of energy in the
electromagnetic calorimeter matched to inner detector tracks. 
The electron candidates must  have $\pt > 20 \gev$  and  $|\eta|<2.47$, excluding the transition region $1.37 < |\eta|< 1.52$ between barrel and endcap electromagnetic calorimeter sections,  and  pass the `medium'  identification criteria described in ref.~\cite{egamma2010}, re-optimized for 2011 conditions.  No additional isolation requirement is applied, since non-isolated electron candidates  are already suppressed by the identification criteria.
Muon candidates are identified as tracks in the inner detector matched and combined with track segments in the muon spectrometer~\cite{muon2010}. They are required to have 
 $\pt > 20 \gev$ and  $|\eta|<2.4$.  In order to achieve  a sufficient rejection of multi-jet events, muons are required to  be isolated: the scalar sum of the transverse momenta of tracks within a cone of $\dr  \equiv \sqrt{(\Delta\phi)^2 + (\Delta\eta)^2} = 0.2$ around the muon candidate  must be less than 10\% of the transverse momentum of the muon. 
All lepton pairs are required to have a separation of $\drll >0.2$.
The $\Zg$  candidates are selected by requiring exactly two
oppositely charged  leptons of the same flavour. Their invariant mass (\mll )  must be within the range \linebreak $66\gev \leq \mll \leq 116\gev$.
With this selection, 1228767~\Zgee\ and 1678500~\Zgmm\ candidate events are identified.

Jets are reconstructed using the \antikt{}
algorithm~\cite{Cacciari:2008gp} with a distance parameter  \linebreak $R = 0.4$.
The  inputs to the jet algorithm are  topological clusters of energy in the calorimeter~\cite{jespaper2010}. 
The energies and directions of reconstructed jets in data and  simulated events are corrected for the presence of additional proton--proton interactions, 
the position of the primary interaction vertex, the measurement biases induced by calorimeter non-compensation, additional dead material, and out-of-cone effects, using detector simulation and a combination of  in-situ methods~\cite{jespaper2010,jesconf2013}. 
Jets are required to have a transverse momentum above $30 \gev$ and a rapidity of $\ayj < 4.4$.  Jets  closer than $0.5$ in \dr\ to a selected lepton are removed. 
In order to  reject jets from additional proton--proton interactions, the   `jet vertex fraction'  is used. 
This is defined as the \pt\ sum of the tracks associated to the jet which are consistent with originating from the primary vertex divided by the \pt\ sum of all tracks associated to the jet.
The jet vertex fraction is required to be greater than 0.75 for jets with $\rm |\eta|<2.4$.
The residual impact of additional proton--proton interactions  on the distribution of the jet observables has been checked to be correctly simulated such that the unfolded cross sections are expected to be independent of the number of additional interactions.
With this definition,    191566 \Zgee\ and  257169  \Zgmm\   candidate events are selected with at least one jet in the final state.

\begin{table}[t]
\centering
\begin{tabular}{|l|l|l|}
\hline\hline
                           &   \Zgee\                                                        &   \Zgmm\ \\
\hline
lepton \pt\         & $\pt >20\gev$                                             &   $\pt >20\gev$\\
lepton $|\eta |$  & $|\eta|<1.37$~or~$1.52<|\eta |<2.47$     &  $|\eta|<2.4$ \\
 \hline
lepton  charges     & \multicolumn{2}{|c|}{opposite charge}\\
lepton separation  \drll\    & \multicolumn{2}{|c|}{$\drll >0.2$ }\\
lepton invariant mass \mll       & \multicolumn{2}{|c|}{$66\gev \leq \mll \leq 116\gev$ }\\ 
\hline\hline
jet \pt\               & \multicolumn{2}{|c|}{$\ptj >  30\gev$ }\\
jet rapidity  \yj\     & \multicolumn{2}{|c|}{$ \ayj <4.4$ }\\
lepton-jet separation \drlj\    & \multicolumn{2}{|c|}{$\drlj >0.5$ }\\
\hline\hline
\end{tabular}
\caption{Summary of \Zgll\  and jet selection criteria.~\label{t:selection}}
 
\end{table}

\section{Background estimation \label{sec:Background}}

The selected data sample is expected to contain  background events  with two isolated leptons (\ttbar , \db\ and \Zgtau\ events), 
with one isolated lepton (\Wen , \Wmn\ and single-top-quark production)  and without isolated leptons (multi-jet events).
The total expected background  fraction increases with the jet multiplicity (\nj )  from 2\% ($\nj \geq 1$) to 20\% ($\nj \geq 6$). It is dominated by
multi-jet processes, \ttbar\ and \db\ events  for  \linebreak \Zjm\ and by \ttbar\ for  larger jet multiplicities.  
The background is estimated  using simulated samples, with the exception of the multi-jet and \ttbar\ background contributions, which are derived from data. 
For these data-driven background estimates, the  shape of the background contribution to each of the measured distributions is derived from a dedicated 
background-enriched  sample in data. 
The  background-enriched samples have been selected and  normalized  as described below.

The multi-jet background contribution in the \Zeejets\ channel is estimated using a multi-jet enriched data template with two electron 
candidates which both pass a `loose'  selection but 
fail to pass the medium identification requirements~\cite{egamma2010}. The dedicated trigger used for the selection of this sample 
requires two clusters of energy  in the electromagnetic calorimeter with $\pt > 20 \gev$.  
This sample is dominated by jets misidentified as electrons in the final state.
The normalization of this sample to the multi-jet background expected with medium requirements  is extracted 
from a template fit in the invariant mass distribution for medium electrons (\mee ) as follows:
A single combined  fit is performed of the multi-jet template  and the standard simulated signal and non-multi-jet background templates to the measured  spectrum  of the 
invariant mass for medium electrons  in the extended mass range  \linebreak $50 \gev < \mee < 150 \gev$  in the inclusive  selection.
Systematic uncertainties are assessed by varying the mass range and the binning in the fit, by using a different generator (\she\ instead of \alp  ) for  the signal template, 
by varying the electron energy scale and resolution in the simulation and by allowing for a modification of the shape of the mass distribution in the multi-jet enriched sample. 
The multi-jet background to the measured inclusive jet multiplicities 
varies between $(0.65\pm0.23)\%$ for $\nj \geq 1$ and $(1.20\pm0.44)\%$ for $\nj \geq 6$.

In the \Zmmjets\ channel, heavy flavour production (with muons originating from $b$- and $c$-quark decays) and decays-in-flight of pions and kaons
 are the primary source of the multi-jet background, which is highly suppressed by the isolation requirement applied to the muon candidates. 
The multi-jet template is derived from a data sample where both muons fail the isolation requirement.
The normalization factor is obtained by fitting the multi-jet template together with a template composed of the simulated signal and the non-multijet 
background events that pass the signal selection to the  spectrum of the invariant mass of  isolated muons  (\mmm ) measured in data in the range  $40 \gev < \mmm < 150 \gev$.
In contrast to the \Zeejets\ channel, the creation of the template and the normalization is performed  separately for $\nj \geq 0 $, $\nj \geq 1 $  and $\nj \geq 2 $. The normalization factor derived for $\nj \geq 2$  is used for all higher jet multiplicities. The systematic uncertainty is assessed by replacing the multi-jet template 
with one formed from muons passing a loose isolation cut but failing the tight cut used to select signal muons. 
Multi-jet fractions vary between $(0.25 \pm 0.04)\%$  for $\nj \geq 1$  and  $(2.2 \pm 2.2)\%$ for $\nj \geq 6$.

The \ttbar\ background contributions in the \Zlljets\ samples are dominated by events where both $W$ bosons decay leptonically. 
Since the kinematic properties of the jets in the final state are independent of the flavours of the two leptons,
 final states with one electron and one muon can be used to model the 
\ttbar\ background contributions to \Zgee\ and \Zgmm\ selections.
The  \ttbar -enriched sample is selected from data in the $e^{\pm}\mu^{\mp}$ final state with  kinematic requirements analogous to the   
\Zlljets\ selection. 
The dedicated trigger used for the selection of this sample requires an electron with $\pt > 10 \gev$ and a
 muon with  $\pt > 6 \gev$. For each of the observables, the number of \Wjets , \Zjets\ and \db\ events 
expected from simulation in the \ttbar -enriched sample is subtracted. The normalization from the $e^{\pm}\mu^{\mp}$  
to the \epem\ and \mpmm\ final states 
is calculated  from \ttbar\ samples generated with \mca +\her\  and with \pow +\pyt  , separately for each jet multiplicity.   Systematic uncertainties
on the normalization arise from the choice of the generator, uncertainty on the lepton trigger, reconstruction and identification efficiency 
 (see section~\ref{sec:Systematics}) and on the electroweak background subtraction.
The \ttbar\ fractions vary between $(0.80 \pm 0.05)\%$ for \Zej\ and  $(18.6 \pm 7.0)\%$ for \Zejjjjjj\  and between $(0.74 \pm 0.03)\%$ for \Zmj\ 
and  $(18.1 \pm 5.3)\%$ for \Zmjjjjjj  .

\section{Detector-level results \label{sec:DetLevel}}

Measured and expected distributions of the  jet observables have been  compared at the reconstruction level, 
separately in the electron and muon channels. 
As an example, figure~\ref{f:ZllSigBak} shows the dilepton invariant mass in events with at least one jet in the final state, as well as the inclusive jet multiplicity.  
For the signal, both \alp\ and \she\  expectations are shown. 
In this figure,  \Wen , \Zgtau\  and \db\ processes are summarized as `electroweak'  background and  \ttbar\ and single-top processes 
are referred to  as `top'  background.
For figures~\ref{f:ZllSigBak}(a) and~\ref{f:ZllSigBak}(b), the selection has exceptionally  been extended beyond the fiducial invariant mass range, 
in order to demonstrate in addition the reasonable agreement between  data and expectations for dilepton mass sideband regimes with larger background fractions. 
Table~\ref{tab:SigBakZll} shows, for the electron and muon channels separately, the observed number
 of events for the different jet multiplicities in the final state compared to expectations for signal (\alp ) and background processes.
The combined statistical and systematic uncertainties on the total expectation increases from  6\% to 30\% with increasing jet multiplicity.
The  data are consistent with  predictions by the generators \alp\ and \she , which 
gives confidence that the simulated samples, which are used in the unfolding, provide a reasonable
description of the event kinematics and of the detector response.

\begin{table}[t]
\centering
\small 
\begin{tabular}{|c|c|c|c|c|c|c|c|c|}
\hline\hline
\multicolumn{9}{|c|}{\Zgee\ channel}\\ 
\hline\hline
                                           & \zj          & \j       &  \jj    & \jjj    & \jjjj  & \jjjjj   &  \jjjjjj        & \jjjjjjj          \\
 \hline
 \Zgee\               &        1230000  &          190000 &           42000 &            9000 &            1800 &             340 &            60 &            10\\
 \Wen\                       &            450 &             140 &            36 &                       9 &            0.5 &          $<0.5$ &          $<0.5$ &          $<0.5$\\
 \Zgtau\                    &            650 &             110 &            24 &                       6 &            1.4 &            0.2 &          $<0.1$ &          $<0.1$\\ 
 \db\                        &           1800 &            1160 &             500 &             110 &            19 &            3.0 &            0.3 &            0.02\\
 \ttbar , single  top     &           2100 &        1700  &           1190 &            510     &          160   &        50   &           13          &  4           \\
 multi-jet                   &           5000 &            1200 &             300 &            70 &            16  &            4 &            0.8 &            0.3\\
 \hline\hline                                                                                                                                                                       
 total expected           &   1240000 &       190000 &      44000 &       10000 &     2000 &      390 &       70 &            14\\
  \hline
data (4.6 \ifb )             &        1228767 &    191566 &      42358 &       8941 &            1941 &             404 &              68 &              17\\
 \hline\hline

 \multicolumn{9}{|c|}{\Zgmm\ channel}\\ 
 \hline\hline
                                          & \zj          & \j       &  \jj    & \jjj    & \jjjj  & \jjjjj   &  \jjjjjj        & \jjjjjjj          \\
 \hline
 \Zgmm       & 1700000 &  260000 &   57000 &   12000 &    2300 &     400 &      80 &      12\\
 \Wmn         &      120 &      42 &      12 &       3 &     $<0.5$ &  $<0.5$ & $<0.5$ & $<0.5$\\
 \Zgtau         &    1070 &     150 &      36 &       8 &     1.6 &     0.3 &     0.1 &     0.1\\
  \db                &    2400 &    1600 &     680 &     150 &      26 &       4 &     0.4 &    0.10\\
  \ttbar ,    single top       &    2700 &    2100 &    1500 &     640 &     190 &      50 &      17 &       7\\
   multi-jet                 &    3900 &     700 &     290 &      80 &      20 &       6 &       2 &     0.2\\
 \hline\hline     
  total expected & 1700000 &  260000 &   59000 &   13000 &    2500 &     500 &      90 &      20\\
 \hline
data (4.6 \ifb )       & 1678500 &  257169 &   56506 &   12019 &    2587 &     552 &     122 &      31\\ 
\hline\hline
\end{tabular}
\caption{Numbers of events expected and  observed in data  that pass the \Zeejets\  and  \Zmmjets\ selections
as a function of the inclusive jet multiplicity. 
The expected numbers are rounded according to the combined statistical and systematic  uncertainty.
\alp\  has been used to simulate the signal events.
\label{tab:SigBakZll}}
\end{table}

\begin{figure}
  \centering
  \subfigure[ ]{
    \includegraphics[width=0.47\textwidth]{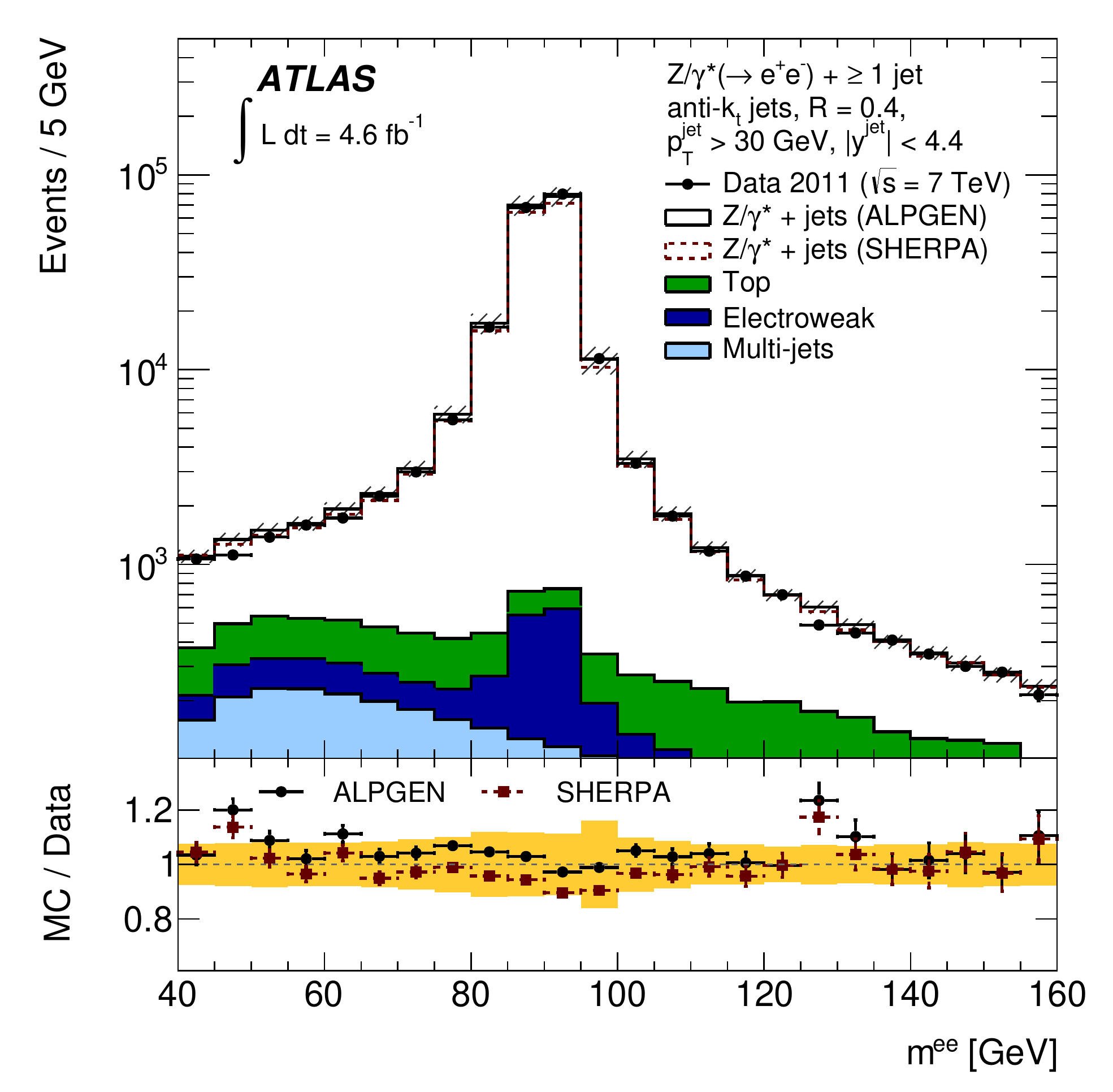}
    \label{fig:Subfigure1}
  }
  \subfigure[ ]{
    \includegraphics[width= 0.47\textwidth]{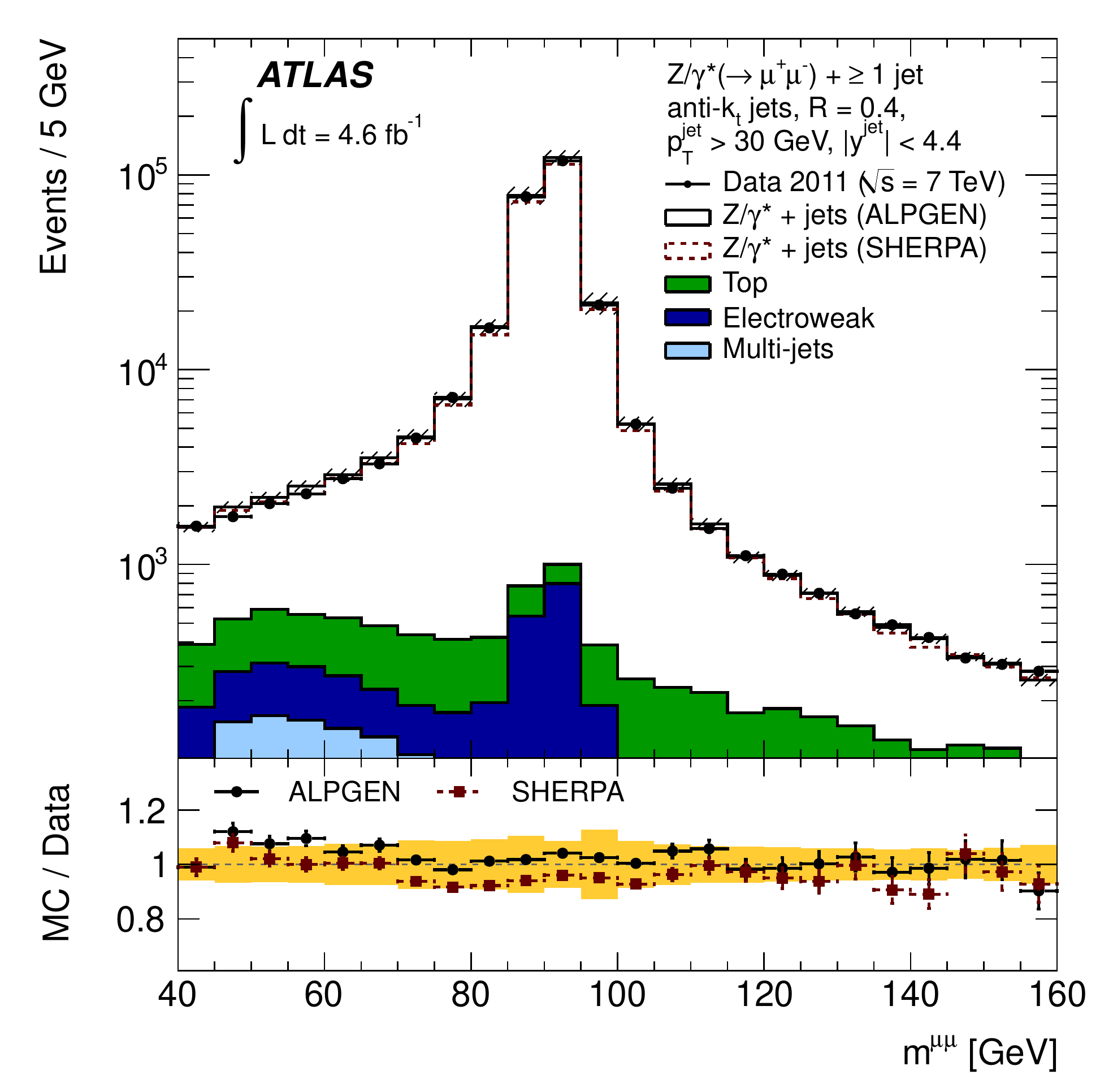}
     \label{fig:Subfigure2}
  }
  \subfigure[ ]{
    \includegraphics[width=0.47\textwidth]{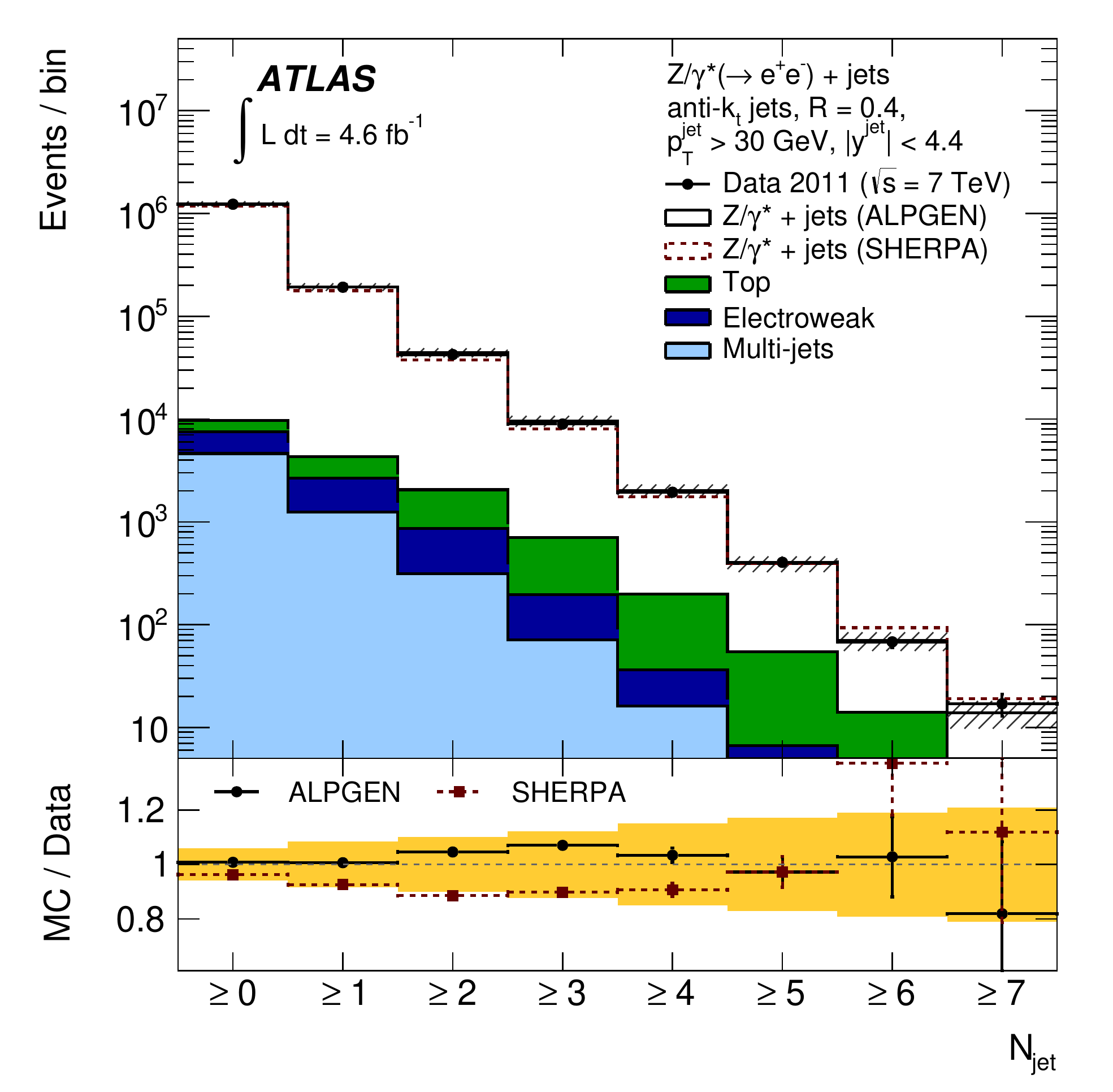}
    \label{fig:Subfigure3}
  }
  \subfigure[ ]{
    \includegraphics[width= 0.47\textwidth]{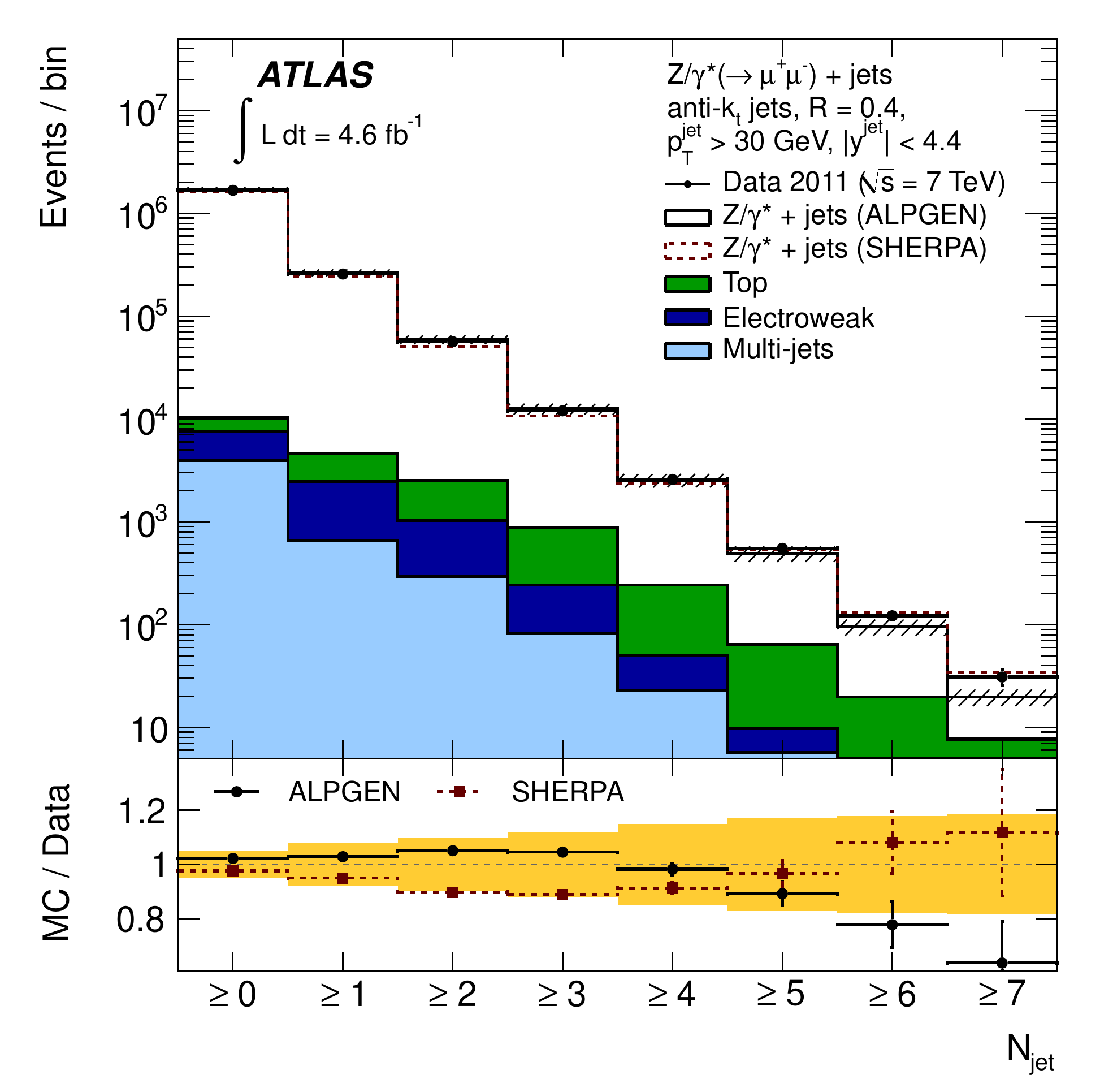}
    \label{fig:Subfigure4}
  }
  \caption{Numbers of events observed in data and predicted in simulation that pass the \hspace{2cm} \Zeejets\  and \Zmmjets\ selection
 as a function of  the invariant mass of the \Zzero\ candidate,  (a) \mee\ and (b) \mmm ,  for events with at least
one jet with  $\ptj > 30 \gev$ and $\ayj <4.4$, and as a function of  the inclusive jet multiplicity, \nj ,  in (c)  di-electron and (d) di-muon events.
The individual contributions of the various backgrounds are also shown, as detailed in the legend.
The hatched band corresponds to the combined  statistical and systematic  uncertainty on the prediction, obtained  using \alp\  to model the \Zjets\ process. 
The error bars on each data point show the statistical uncertainty. The bottom panel shows the corresponding  MC/data ratio. The shaded band corresponds 
to the total systematic  uncertainty and the error bars to the statistical uncertainty on the MC/data ratio.
\label{f:ZllSigBak}}
\end{figure}

\section{Correction for detector effects and combination of channels \label{sec:Unfolding}}

The cross sections in this article are quoted at  the   particle level,  which corresponds to  `dressed'  muons and electrons, 
calculated using  final-state leptons from the Z decay for which collinear radiation in a cone of  $\Delta R < 0.1$  is added to the lepton four-momentum.
Particle  jets are clustered from all final-state particles
(decay length $\rm c\tau > 10~mm$) excluding the dressed \Zzero\ decay products. 
The  phase-space requirements are the same as in the selection at reconstruction level (see table~\ref{t:selection}).

After subtracting the expected background contributions, the  data distributions in each channel are unfolded to the particle level using an iterative 
technique~\cite{DAgostini}.
Response matrices are calculated for each observable, using \Zjets\ samples generated with \alp   .
Before entering the iterative process, the  data are corrected for the fraction of reconstructed events in  the \alp\ sample
which do not match to a  particle-level equivalent.
The number of iterations, typically two or three,  is optimized for each observable using a $\chi^2$ comparison of generated 
and unfolded  reconstructed \Zjets\ events from the generators \she\  and \mca  .  

The uncertainties from the limited number of events in data are propagated
into the particle-level cross sections using a Monte Carlo
method. One thousand pseudo-experimental spectra are generated by fluctuating the
content of each bin according to the statistical uncertainty. The unfolding procedure
is applied to each pseudo-experiment, and the r.m.s. of the results is taken as the
statistical uncertainty.
Systematic uncertainties arising from  the unfolding procedure are estimated by comparing 
with an iterative unfolding based on response matrices and corrections derived from \she  .
The statistical uncertainties of the response matrices are propagated
into systematic uncertainties on the unfolded cross sections using pseudo-experiments.

The cross sections measured in the electron and muon channels are
extrapolated to a common phase-space  region, derived from  table~\ref{t:selection} by extending 
the $\eta$ range of the leptons  to  $|\eta^{\rm{lep}}|<2.5$, using global 
acceptance corrections  derived from \alp\ \Zjets{} Monte Carlo samples, reweighted to the \ctt\ PDF set.  The corrections are of the order of 
14\%  and 5\% for the electron  and muon channel, respectively.
Systematic uncertainties are estimated by comparing with corrections obtained using  the corresponding \she\  \Zjets{} sample and  the original  \alp\ sample. 
Total uncertainties on the corrections are calculated as the quadratic sum of the statistical and systematic uncertainties and amount to 0.2--0.3\%.
The extrapolated cross sections measured in the electron and muon channels are in agreement.
\par
For each observable, the extrapolated cross sections are combined using the averaging procedure introduced in ref.~\cite{comb},
which accounts for systematic uncertainties (bin-to-bin correlated and uncorrelated)  proportional to the 
central values  of the respective cross sections. The weights of the individual cross-section measurements
($\mu^i_k$) in channel $k$ ($ee$ or $\mu\mu$) and  bin $i$ in the combined cross sections ($m^i$ )   
are derived by minimizing the following $\chi^2$ function~\cite{comb}:
\begin{equation}
  \chi^2({\bf m,b}) = \sum_{k,i}\frac{[m^i - \sum_j \gamma^i_{j,k}m^i b_j - \mu^i_k]^2}{(\delta^i_{{\rm stat},k})^2\mu^i_k(m^i - \sum_j \gamma^i_{j,k}m^ib_j) + (\delta^i_{{\rm uncor},k}m^i)^2} + \sum_j b_j^2, 
  \label{eq:combChi2}
\end{equation}
where   $b_j$  denote the shift introduced by a correlated systematic error source $j$ normalized to its respective standard deviation.
The relative statistical and  uncorrelated systematic uncertainties on $\mu^i_k$ are denoted by $\delta^i_{\rm{stat},k}$ and $\delta^i_{\rm{uncor},k}$ and the variable  $\gamma^i_{j,k}$ quantifies the influence of the correlated systematic error source $j$ on the measurement $i$ in the channel $k$.

The following bin-to-bin correlated systematic sources are taken into account:
normalization of the multi-jet background,  lepton  energy scale and resolution,
lepton reconstruction, identification and trigger efficiencies and  normalization
of \ttbar{}, electroweak and single-top background contributions, the latter three treated as correlated between the channels.
Bin-to-bin correlated systematic sources which have the  same impact in both channels do not enter in the combination procedure. These are  the individual components of the  jet energy scale,
 the jet energy resolution, the luminosity, the unfolding procedure, and the extrapolation factor. The uncertainties from these sources on the combined result are taken as the weighted average of the corresponding uncertainties on the electron and muon measurements.

\section{Systematic uncertainties \label{sec:Systematics}}

The kinematic ranges and the binning are chosen such that the statistical uncertainty of the measurement is comparable to or 
smaller than the  systematic uncertainty.
The relative systematic uncertainties on the cross sections measured in each channel are derived for each observable by 
propagating systematic shifts from a set of  independent sources through the response matrices and the subtracted background contributions 
into the unfolded data. The resulting systematic uncertainties for each source in each channel are symmetrized in order to 
mitigate the impact of statistical  fluctuations and  are combined in the averaging  procedure.

The uncertainty on the jet energy scale (JES), determined from the combination of methods based on MC and in-situ techniques used to determine the scale, 
 constitutes the dominant component of the total systematic uncertainty.
It is propagated through the analysis using 14 independent components 
fully correlated in \ptj\  \cite{jespaper2010,jesconf2013}. They account for uncertainties on the different in-situ measurements which enter the jet calibration, 
on the jet flavour and  on the impact of pile-up and  close-by jets.
 The uncertainty on the jet energy resolution, derived from a
comparison of the resolution obtained in data and in simulated dijet events, is  propagated into 
the final cross section  by varying the energy resolution of the simulated jets.
Uncertainties on the normalization of the background expectations, 
for simulated and data-driven background contributions respectively,
are  treated as correlated between bins and  are propagated to the measured cross sections 
by unfolding the data distributions after the subtraction of the systematically shifted background. 
The statistical uncertainties of the background contributions are added quadratically to the statistical uncertainties of the data. 
The uncertainty from the unfolding process is derived from the different components 
discussed in section~\ref{sec:Unfolding}, which are considered to be uncorrelated.
Systematic uncertainties on electron and muon trigger efficiencies, energy scale, resolution, reconstruction and identification  
efficiencies are derived from the comparison of tag-and-probe results in data and simulated events~\cite{egamma2010,muon2010}.

Table~\ref{tab:SysSummary} summarizes
the systematic uncertainties on the \Zjets\ cross sections as a
function of the inclusive jet multiplicity and of \ptj\ of the leading jet
separately for the electron and muon channels.  
The uncertainty on the integrated 
 luminosity of  1.8\% translates into comparable uncertainties on the measured cross sections.
The total uncertainties on the inclusive jet cross sections
range from 8\% for $\nj \geq 1$ to 16--17\%  for $\nj \geq 4$,  dominated by the JES uncertainty.

The uncertainty on cross-section ratios, \Rinmulti \footnote{For simplicity,  $n$ is used in the subscript instead of \nj } , for successive jet multiplicities $n$ is significantly reduced 
due to the strong correlations between  the lepton and jet reconstruction and calibration  uncertainties in 
neighbouring jet bins and amounts to a total of 3--4\%  for \Rintwoone\  and higher multiplicities,  which are of interest in this article, 
dominated by the residual  JES uncertainty.
The large JES uncertainties in the forward region propagate into uncertainties on the unfolded cross sections  at the level of 20\%  (30\%) 
for jet rapidities of $\ayj = 3.0~(4.0)$. This is reflected in large jet energy scale uncertainties on the cross section for events with large rapidity 
distance ($\dyjj$)  between the leading jets,
which combine with the unfolding uncertainties to total uncertainties of 20\% (50\%) for $\dyjj = 3.0~(4.0)$.

\begin{table}
\begin{tabular}{|l||c|c|c|c||c|}
\hline\hline
     \Zgee                         & \j         & \jj        &  \jjj     &  \jjjj        & \ptj\ in [30--500 \gev] \\
\hline\hline
         electron reconstruction  &       2.8\% & 2.8\% &       2.8\% &       2.8\% & 2.6--2.9\%\\
    jet energy scale, resol.   &       7.4\% & 10.1\% &        13\% &        17\% & 4.3--9.0\%\\
                 backgrounds      &       0.26\% & 0.34\% &       0.44\% &       0.50\% & 0.2--3.2\%\\
                       unfolding   &       0.22\% & 0.94\% &       1.2\% &       1.9\% & 1.4--6.8\%\\
\hline
                       total       &       7.9\% & 10.5\% &        13\% &        17\% & 5.5--12.0\%\\

\hline\hline
 \Zgmm        & \j      &  \jj   &  \jjj    &  \jjjj   & \ptj\  in [30--500 \gev]     \\
\hline\hline
muon reconstruction          & 0.86\% & 0.87\% & 0.87\% & 0.88\% & 0.8--1.0\% \\
jet energy scale, resol. & 7.5\% & 9.9\% & 13\% & 16\% & 3.2--8.7\% \\
backgrounds                  & 0.093\% & 0.20\% & 0.41\% & 0.66\% & 0.1--1.9\% \\
unfolding                    & 0.30\% & 0.68\% & 0.52\% & 1.3\% & 0.5--6.2\% \\
\hline
total                        & 7.6\% & 10.0\% & 13\% & 16\% & 4.4--10.2\% \\
\hline\hline
\end{tabular}
\caption{Systematic uncertainties on the cross sections for \Zeejets\  and \Zmmjets\ as a function of the inclusive
jet   multiplicity  and as a function of the transverse momentum, \ptj ,  of the leading jet  for events with at least one jet  with  $\ptj > 30 \gev$ and $\ayj <4.4$.
The rows labelled `electron reconstruction' and `muon reconstruction' include uncertainties on 
trigger, reconstruction and identification, energy scale and resolution. 
\label{tab:SysSummary}}
\end{table}

\section{Theoretical predictions \label{sec:Theory}}

Fixed-order calculations at NLO pQCD for the production of \Zj\  up to \linebreak  \Zjjjj\ are computed using the \bhs\ 
program~\cite{blackhat,bhz3jets,bhz4jets}.
\ctt\ PDFs~\cite{cteq10} are employed and renormalization and factorization scales are set to $\htj /2$, 
where \htj\  is defined event-by-event as the scalar sum of the \pt\ of all stable particles/partons. 
The \antikt\ algorithm with $R=0.4$ is used to reconstruct jets at the parton level.
Systematic uncertainties on the predictions related to PDF uncertainties are computed from the 52 \ctt\ eigenvectors 
at 68\% confidence level~\cite{cteq10}. The uncertainties on the cross sections increase from 1\% for ($\nj \geq 1$) to 3\% for ($\nj \geq 4$) 
and from 1\% to 5\%  with  \ptj\ of the leading jet between 30$ \gev$ and  500$ \gev$. 
Additional changes in the PDFs due to the variation of the input value for the strong coupling constant \as\ at the Z-boson mass scale by $\pm$0.001 
around its nominal value $\as (m_{\rm Z}) = 0.118$ introduce uncertainties on the predicted cross sections in the range of 1\% to 3\% 
for \ZjZjjjj . These are added in quadrature to the PDF uncertainties. Scale uncertainties are estimated by variations 
of the renormalization and factorization scales to one  half and two times the nominal scale. The scale uncertainties for different parton multiplicities are assumed
to be uncorrelated. 
For inclusive calculations, the scale variations translate into variations of the cross section by 4\% to 13\% as \nj\ 
increases and by 2\% to 18\%  with increasing \ptj\ of the leading jet. 
For exclusive final states, the scale uncertainties are 
calculated using the prescription of ref.~\cite{exclscale}. For comparison, the theory/data  ratios presented in section~\ref{sec:Results}
also show   the scale uncertainty resulting from a simple variation
of the renormalization and factorization scales by a factor of two,
assuming the uncertainties to be correlated for different parton
multiplicities. The scale uncertainties constitute  
the dominant uncertainties in most kinematic regions.

The NLO fixed-order calculations at the parton level are corrected to the
particle level  for the underlying event  and for effects of fragmentation and of QED final-state radiation (QED-FSR). 
Parton-to-hadron correction factors ($\rm \delta^{had}$)  approximately account for non-perturbative
contributions from the underlying event and fragmentation into particles. For each observable, the
correction factor is estimated using simulated \Zjets\ samples, produced with  \alp\  with the \her\  cluster fragmentation  in which \jim\
models  the underlying event  using  the \aue -\cts ~\cite{auet2lo} tune. It is calculated as the bin-by-bin ratio
 of the nominal distribution at the particle level to the one obtained by turning off both the interactions between proton remnants 
and the fragmentation in the simulated samples. 
The non-perturbative corrections are also computed using \alp\  samples, this time interfaced to \pyt , where the correction
corresponds to the combined effect of string fragmentation and of the underlying event predicted  by the \per ~\cite{perugia2011} 
tune. The difference is taken as a systematic uncertainty.
The combined nominal correction  is  7\% in the low \ptj\ region  and decreases with increasing 
\ptj\ towards zero. 
 The correction factors for the inclusive \nj\ distributions are about 3--4\%. 
Nonperturbative corrections for quantities calculated with several jets
include implicitly the corrections for all jets.
The statistical and the symmetrized systematic uncertainties on
$\rm \delta^{had}$ are added in quadrature
to the total uncertainty from the \bhs\  calculation.

The QED-FSR correction factors ($\rm \delta^{QED}$) are determined using \Zjets\ samples produced with 
the  \alp\  generator,
interfaced to  \pho ~\cite{photos}, by calculating the expected
cross sections both with  the lepton four-momentum before final-state photon
radiation  (`Born level'), and with dressed leptons.
The correction factors are about 2\% for the
electron and muon channels. They do not show a significant \nj\
dependence and are stable with respect to  the jet 
rapidity and for large jet transverse momentum.
Systematic uncertainties are derived by  comparing with $\rm \delta^{QED}$ obtained using a \Zjets\ 
sample produced with the \she\  generator~\cite{sherpa} which generates QED-FSR using the YSF method~\cite{yfsm}.
The differences between the two  predictions are usually at the per-mille level.

\section{Results and discussion  \label{sec:Results}}

For each observable, the spectrum measured  in data  is  unfolded to the particle level. 
After extrapolation and combination of electron and muon channels, the results are  compared with calculations from \bhs , corrected to the particle level,
 and with  predictions by \alp , \she\  and \mca  .
Both \alp\ and \she\  employ matrix elements for up to five partons. Higher multiplicities are generated by the parton shower.
In contrast, \mca\  generates the Drell--Yan process at NLO precision, which includes the real emission of one additional parton. 
All higher parton multiplicities are generated by the parton shower.
Inclusive and differential cross sections for   \linebreak \Zlnj\  are compared with \bhs\   
fixed-order pQCD calculations for  \Znp , which provide a NLO estimate for the respective parton multiplicity, including 
the real emission of one additional parton.
Measured cross sections as a function of the jet multiplicity and their ratios are detailed in table~\ref{t:DataTheoMulti}.
Tabulated values of all observed results are available in the Durham HEP database~\cite{hepdata}.

\subsection{Jet multiplicities\label{sec:jetmulti}}

Figure~\ref{fig:ElMuComb0}(a) presents the absolute cross sections for inclusive jet multiplicities
 for up to seven hadronic jets in the final state. 
The  ratios \Rinmulti\  of cross sections for two successive multiplicities, presented in  figure~\ref{fig:ElMuComb0}(b),  provide a more precise 
measurement of the QCD process, due to the cancellation  of part of the systematic uncertainty.
The  data are consistent with  \bhs\  calculations and with
predictions of the generators \alp\ and \she  .
 The \mca\  parton shower underestimates the observed rate for additional 
jet emission by a factor of two,  which leads to large offsets to the data for higher jet multiplicities.
For this reason, in subsequent figures the \mca\  predictions are only shown for \Zlj\ selections, where the parton corresponding to  the NLO
real emission can be expected to yield a reasonable description of the kinematics.

Exclusive jet multiplicities at the LHC are expected to be described by means of two benchmark patterns, 
 `staircase scaling' with \Rmulti\ constant and `Poisson scaling'  with \Rmulti\ inversely proportional  to $n$~\cite{scaling89,scaling12}, which provide 
limiting cases for certain kinematic conditions. 
While for high multiplicities a flat exclusive jet multiplicity ratio is derived from the non-abelian nature of QCD FSR, 
at low multiplicity the jet multiplicity ratio is flat due to the  combined effect of a Poisson-distributed multiplicity distribution and parton density suppression~\cite{scaling12}.  
The emission of the first parton should be suppressed more strongly than the subsequent parton emissions.
The underlying Poisson scaling is expected to emerge after introducing  large scale differences between the 
core process (\Zjex ) and the \ptj\ of the second leading jet.
Two selections are chosen to test the two benchmark scenarios: (a) the standard \Zjets\ selection and (b)  
events where the leading jet has a transverse momentum in excess of $150\gev$. 

 Figure~\ref{fig:ElMuComb1}(a) presents the ratios \Rmulti\  of cross sections for two successive exclusive multiplicities
for the standard \Zjets\ selection.  The comparitatively large scale uncertainties on the pQCD predictions result from 
 the prescription of ref.~\cite{exclscale}, assuming  the scale variations to be uncorrelated across the jet multiplicities. 
For comparison,  the total uncertainty calculated using a naive scale variation, and a reduced uncertainty 
that does not include any scale uncertainty are also shown.
The data are consistent with the central values of the \bhs\  calculations 
and with predictions by the  generators  \alp\ and \she .   The cross-section ratios show an approximately linear dependence on the jet multiplicity with a small slope. 
A linear fit $\Rmulti  = R_0+\frac{{\rm d}R}{{\rm d}n}\cdot n$  of the  observed multiplicity ratio  starting with \Rtwoone\   yields 
$R_0 = 0.232 \pm 0.009 $   and   ${\rm d}R/{\rm d}n = -0.011  \pm  0.003$. The uncertainties include a systematic contribution,  derived from  a series of  fits 
to systematic variations of the multiplicity ratio.
The flat staircase pattern provides an  acceptable  approximation  of the observed scaling behaviour for the standard \Zjets\ selection.
The observation is consistent with results presented in~\cite{vjetscms} on the smaller data set collected in 2010.

\begin{figure}[t]
\begin{center}
  \subfigure[ ]{
    \includegraphics[width=0.47\textwidth]{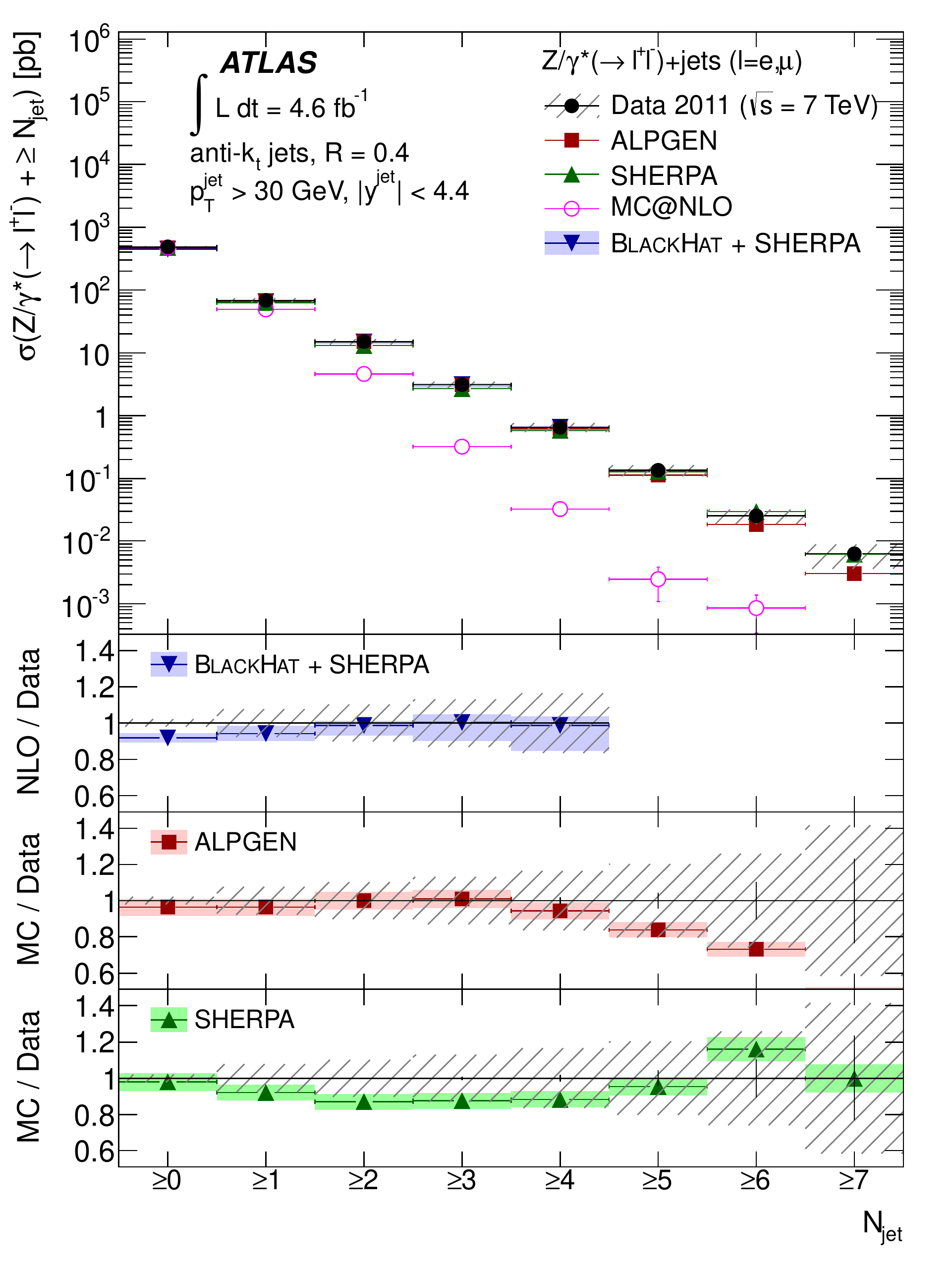}
    \label{fig:ElMuComb0_a}
  }
\subfigure[ ]{
  \includegraphics[width= 0.47\textwidth]{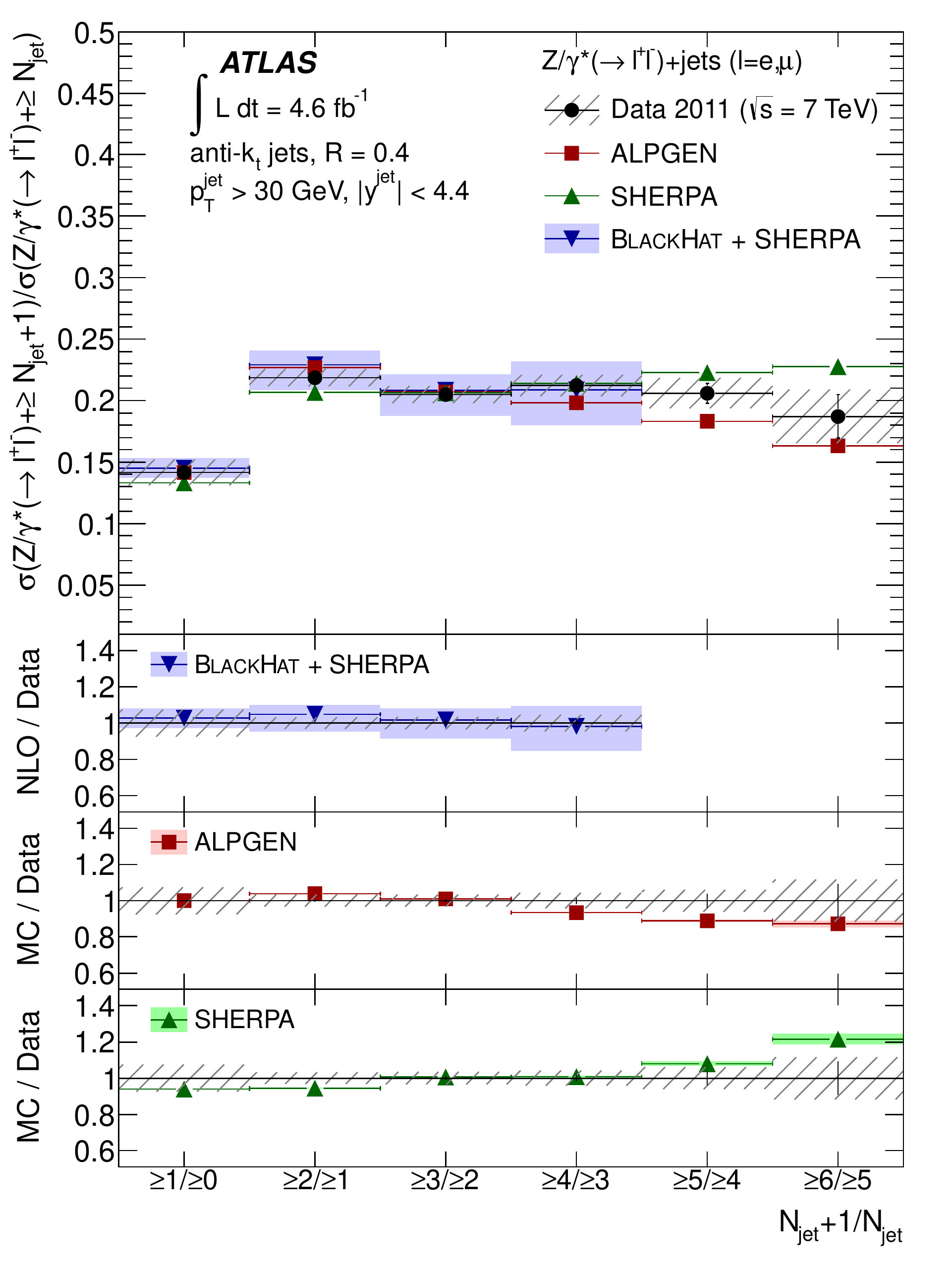}
   \label{fig:ElMuComb0_b}
  }
\end{center}
\caption{(a) Measured cross section for \Zlljets\ as a function of the inclusive jet multiplicity, \nj , 
 and (b)  ratio of cross sections for successive inclusive jet multiplicities.
The data are compared to NLO pQCD predictions from \bhs\  corrected to the particle
level, and the \alp , \she\  and \mca\  event generators (see legend for details). The error bars indicate the
statistical uncertainty on the data, and the hatched (shaded) bands the statistical and systematic
uncertainties on data (prediction) added in quadrature.
\label{fig:ElMuComb0}}
\end{figure}

Figure~\ref{fig:ElMuComb1}(b)  presents the exclusive jet multiplicity ratio for events where the leading jet has 
a transverse momentum in excess of $150\gev$. 
The observed ratio  \Rmulti\ is now steeply increasing towards low jet
multiplicities, a pattern described by the central values of the \bhs\ calculations,
by the generator  \alp\ and approximately also by \she  .
The observed cross-section ratios have been fitted with a  pattern expected from a Poisson-distributed jet multiplicity with the expectation value $\bar{n}$, 
$\Rmulti = \frac{\bar{n}}{n}$.
The Poisson scaling provides a good overall description of the jet multiplicity observed in data for the selected kinematic regime, with
$\bar{n} = 1.02 \pm 0.04$, where the uncertainty includes statistical and systematic components.

The scaling pattern is also investigated for a  preselection  typically employed in the selection of particles produced 
via vector boson fusion (VBF).
Figure~\ref{fig:ElMuComb3} presents the absolute cross section as a function of the exclusive jet
multiplicity and \Rmulti\ after requiring two jets with   $\mjj >350\gev$
and $\dyjj >3.0$, in the following referred to as `VBF preselection'.
The data are consistent with the \bhs\  prediction.  
 \she\  describes the multiplicity well whereas \alp\ overestimates  \Rthreetwo . 


\begin{figure}[t]
\begin{center}

 \subfigure[]{
   \includegraphics[width=0.47\textwidth]{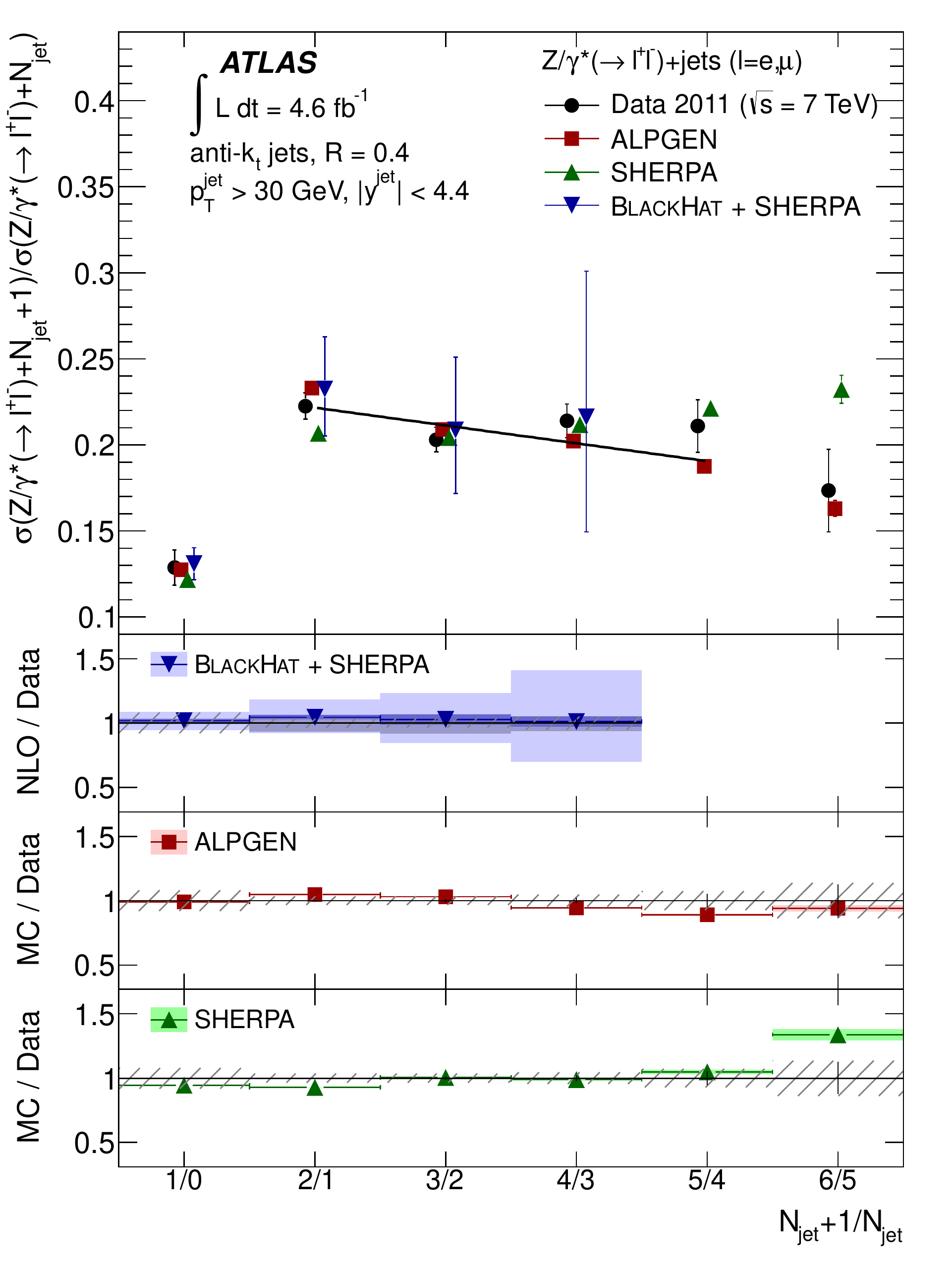}
    \label{fig:ElMuComb3_a}
  }
 \subfigure[]{
   \includegraphics[width=0.47\textwidth]{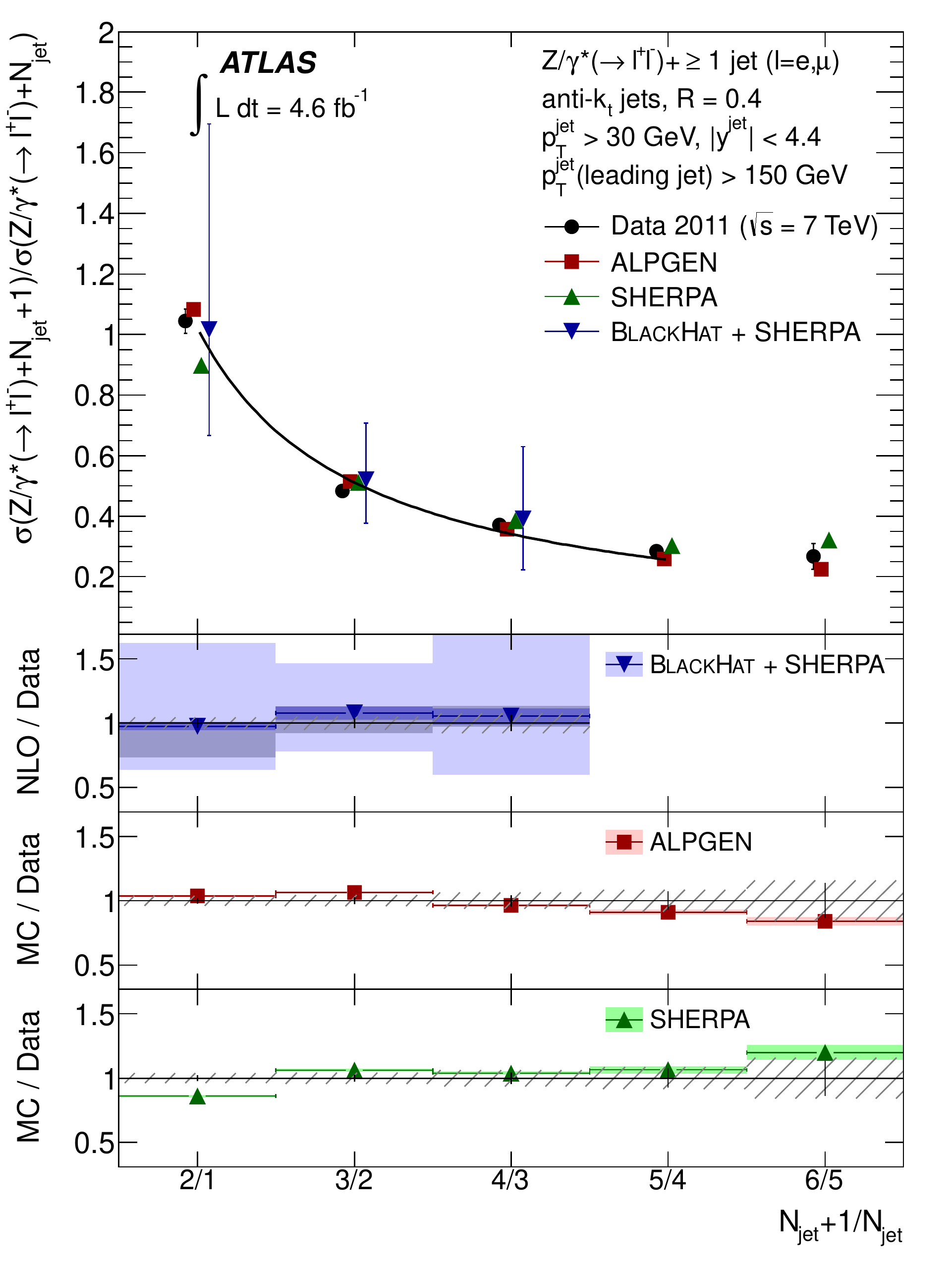}
    \label{fig:ElMuComb3_b}
  }
\end{center}
\caption{(a) Ratio of cross sections for successive exclusive jet multiplicities, \nj ,  in  events selected with
 the standard selection and (b)  in events with at least one jet with  $\ptj > 150 \gev$ and $\ayj<4.4$.
The data are compared to NLO pQCD predictions from \bhs\  corrected to the particle
level, and the \alp , \she\  and \mca\  event generators (see legend for details). The error bars indicate the
statistical uncertainty on the data, and the hatched (shaded) bands the statistical and systematic
uncertainties on data (prediction) added in quadrature.
The shaded bands  on the theory calculations show the 
 systematic uncertainty excluding the scale uncertainty (dark shaded) and the total systematic  uncertainties using the 
naive approach (medium shaded) and the nominal approach (light shaded) to derive the scale uncertainty (see section~\ref{sec:Theory}). 
The figures include (a) a linear fit $\Rmulti  = R_o+\frac{{\rm d}R}{{\rm d}n}\cdot n$ 
in the range $R_{2/1} <R_{(n+1)/n}<R_{5/4} $ and (b) a Poisson fit  $\Rmulti = \frac{\bar{n}}{n}$ to the data points, with the free parameters
$R_o$, $\frac{{\rm d}R}{{\rm d}n}$ and $\bar{n}$.
\label{fig:ElMuComb1}}
\end{figure}

\begin{figure}[t]
\begin{center}

 \subfigure[]{
   \includegraphics[width=0.47\textwidth]{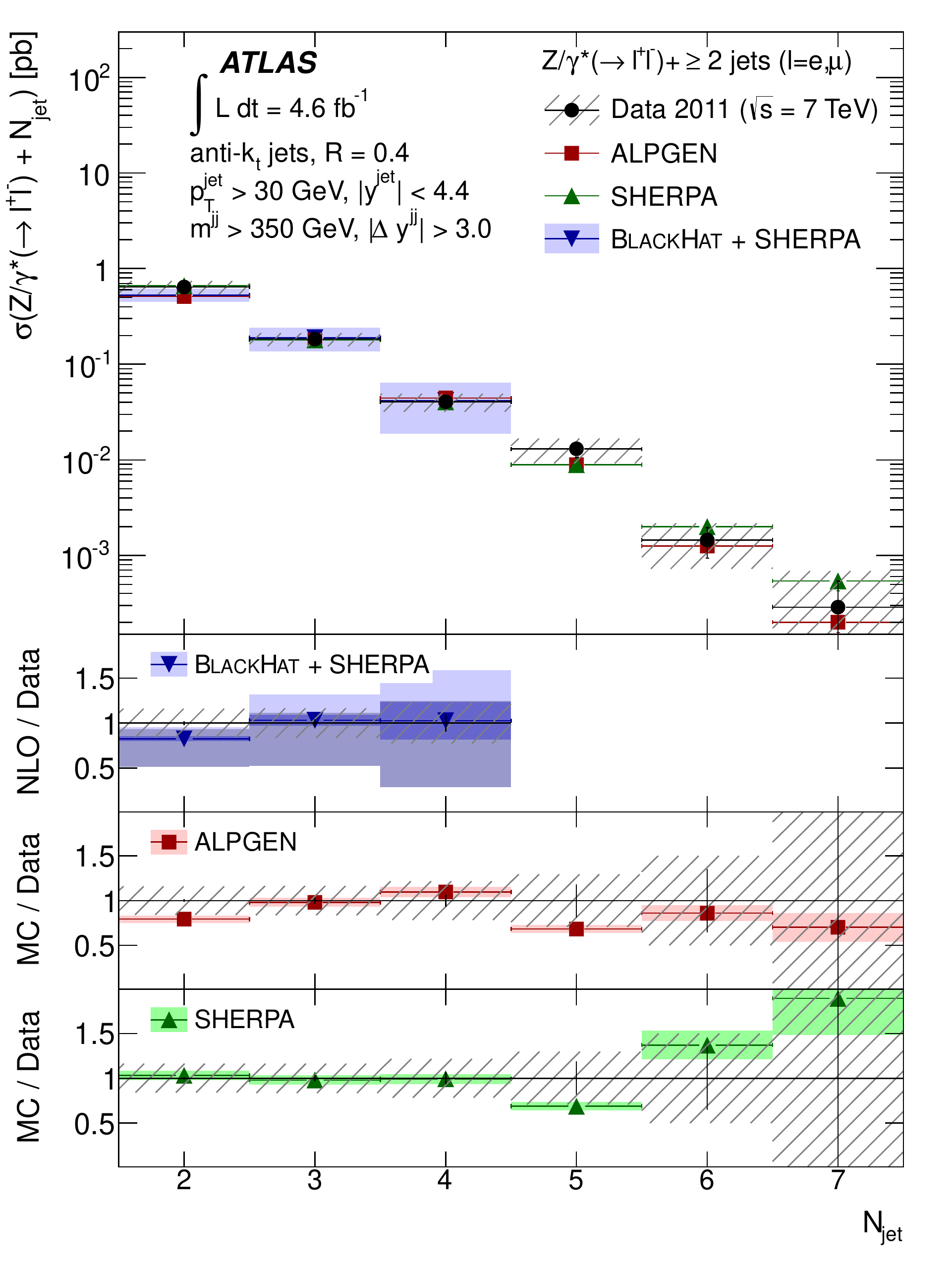}
    \label{fig:ElMuComb3_a}
  }
 \subfigure[]{
   \includegraphics[width=0.47\textwidth]{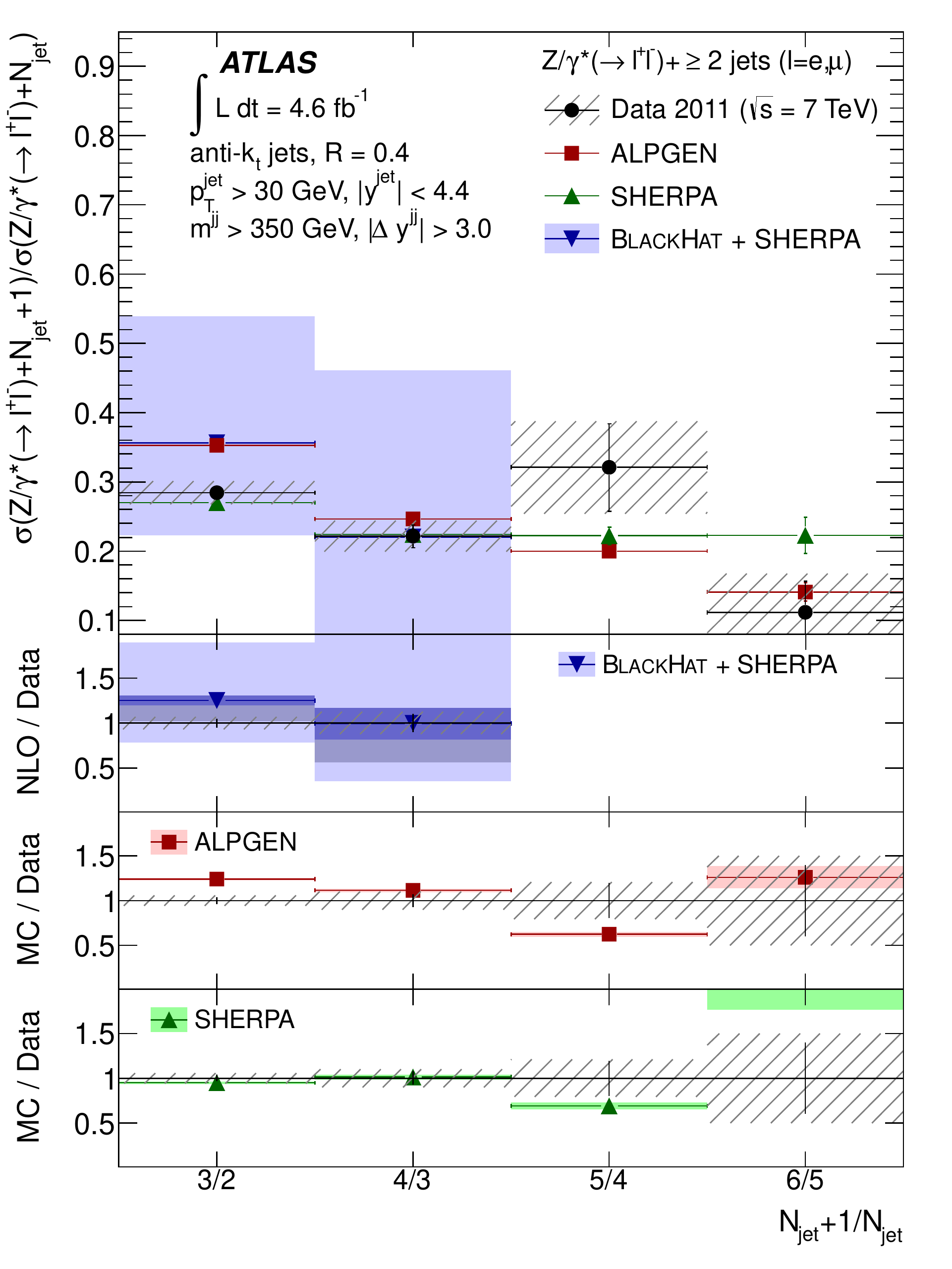}
    \label{fig:ElMuComb3_b}
  }
\end{center}
\caption{(a) Measured cross section for \Zlljets\ as a function of the exclusive jet multiplicity, \nj ,   and (b) 
ratio of the cross sections for two  successive  multiplicities,  in events passing the VBF
preselection (at least two jets with  $\ptj > 30 \gev$ and $\ayj<4.4$ and $\mjj >350\gev$
and $\dyjj >3.0$ for the two leading jets).
The other  details are as in Figure~\ref{fig:ElMuComb1}.
\label{fig:ElMuComb3}}
\end{figure}


\subsection{Jet transverse momentum} 

Differential cross sections with respect to the jet transverse momentum,   \ptj , provide a test of pQCD over  a large
kinematic range. 
In particular, when \ptj\ exceeds the scale given by the gauge boson mass,  NLO/LO  K-factors  can be large due to 
the presence of QCD corrections of the order of  $\alpha_s\ln^2(\ptj /\mz)$~\cite{gkfactors}.
In addition,  higher-order electroweak corrections are expected to  
reduce the cross section  with increasing transverse momentum of the
\Zzero\ boson candidate, by 5--20\%  for $100\gev < \ptll\
<500\gev$~\cite{ewcorrection}.

Figures~\ref{fig:ElMuComb4a}  and~\ref{fig:ElMuComb4b}  show the cross section as a function of \ptj\ 
of the first, the second, the third and the fourth leading jet (in descending order of \ptj ) 
for events with at least one, two, three and four jets in the final state, respectively. The cross sections  are normalized to the inclusive 
\Zgll\ cross section, which reduces the systematic uncertainties connected to lepton identification and integrated luminosity. 
The fixed-order NLO predictions by \bhs\  are consistent with the data for all jet multiplicities.

\begin{figure}[t]
\begin{center}
\subfigure[ ]{
  \includegraphics[width=0.47\textwidth]{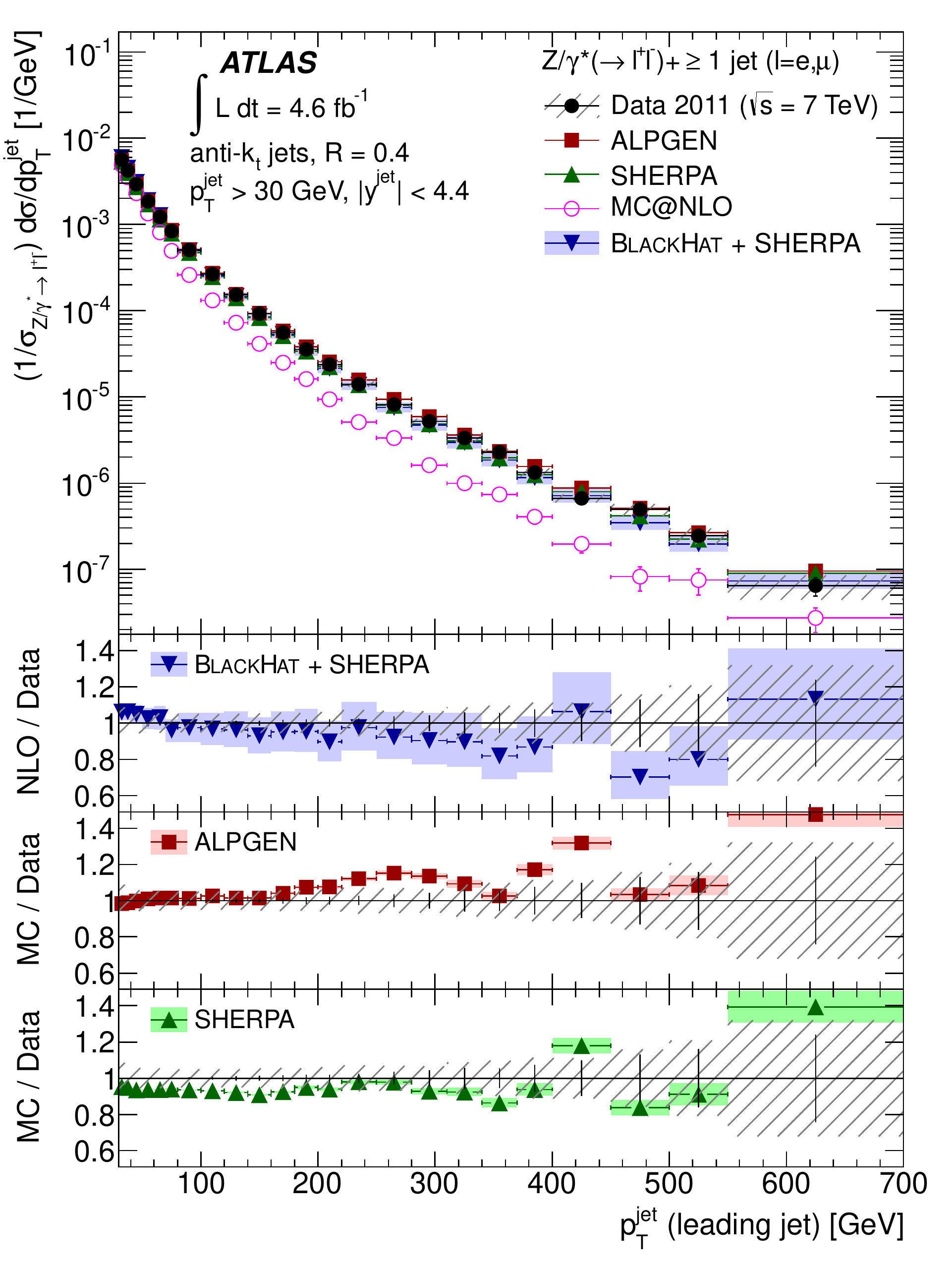}
  \label{fig:Subfigure1}
}
\subfigure[ ]{
  \includegraphics[width= 0.47\textwidth]{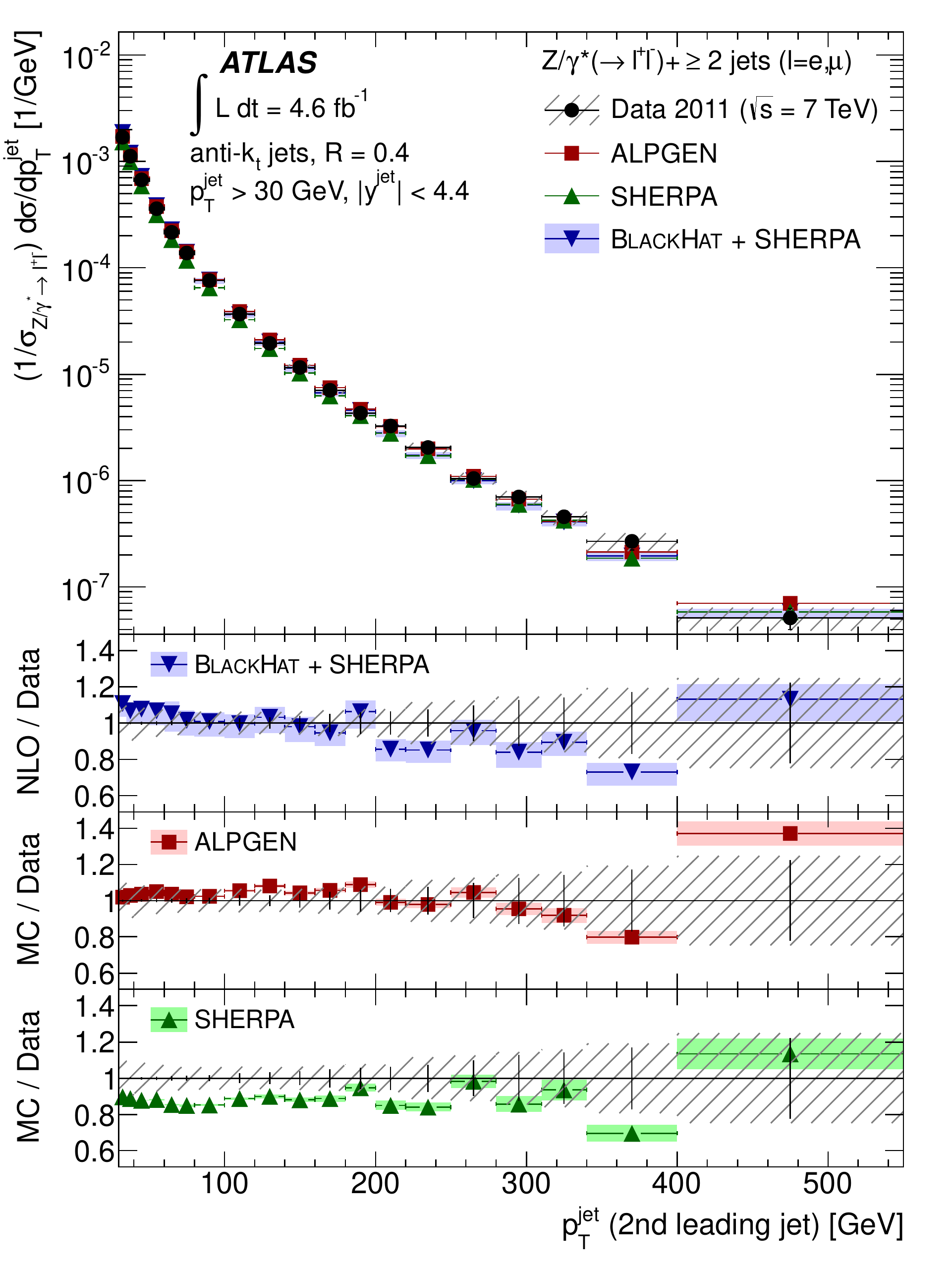}
  \label{fig:Subfigure2}
}
\end{center}
\caption{(a) Measured cross section for \Zlljets\ as a function of the
  transverse momentum,  \ptj ,  of the leading jet for events with at least one jet with 
 $\ptj > 30 \gev$ and $\ayj<4.4$ in the final state and  (b)  as a function of  \ptj\  of the second  leading jet  for events with at least two jets.
The cross sections are normalized to the inclusive \Zgll\ cross section.
The other  details are as in Figure~\ref{fig:ElMuComb0}.
\label{fig:ElMuComb4a}}
\end{figure}


For the leading jet,  the precision of the measurement exceeds the precision of the theory prediction.
While \alp\ predictions for the \ptj\ spectrum of the second to fourth leading  jet are consistent with
the data, the \ptj\ spectrum of the leading jet is predicted to be too hard for larger values of \ptj .
 \she\  is characterized by offsets to the data at the level of 5--15\%, consistent with the observations
presented in figure~\ref{fig:ElMuComb0}(a)  for the inclusive jet cross section.
\mca\  predicts a too soft  \ptj\  spectrum, resulting in a discrepancy with the data  by one order of magnitude for large \ptj .
This is attributed to the fact that the  fraction of events with a second resolved jet, which in \mca\   is modelled via the parton shower,
increases considerably with \ptj\ of the leading jet (see  figures~\ref{fig:ElMuComb1}(a) and~\ref{fig:ElMuComb1}(b) for small and larger \ptj\ (leading jet)). 
A too soft \ptj\ spectrum of the parton shower  will hence result  in an increasing  discrepancy between the \mca\  prediction and  the data.

\begin{figure}[t]
\begin{center}
\subfigure[ ]{
  \includegraphics[width=0.47\textwidth]{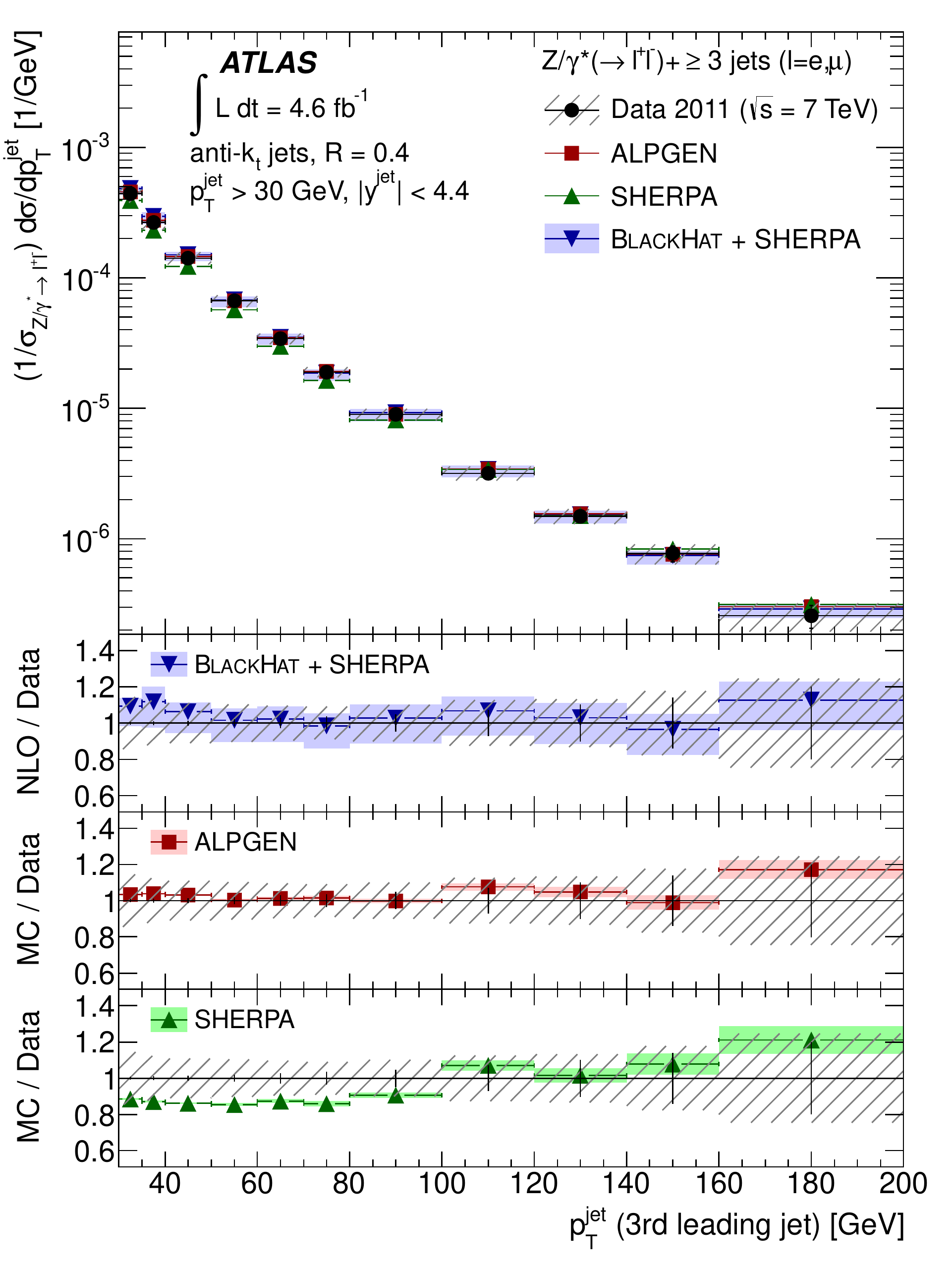}
  \label{fig:Subfigure1}
}
\subfigure[ ]{
  \includegraphics[width= 0.47\textwidth]{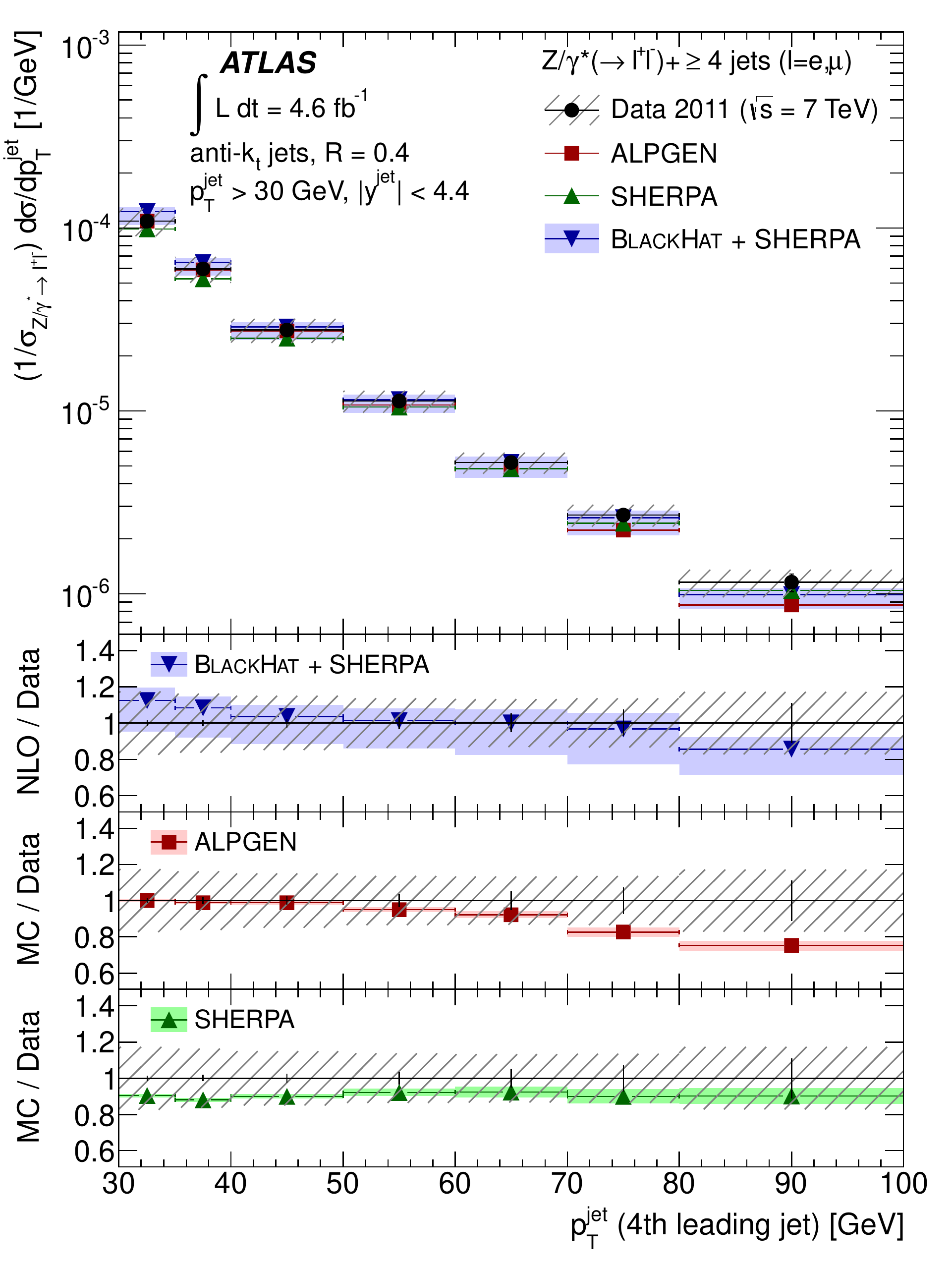}
  \label{fig:Subfigure2}
}
\end{center}
\caption{(a) Measured cross section for \Zlljets\ as a function of the transverse momentum, \ptj , of the third leading jet  for events with at least three jets with 
 $\ptj > 30 \gev$ and $\ayj<4.4$ in the final state and  (b)  as a function of  \ptj\  of the fourth  leading jet  for events with at least four jets.
The cross sections are normalized to the inclusive \Zgll\ cross section.
The other  details are as in Figure~\ref{fig:ElMuComb0}.
\label{fig:ElMuComb4b}}
\end{figure}

\begin{figure}[t]
\begin{center}
\subfigure[]{
  \includegraphics[width=0.47\textwidth]{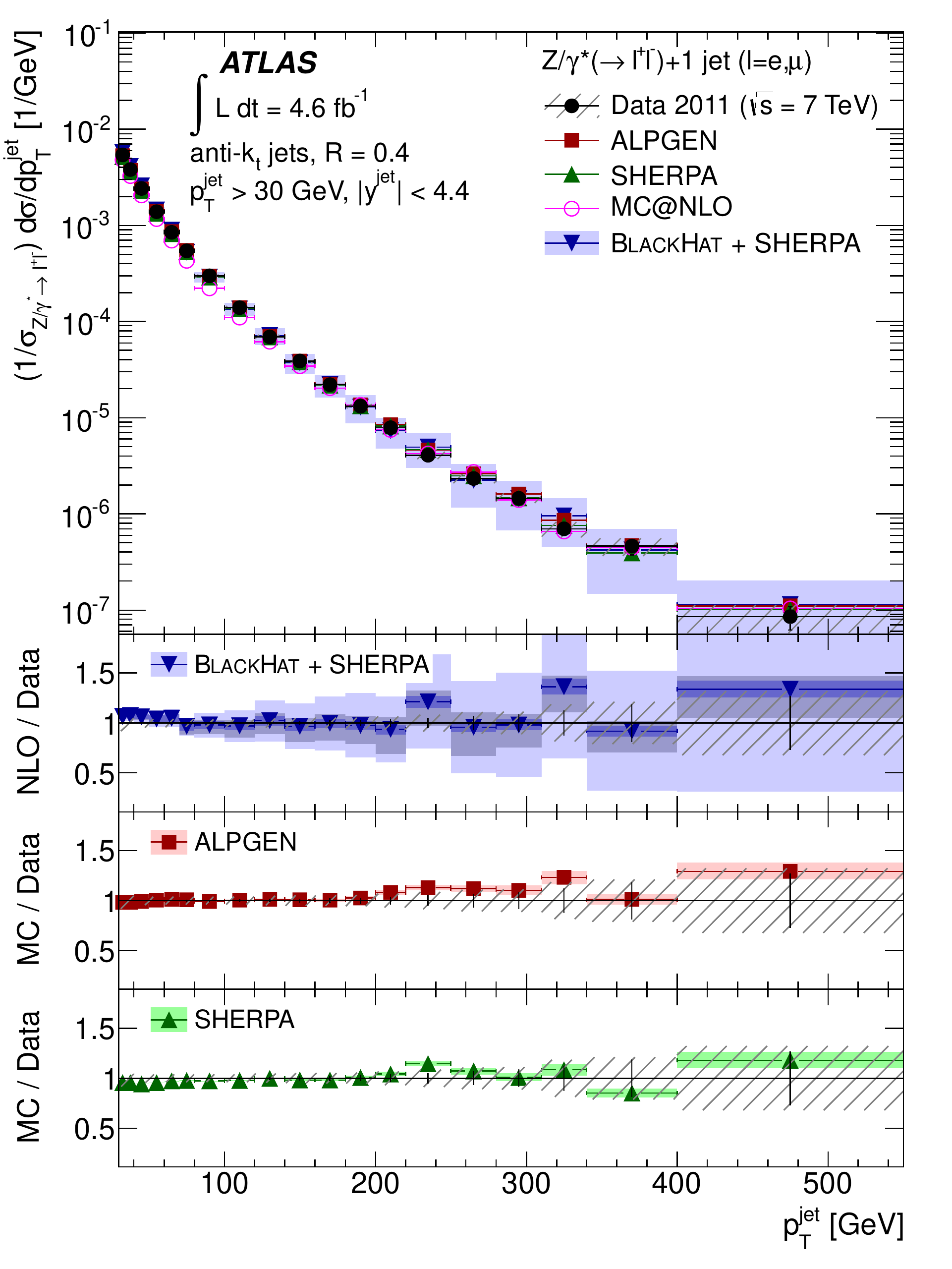}
  \label{fig:Subfigure1}
}
\subfigure[]{
  \includegraphics[width= 0.47\textwidth]{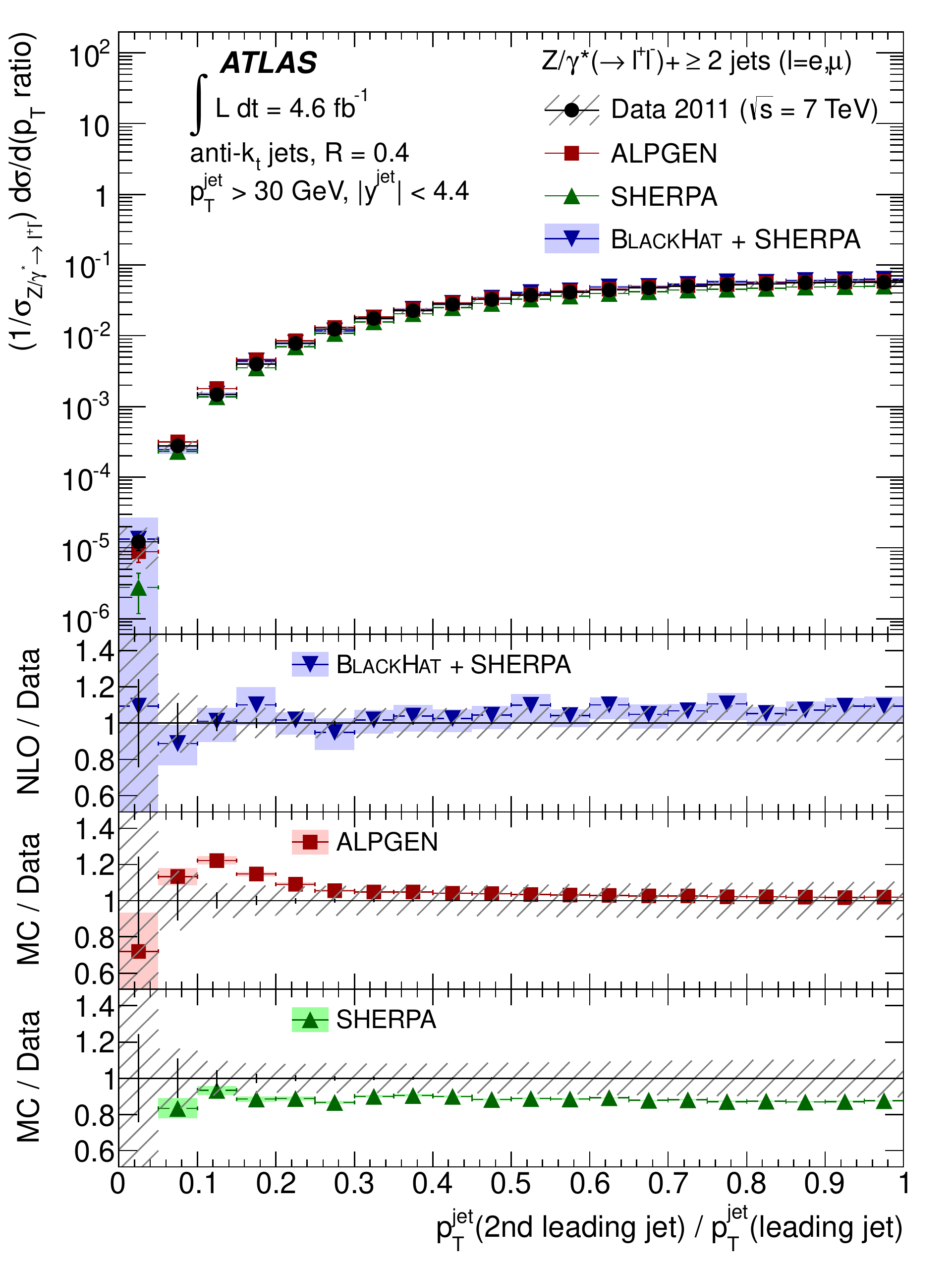}
  \label{fig:Subfigure2}
}
\end{center}
\caption{(a) Measured cross section for \Zlljets\ as a function of the jet  transverse momentum,  \ptj ,  for events with exactly one jet with 
 $\ptj > 30 \gev$ and $\ayj<4.4$ in the final state and  (b) as a function of the  ratio of \ptj\  of the  second leading jet  to \ptj\ of the  leading jet
for events with at least two jets. 
The cross sections are normalized to the inclusive \Zgll\ cross section.
The other  details are as in Figure~\ref{fig:ElMuComb1}.
\label{fig:ElMuComb5a}}
\end{figure}

Figure~\ref{fig:ElMuComb5a}(a) shows the cross section as a function of \ptj\ of the leading jet, normalized to the inclusive \Zgll\ cross section,  when a veto on a second jet is applied.
A better agreement between the predicted and observed cross-sections is observed.
For  events with at least two jets, 
figure~\ref{fig:ElMuComb5a}(b) shows cross section as a function of  the \ptj\ ratio of the two leading jets, normalized to the inclusive \Zgll\ cross section.
 \alp\ overestimates the cross section for events 
with a \ptj\ ratio of the leading jets in the range of 0.1--0.2.  \she\  underestimates the  cross section as a function of the \ptj\ ratio  by $\approx$15\%,
consistent with the results presented in figure~\ref{fig:ElMuComb0}(a).

In a  complementary approach,  the cross section is measured as a function of the \pt\ of the
recoiling \Zzero\  boson,  reconstructed from the  momenta of the two leptons.
The results are  presented in figure~\ref{fig:ElMuComb5b}  for both the inclusive and the exclusive  \Zjex\ selection, normalized to the inclusive \Zgll\ cross section.
Both \alp\  and \she\ predict a too hard \ptll\ spectrum, in particular in the inclusive case.
The discrepancy with the data is comparable to 
 the expected  higher-order electroweak corrections~\cite{ewcorrection} although higher-order QCD corrections could 
equally account for this.
The \bhs\   \Zj\ fixed-order calculation for the inclusive final state  is too soft 
whereas for the exclusive final state  the central predictions are closer to the observed spectrum.
This result is attributed to missing higher jet multiplicities in the fixed-order calculation and will be discussed in more detail in section~\ref{sec:inclusive}.
The comparison with \bhs\    yields  no indication for missing higher-order electroweak corrections in the large-\ptll\  region.
Consistent with the results presented for the \ptj\ spectrum of the leading jet,  \mca\  describes the exclusive \Zjex\ final state better than the corresponding inclusive final state.

\begin{figure}[t]
\begin{center}
\subfigure[]{
  \includegraphics[width=0.47\textwidth]{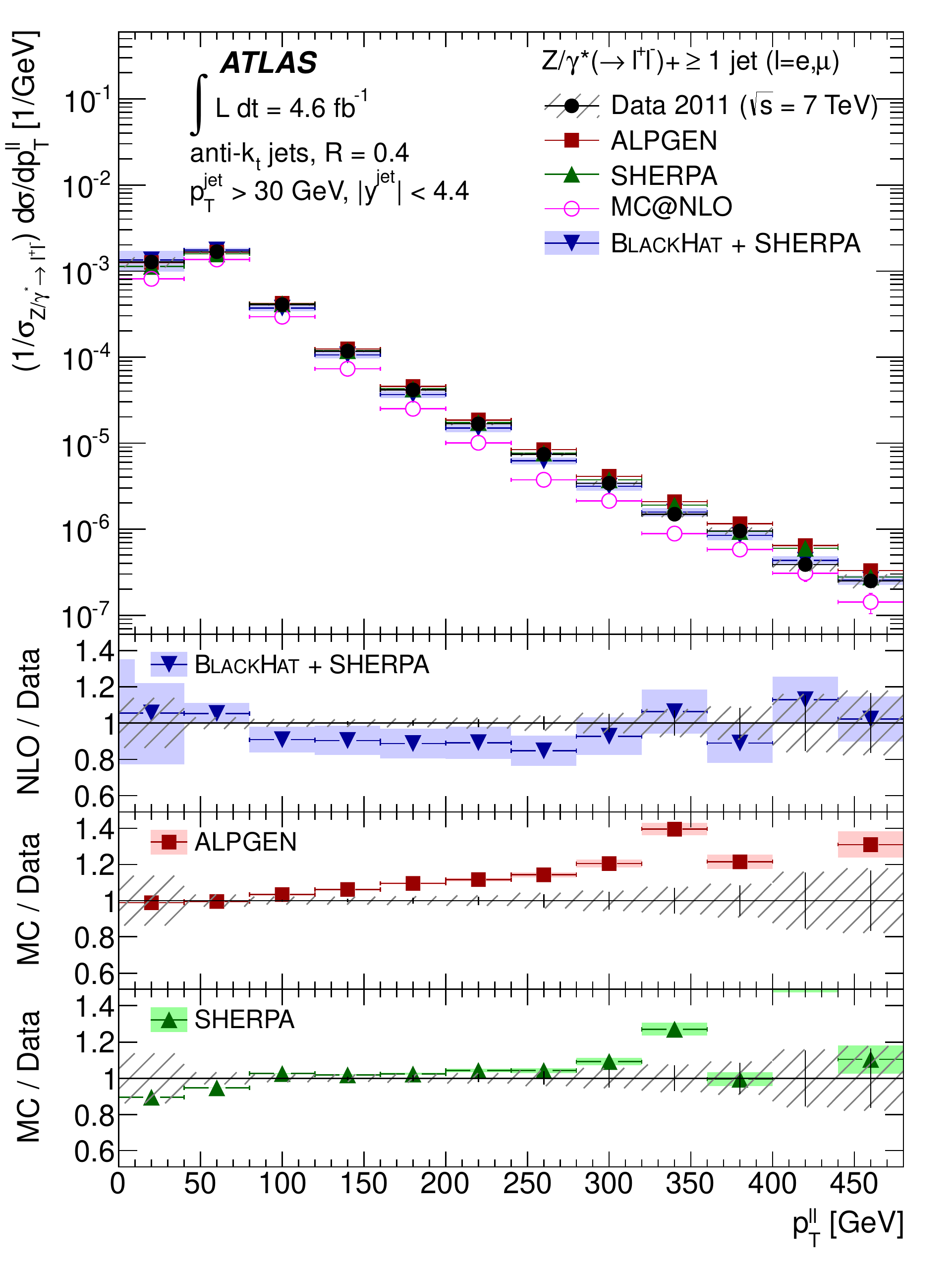}
  \label{fig:Subfigure1}
}
\subfigure[]{
  \includegraphics[width= 0.47\textwidth]{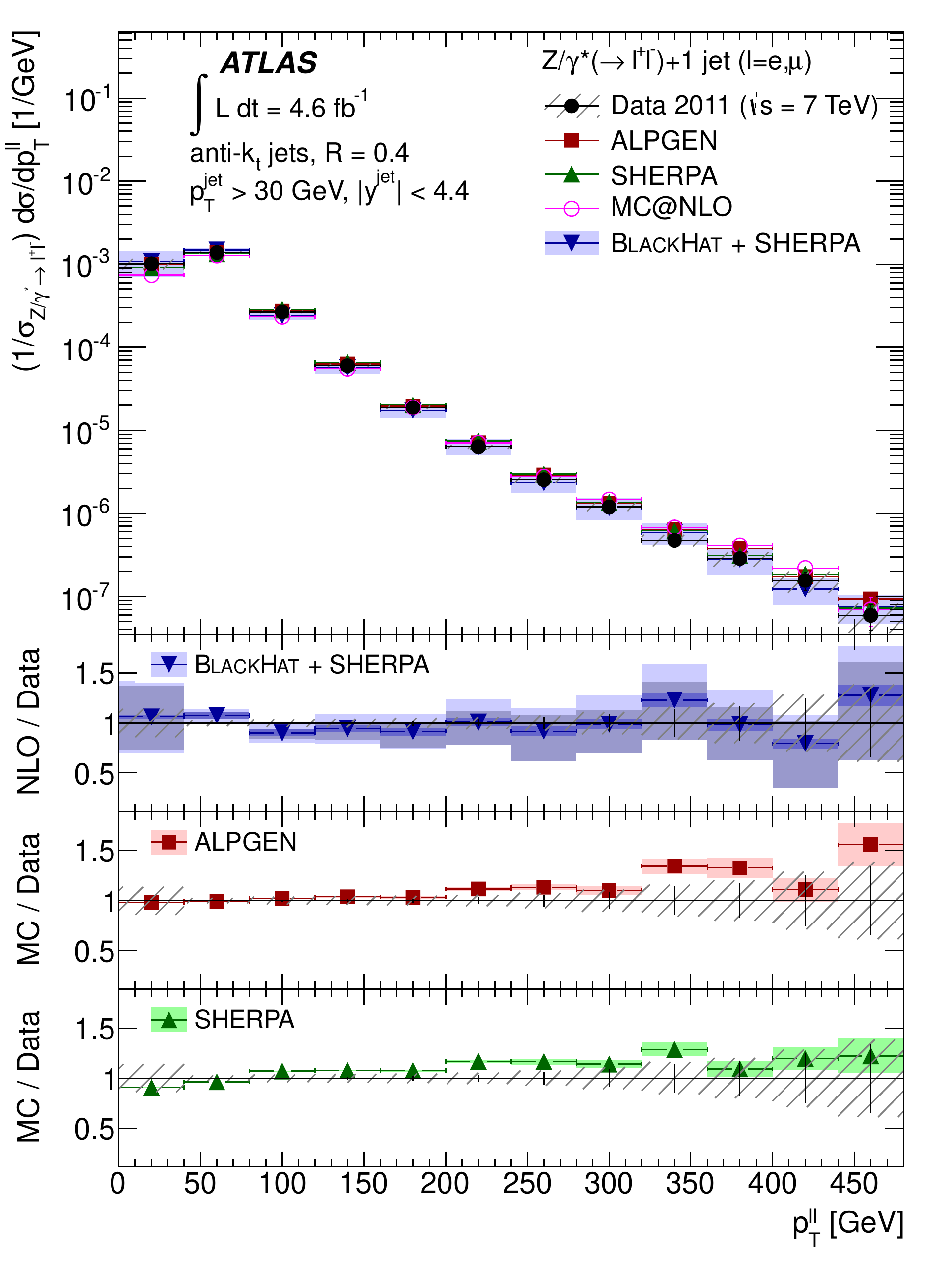}
  \label{fig:Subfigure2}
}
\end{center}
\caption{(a) Measured cross section for \Zlljets\ as a function of the transverse momentum of the \Zzero\ candidate, \ptll ,  in events with at least one jet  with 
 $\ptj > 30 \gev$ and $\ayj<4.4$ in the final state and (b)  as a function of \ptll\ in events with exactly one jet.  
The cross sections are normalized to the inclusive \Zgll\ cross section.
The other  details are as in Figure~\ref{fig:ElMuComb1}.
\label{fig:ElMuComb5b}}
\end{figure}


\subsection{Angular distributions}

Figures~\ref{fig:ElMuComb6a} and~\ref{fig:ElMuComb6b}   show the rapidity spectrum of the four
leading jets,  normalized to the inclusive \Zgll\ cross section.  Both  \bhs\  and \she\  predict rapidity spectra for the leading jet that are somewhat wider than observed in the data.
 \alp\  predictions are  compatible with the measurements.

Figure~\ref{fig:ElMuComb7a} presents  the separation in rapidity,   \dyjj , and the invariant mass,  \mjj ,  
of the two leading jets, normalized to the inclusive \Zgll\ cross section.
The predictions by  \bhs\  and \alp\ are consistent with the data.
 \she\  overestimates the cross section for large \dyjj , consistent with the too wide rapidity spectra.

\begin{figure}[t]
\begin{center}
\subfigure[]{
  \includegraphics[width=0.47\textwidth]{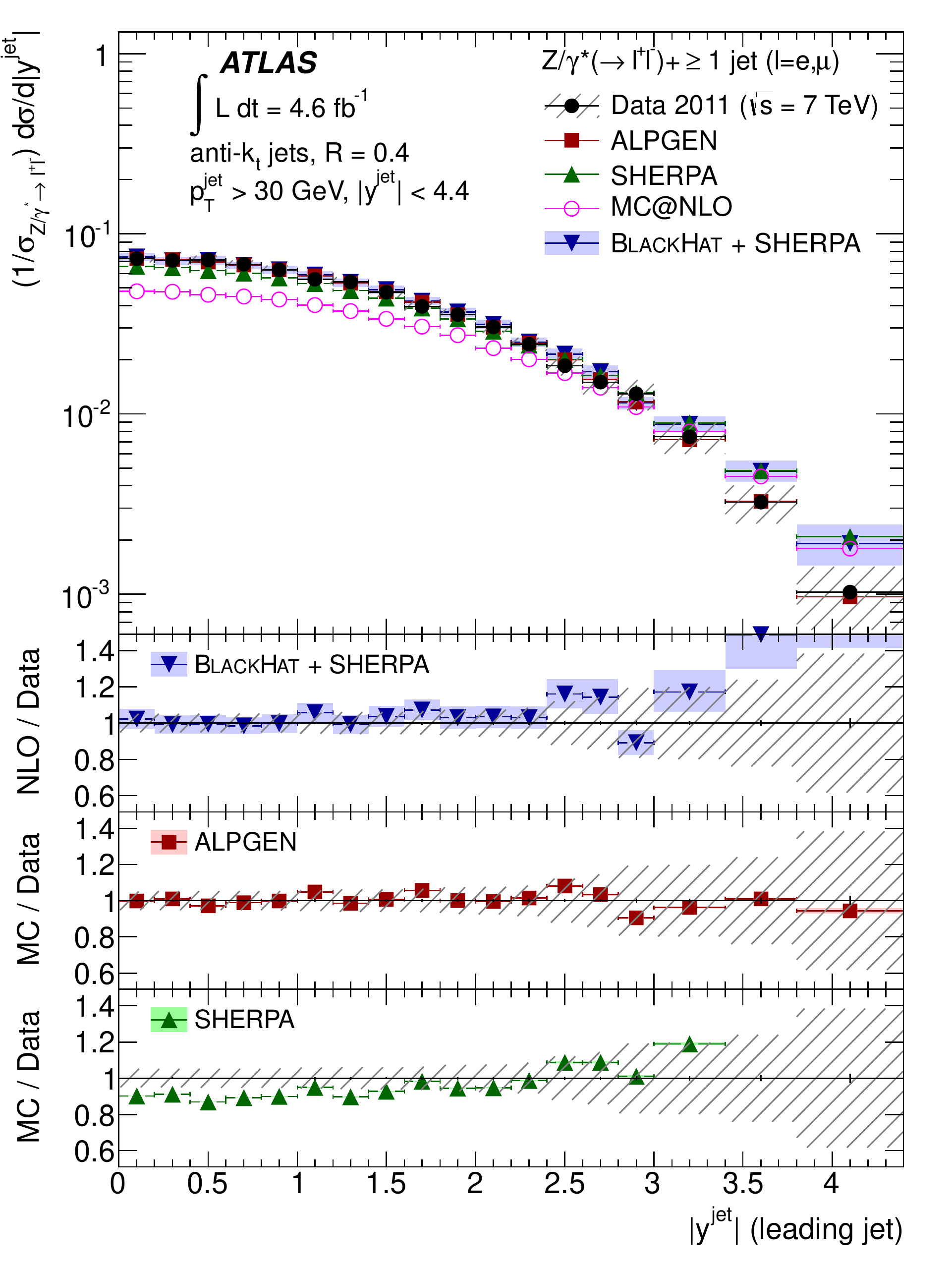}
  \label{fig:Subfigure1}
}
\subfigure[]{
  \includegraphics[width= 0.47\textwidth]{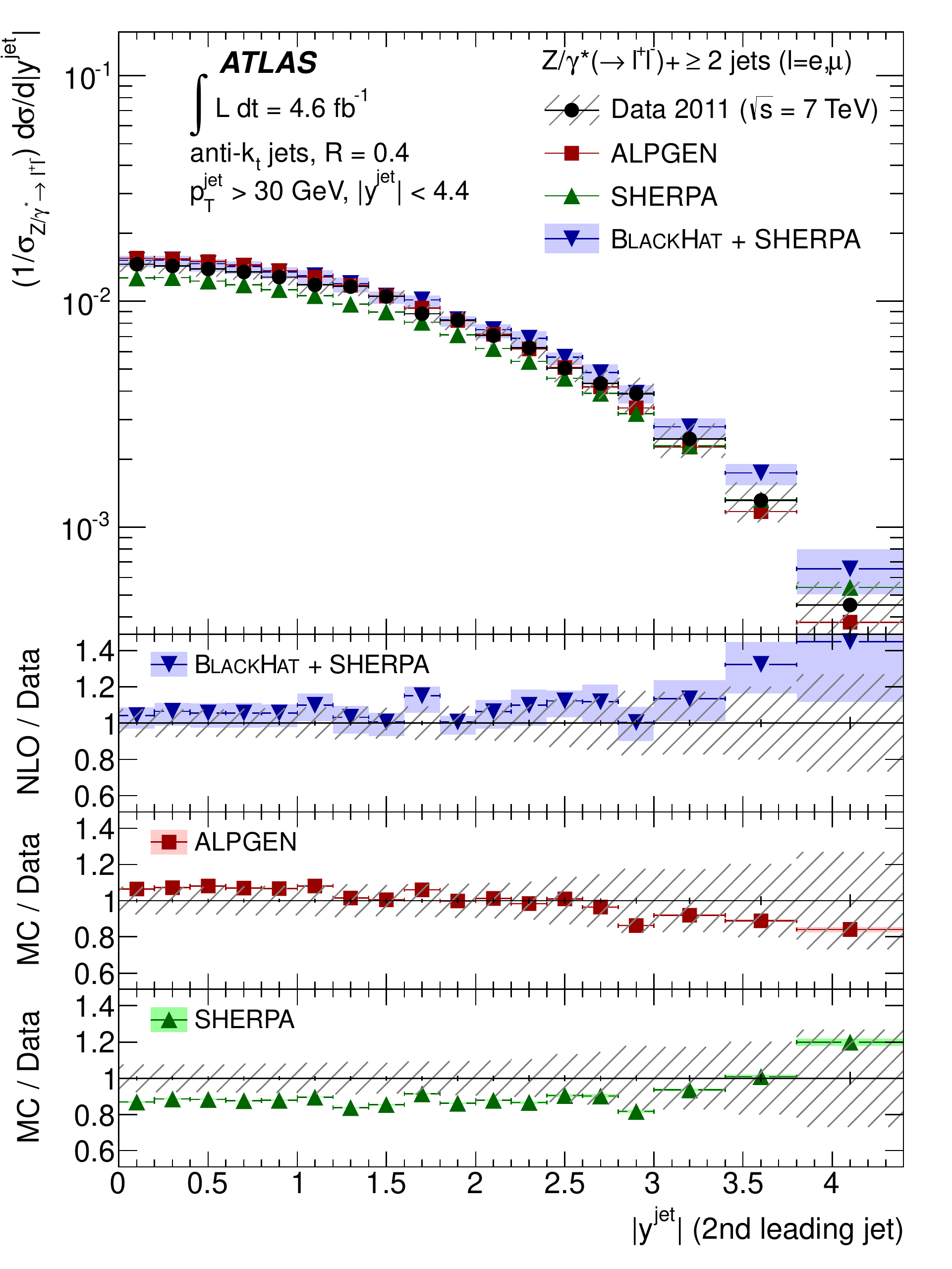}
  \label{fig:Subfigure2}
}
\end{center}
\caption{(a) Measured cross section for \Zlljets\ as a function of the absolute value of the rapidity,  \ayj ,  of the leading jet  for events with at least one jet with 
 $\ptj > 30 \gev$ and $\ayj <4.4$ in the final state and  (b)  as a function of  \ayj\  of the second leading  jet  for events with at least two jets.
The cross sections are normalized to the inclusive \Zgll\ cross section.
The other  details are as in Figure~\ref{fig:ElMuComb0}.
\label{fig:ElMuComb6a}}
\end{figure}

Differential jet cross sections as a function of  angular distances (\dpjj\ and \drjj )   between the two leading jets are presented in  
figures~\ref{fig:ElMuComb7b}(a) and~\ref{fig:ElMuComb7b}(b),  respectively,  normalized to the inclusive \Zgll\ cross section.
The azimuthal distance is well modelled by \alp\ and by \bhs .
The tendencies observed  in the modelling of the distance in $\phi$ and in rapidity are  reflected 
in the measurement of the \dr\ spectrum of the leading jets.
 \she\  models a too flat spectrum for both $\Delta \phi$  and   \dr .
The offset of 15\%  of the \she\ prediction from the observed cross section in the bulk of the data in figures~\ref{fig:ElMuComb7a} and~\ref{fig:ElMuComb7b} is consistent with 
 the results presented in figure~\ref{fig:ElMuComb0}(a) for the inclusive \Zljj\ cross section.

\begin{figure}[t]
\begin{center}
\subfigure[]{
  \includegraphics[width=0.47\textwidth]{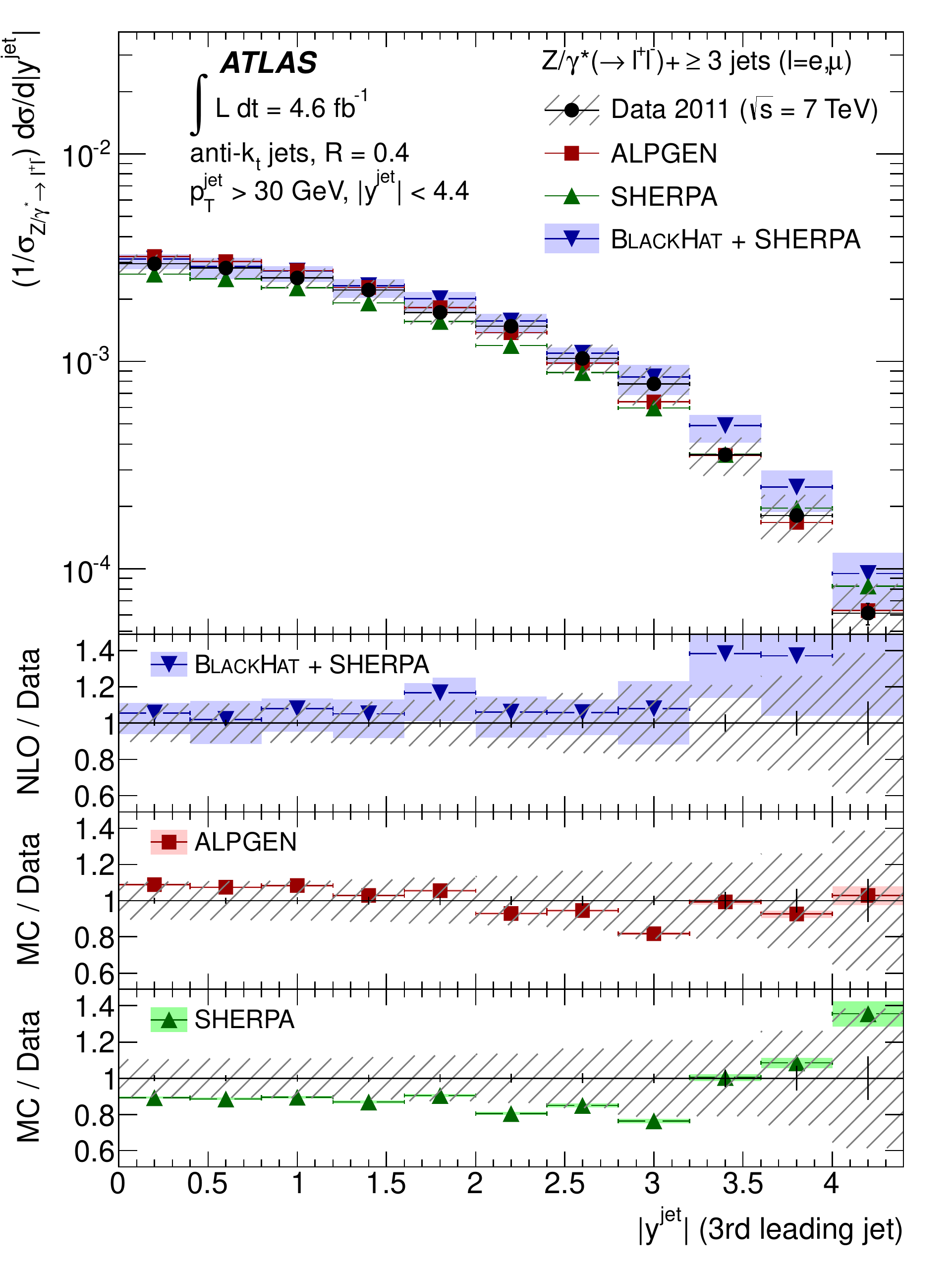}
  \label{fig:Subfigure1}
}
\subfigure[]{
  \includegraphics[width= 0.47\textwidth]{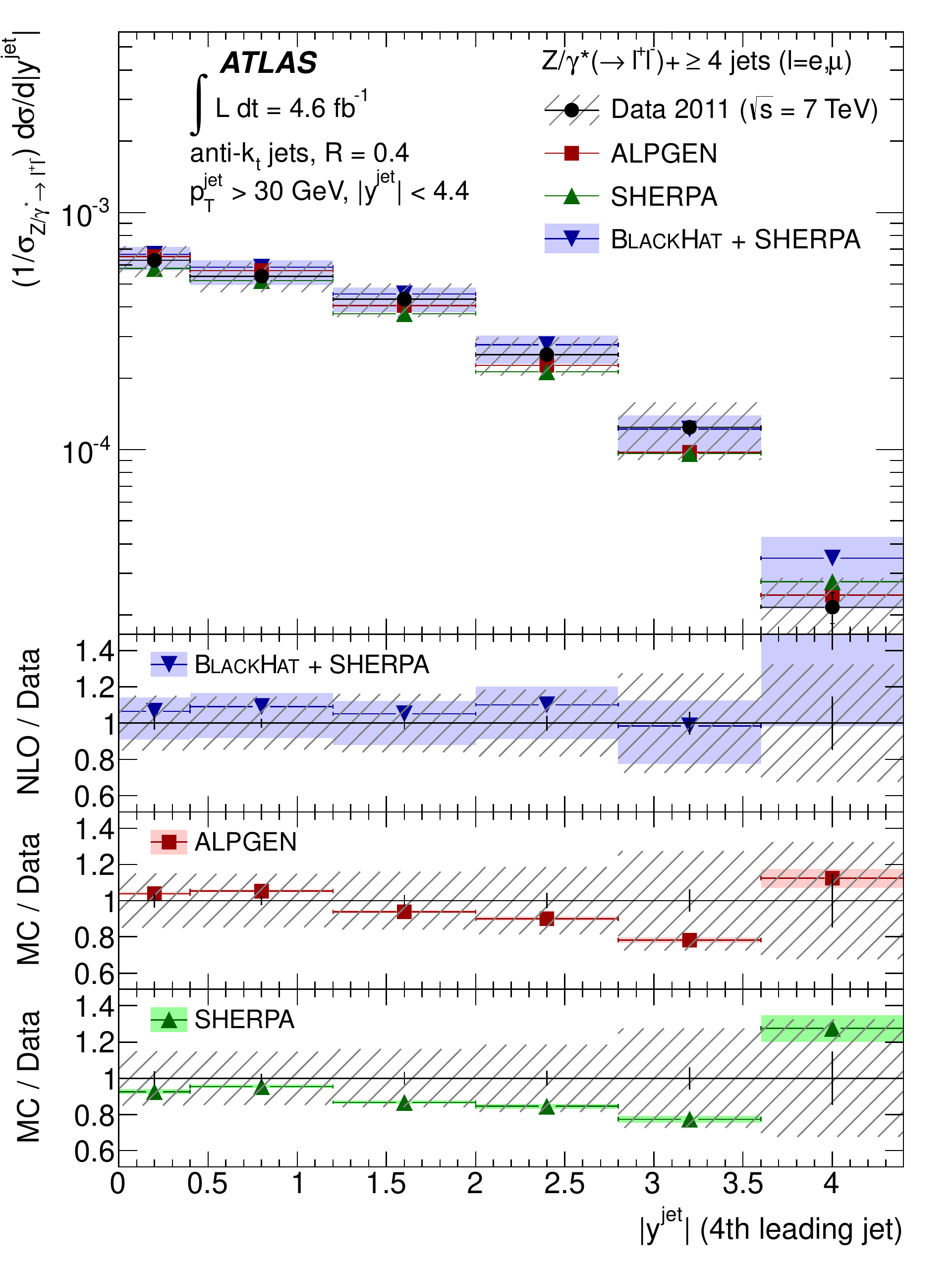}
  \label{fig:Subfigure2}
}
\end{center}
\caption{(a) Measured cross section for \Zlljets\ as a function of the absolute value of the rapidity,  \ayj ,  of the third jet  for events with at least three jets with 
 $\ptj > 30 \gev$ and $\ayj <4.4$ in the final state and  (b)  as a function of \ayj\ of the fourth  jet  for events with at least four jets.
The cross sections are normalized to the inclusive \Zgll\ cross section.
The other  details are as in Figure~\ref{fig:ElMuComb0}.
\label{fig:ElMuComb6b}}
\end{figure}

\begin{figure}[t]
\begin{center}
\subfigure[]{
  \includegraphics[width=0.47\textwidth]{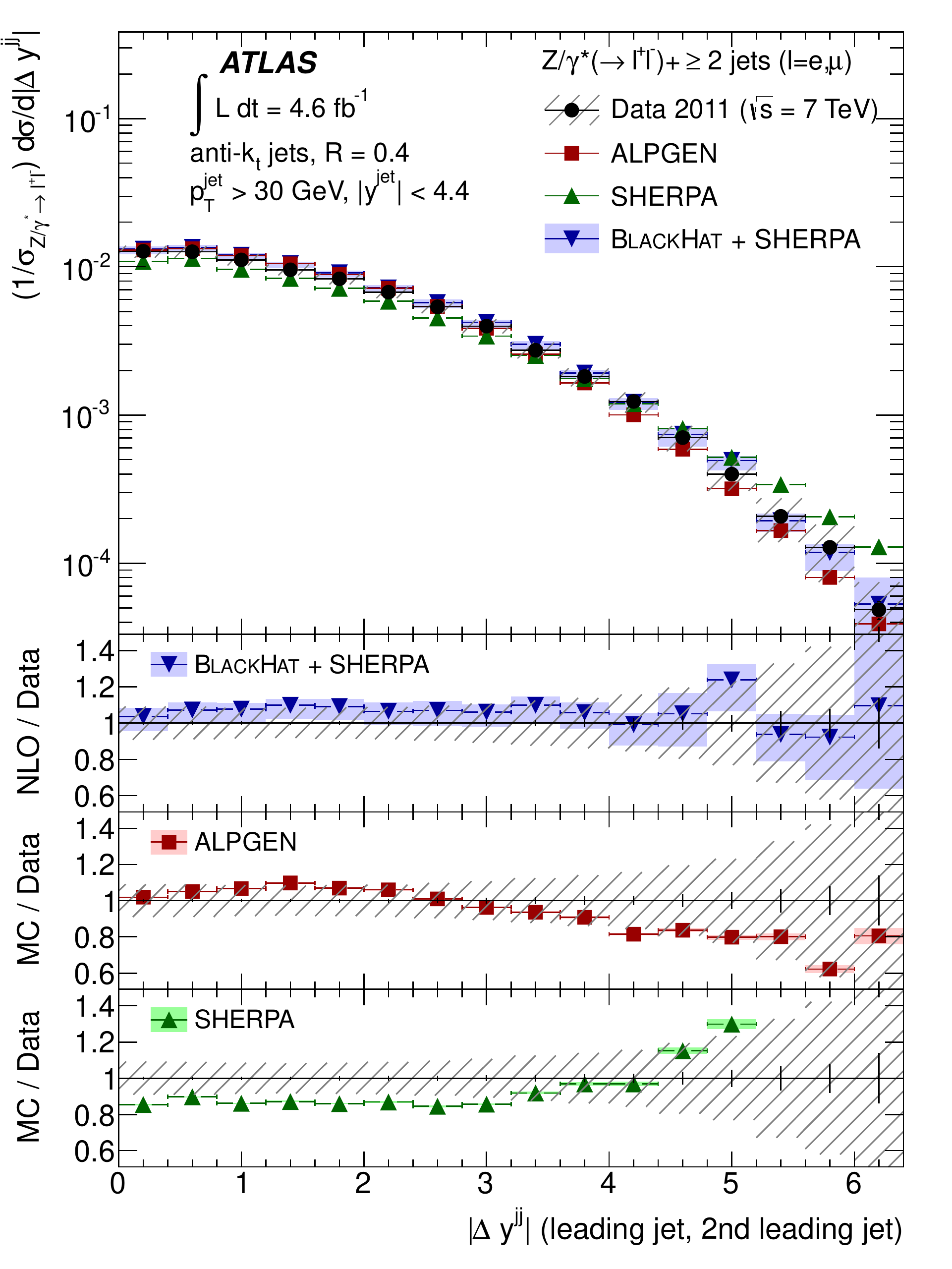}
  \label{fig:Subfigure1}
}
\subfigure[]{
  \includegraphics[width= 0.47\textwidth]{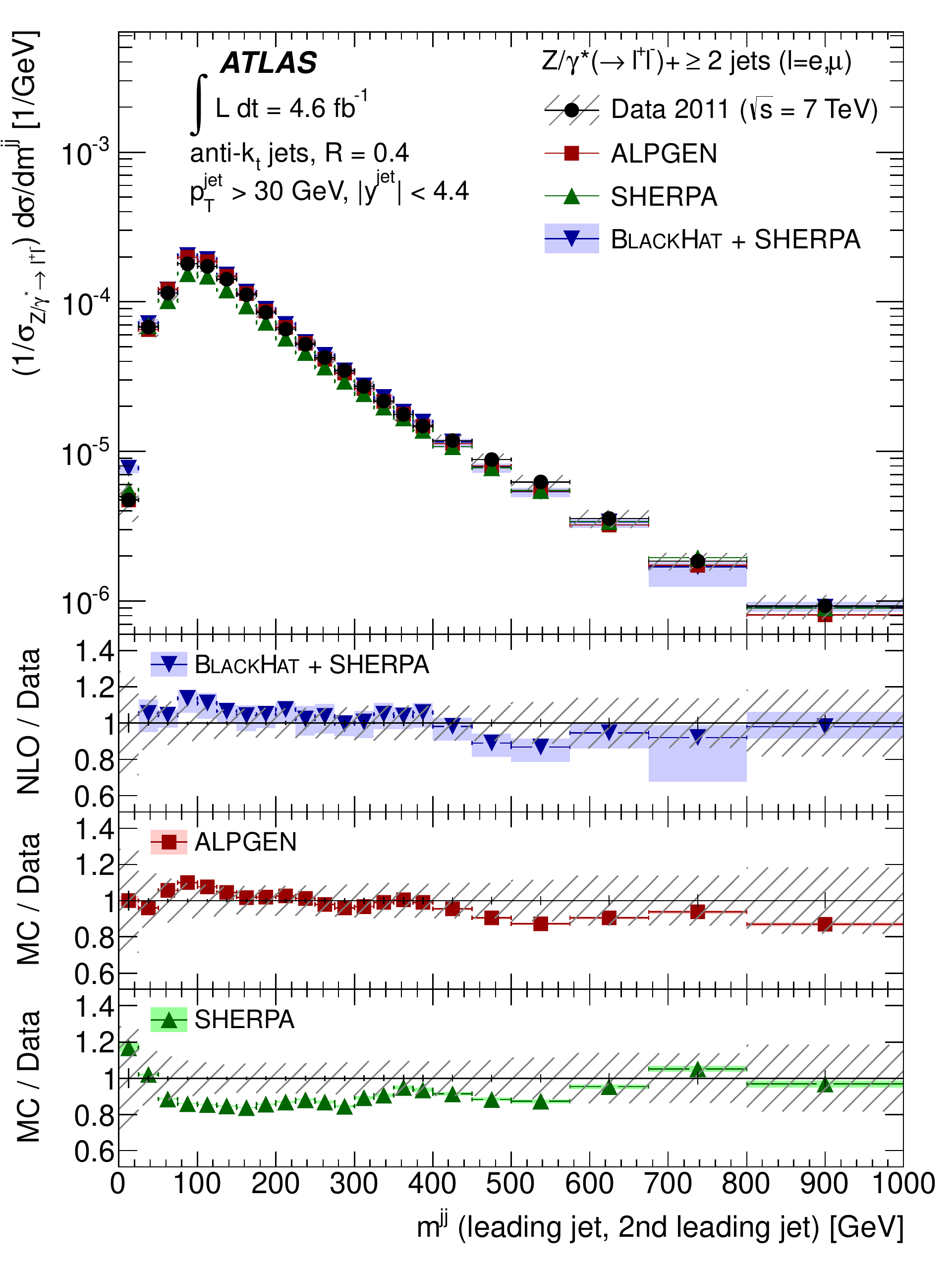}
  \label{fig:Subfigure2}
}
\end{center}
\caption {(a) Measured cross section for \Zlljets\ as a function of the separation in rapidity, \dyjj ,  between the two leading jets and 
(b) as a function of the invariant mass of the two leading jets, \mjj ,
for events with at least two jets with $\ptj > 30 \gev$ and $\ayj <4.4$ in the final state.
The cross sections are normalized to the inclusive \Zgll\ cross section.
The other  details are as in Figure~\ref{fig:ElMuComb0}.
\label{fig:ElMuComb7a}}
\end{figure}

\begin{figure}[t]
\begin{center}
\subfigure[]{
  \includegraphics[width=0.47\textwidth]{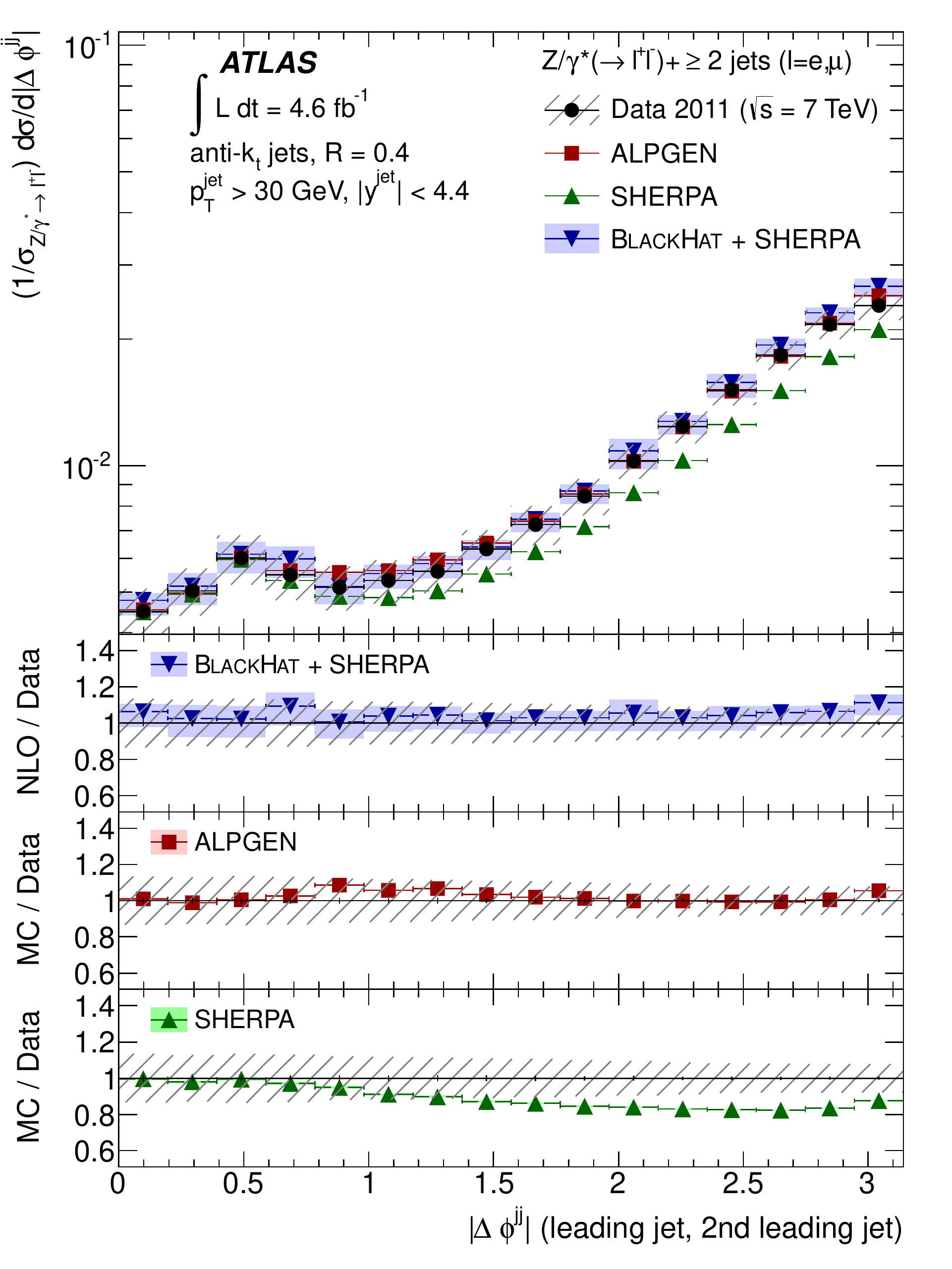}
  \label{fig:Subfigure3}
}
\subfigure[]{
  \includegraphics[width= 0.47\textwidth]{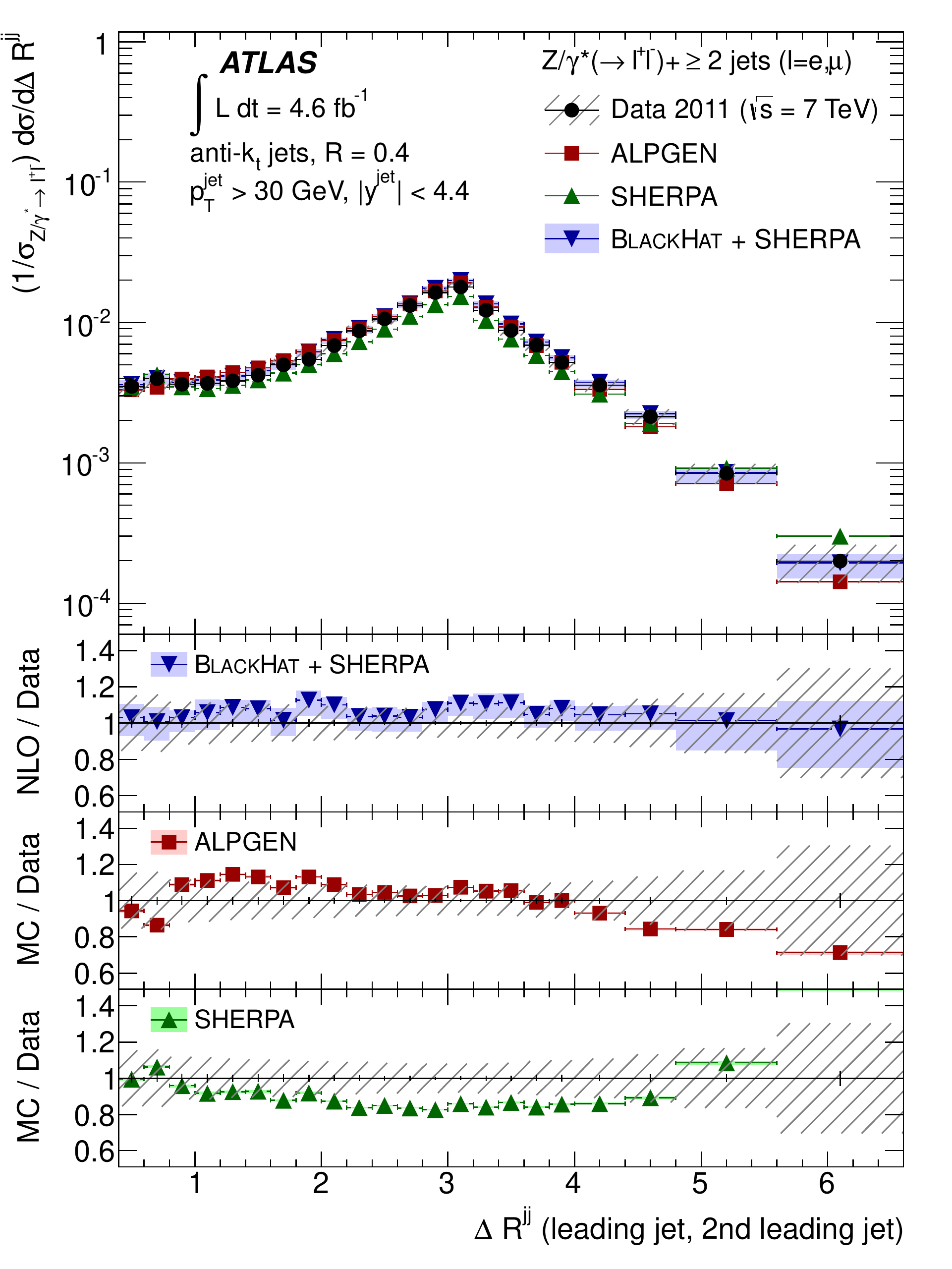}
  \label{fig:Subfigure4}
}
\end{center}
\caption{(a) Measured cross section for \Zlljets\ as a function of the distance in $\phi$  between the two leading jets, \dpjj ,  and 
(b) as a function of the distance \drjj\ between the two leading jets,  for events with at least two jets
with $\ptj > 30 \gev$ and $\ayj <4.4$ in the final state.
The cross sections are normalized to the inclusive \Zgll\ cross section.
The other  details are as in Figure~\ref{fig:ElMuComb0}.
\label{fig:ElMuComb7b}}
\end{figure}


\subsection{Distributions after VBF preselection}

A veto on a third jet is used to reject  \Zjets\ background in selections of Higgs boson candidates  produced by VBF.
Figure~\ref{fig:ElMuComb11}  shows the transverse momentum and rapidity distributions of the third jet after the VBF preselection,
as defined in section~\ref{sec:jetmulti}, normalized to the inclusive \Zgll\ cross section.
The predictions by  \bhs , \alp\ and \she\  are consistent with the measurements.
Figure~\ref{fig:jvetoeff}  shows the fraction of events which have  fulfilled the requirements of a  VBF preselection that pass in addition a veto on a  third jet 
in the central region ($|\eta| < 2.4$) as a function of the minimum transverse momentum of the veto  jet, 
referred to as `jet veto efficiency'  in the following. 
The results are shown at detector level, separately for the  \Zee\ and the \Zmm\ channel.
The overestimate of  \Rthreetwo\  in \alp\  (see figure~\ref{fig:ElMuComb3})
leads to an underestimate of the  veto efficiency,   particularly for the low-\ptj\ regime.
 \she\  predicts the veto efficiencies better. \\

\begin{figure}[t]
\begin{center}
\subfigure[]{
  \includegraphics[width=0.47\textwidth]{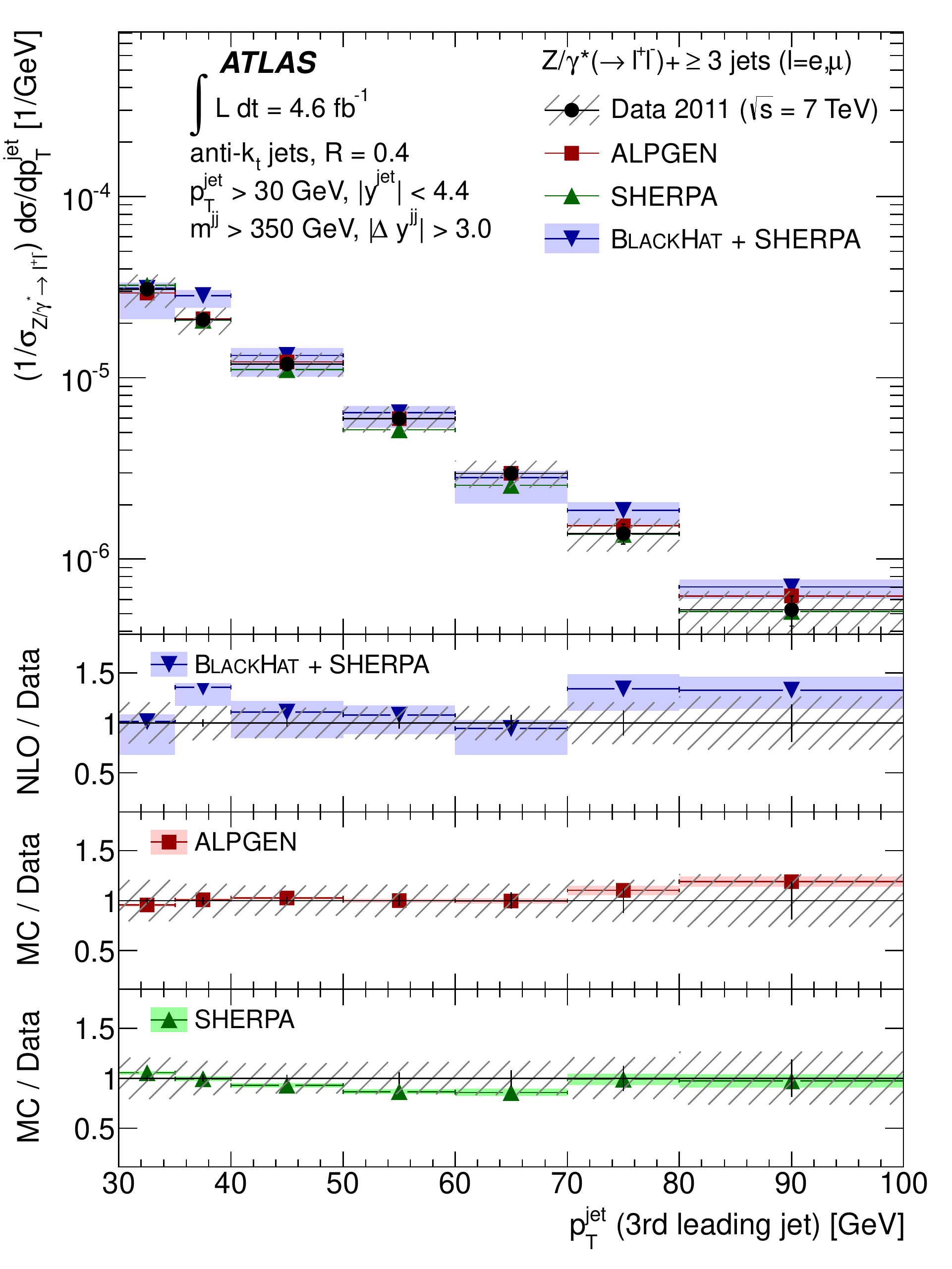}
  \label{fig:Subfigure1}
}
\subfigure[]{
  \includegraphics[width= 0.47\textwidth]{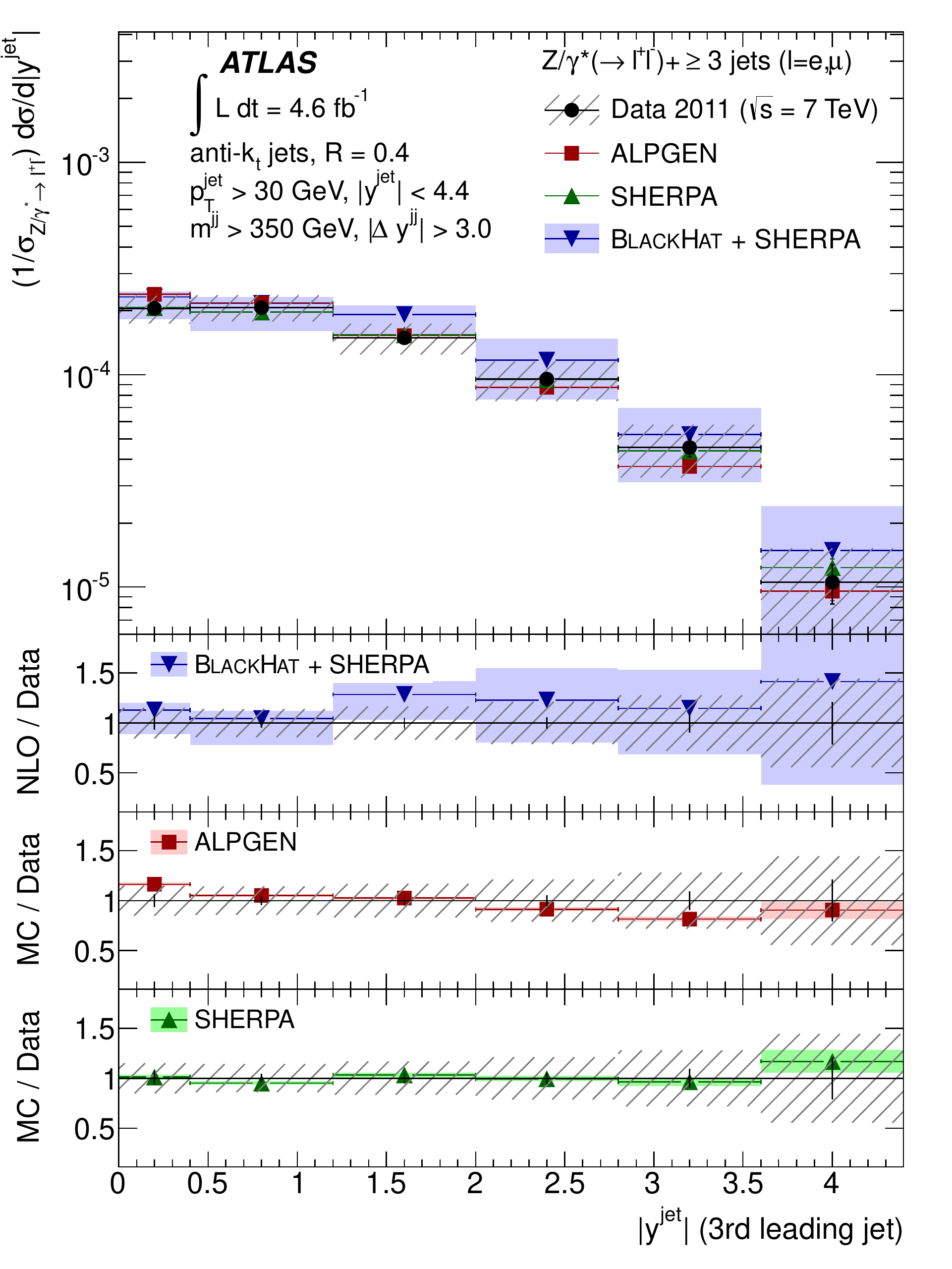}
  \label{fig:Subfigure2}
}
\end{center}
\caption{(a) Measured cross section for \Zlljets\ as a function of the transverse momentum, \ptj ,   of the third  jet   and (b) 
as a function of the absolute value of the rapidity, \ayj,  of the third jet,  in events passing the VBF preselection 
(at least two jets with  $\ptj > 30 \gev$ and $\ayj <4.4$ and $\mjj >350\gev$
and $\dyjj >3.0$ for the two leading jets).
The cross sections are normalized to the inclusive \Zgll\ cross section.
The other  details are as in Figure~\ref{fig:ElMuComb0}.
\label{fig:ElMuComb11}}
\end{figure}

\begin{figure}[t]
\begin{center}
  \subfigure[]{
   \includegraphics[width=0.47\textwidth]{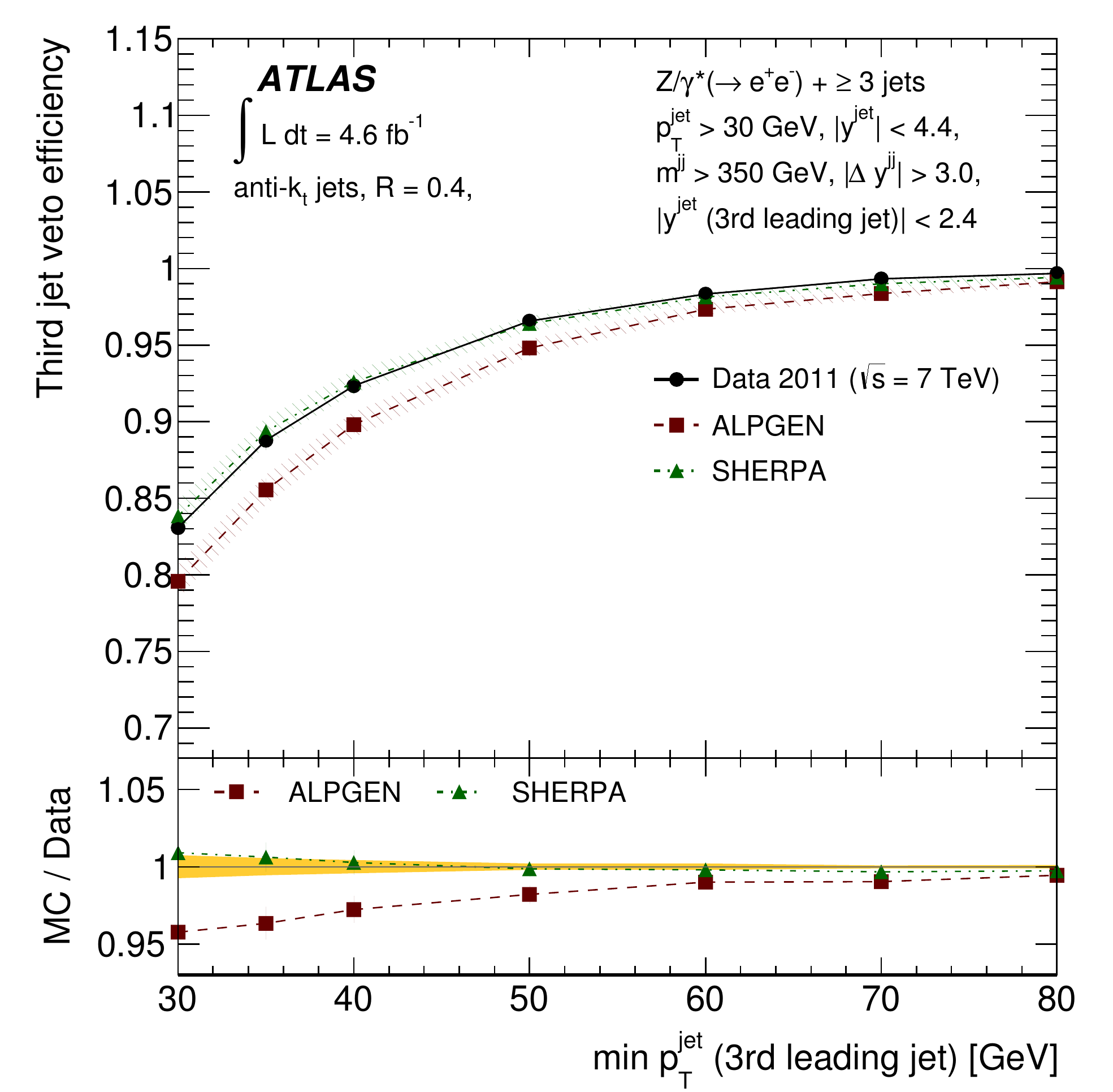}
   \label{fig:Subfigure1}
 }
 \subfigure[]{
   \includegraphics[width= 0.47\textwidth]{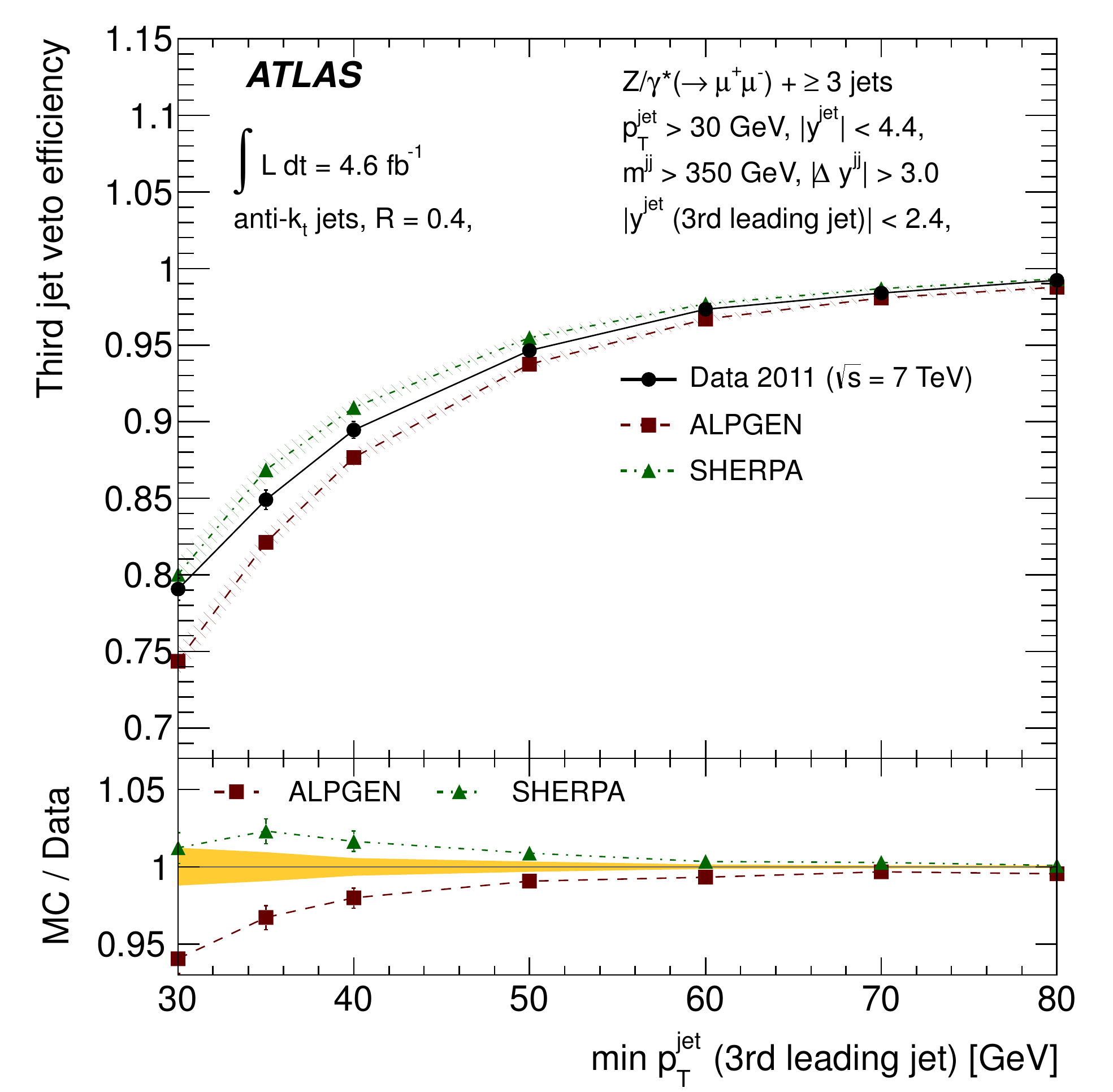}
  \label{fig:Subfigure2}
 }
\end{center}
\caption{Fraction of events that pass  a veto on a  central ($|\eta| < 2.4$) third jet  after VBF preselection 
(at least two jets with  $\ptj > 30 \gev$ and  $\ayj <4.4$, $\mjj >350\gev$
and $\,\,\,\,\,\,\,\,\,$ $\dyjj >3.0$ for the two leading jets) as a function of the third jet \ptj\ threshold, min \ptj ,   (a) in 
 the electron channel and (b) in the muon channel,  measured in data and predicted by the generators 
\alp\ and \she\ (see legend for details). 
The data points indicate the measured distribution after subtraction of electroweak and multi-jet background.
The hatched bands correspond to the combined  statistical and systematic uncertainty on the \Zjets\ prediction, using \alp\ to derive the systematic uncertainties. 
The error bars on each data point show the combined statistical and systematic uncertainty on the data.  The bottom panel shows the MC/data ratio. 
The shaded band corresponds to the total systematic  uncertainty and the 
 error bars to the statistical uncertainty on the MC/data ratio.
\label{fig:jvetoeff}}
\end{figure}

\subsection{Inclusive quantities\label{sec:inclusive}} 

Quantities based on inclusive \pt\ sums of final-state objects, such as
\htj\ or \stj ,  are often employed in searches in order to
enrich final states resulting from the decay  of heavy particles.
Reference \cite{wjetsatlas} reports a  discrepancy between  fixed-order pQCD calculations and data 
for moderate energy regimes  in \Wjets\ events
which can be mitigated by including  higher jet multiplicities in the 
theoretical calculations by means of `exclusive sums'~\cite{lesHouches11}.

Differential cross sections of \Zj\ events as a function of 
 \htj\ and \stj , normalized to the inclusive \Zgll\ cross section, 
are presented in  figure~\ref{fig:ElMuComb9}. 
 \alp\ predicts  slightly too hard spectra for both variables in line with 
the too hard spectrum for \ptj .  \she\ predictions show an offset of 10--15\% to the data.
The softer spectra from \bhs ,  based on a
\Zj\  fixed-order NLO calculation,  deviate increasingly from  the data  for larger values of 
\htj\ and  \stj  , which confirms and extends the 
results in reference~\cite{wjetsatlas}  to  a higher energy regime.
The discrepancy  is attributed to missing higher jet  multiplicities  in the fixed-order calculation.
This interpretation is investigated further in what follows.

\begin{figure}[t]
\begin{center}
\subfigure[]{
  \includegraphics[width=0.47\textwidth]{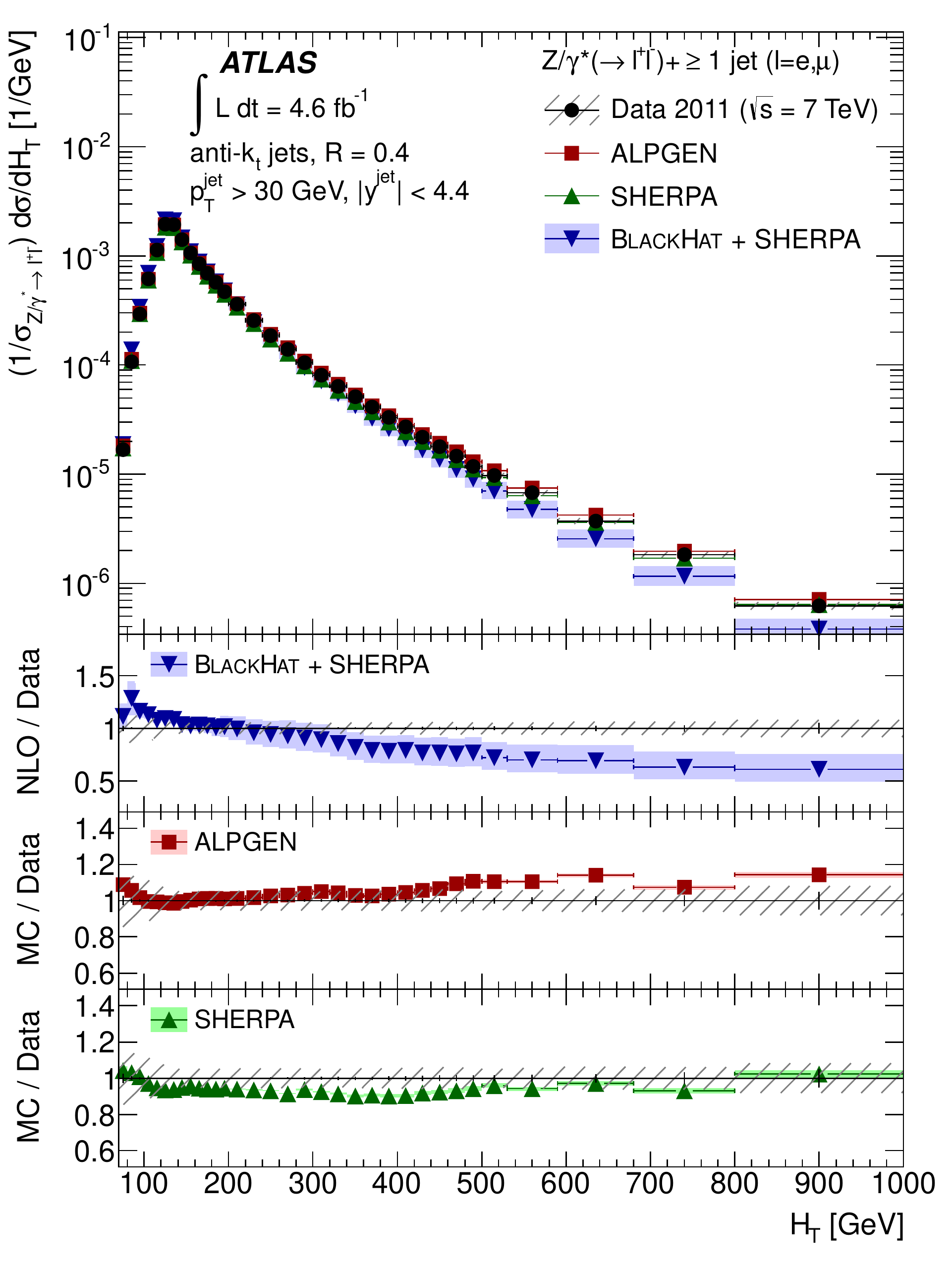}
  \label{fig:Subfigure1}
}
\subfigure[]{
  \includegraphics[width= 0.47\textwidth]{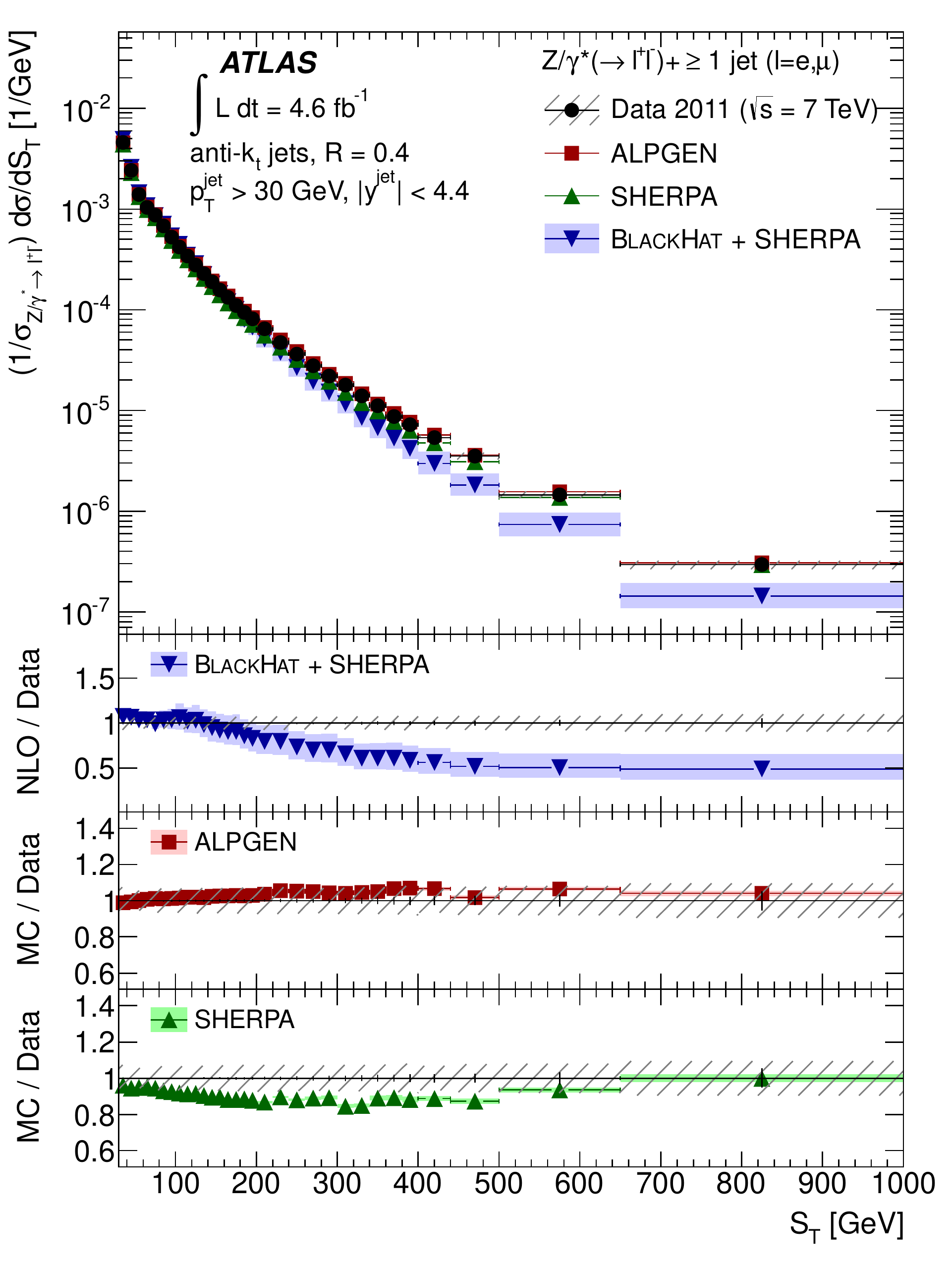}
  \label{fig:Subfigure2}
}
\end{center}
\caption{(a) Measured cross section for \Zlljets\ as a function of the scalar \pt\ sum of the leptons and the jets,  \htj , and 
(b) as a function of  the  scalar \pt\ sum of the jets, \stj , in  events with at least one jet
with $\ptj > 30 \gev$ and $\ayj <4.4$ in the final state.
The cross sections are normalized to the inclusive \Zgll\ cross section.
The other  details are as in Figure~\ref{fig:ElMuComb0}.
\label{fig:ElMuComb9}}
\end{figure}

Figure~\ref{fig:ElComb10}(a) shows, at reconstruction level, the average jet multiplicity 
as a function of \htj\  for the \Zee\  channel. Compatible results have been  
obtained in the muon channel.
Predictions  by  \alp\ and  \she\  are consistent with the data.
For values of \linebreak $\htj \approx\ 350 \gev$, where data and NLO calculation start to deviate significantly,
the average jet multiplicity exceeds two.
A similar measurement  is performed as a function of the \ptll\ in the 
\Zmm\  channel  and shown in figure~\ref{fig:ElComb10}(b). Compatible results have been
obtained in the electron channel. For values of $\ptll \approx\ 200 \gev$,  where the NLO predictions 
underestimate the measured cross section (see figure~\ref{fig:ElMuComb5b}),  on  average  two jets are resolved, typically 
one hard jet that carries most of the \Zzero\ recoil, accompanied by a soft jet.
In both cases, the kinematic regions where the NLO fixed-order
calculations perform poorly are characterized by  average jet multiplicities in excess of the fixed order
used in the NLO calculation.

\begin{figure}[t]
\begin{center}

  \subfigure[]{
    \includegraphics[width=0.47\textwidth]{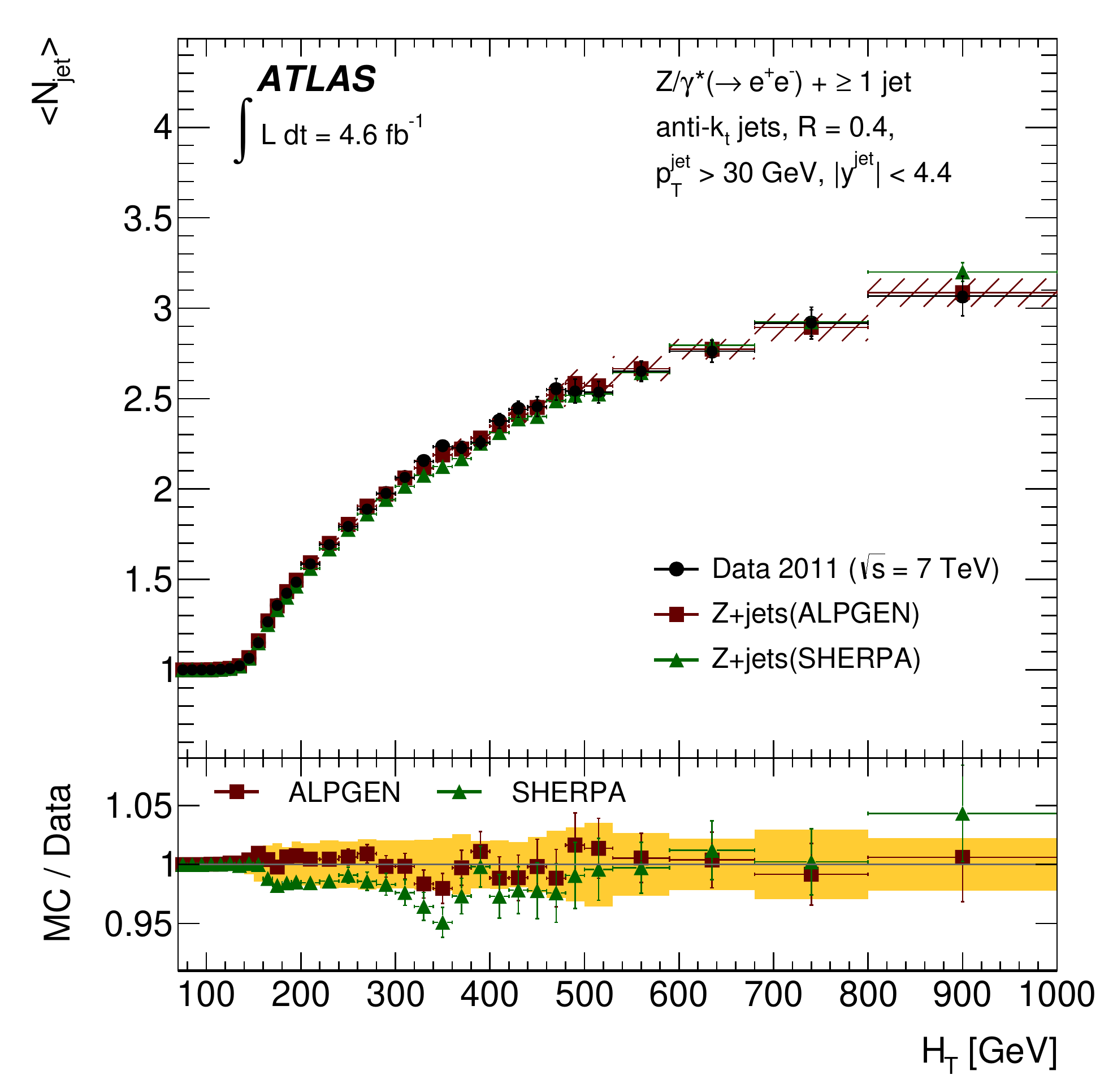}
    \label{fig:Subfigure1}
  }
 \subfigure[]{
   \includegraphics[width= 0.47\textwidth]{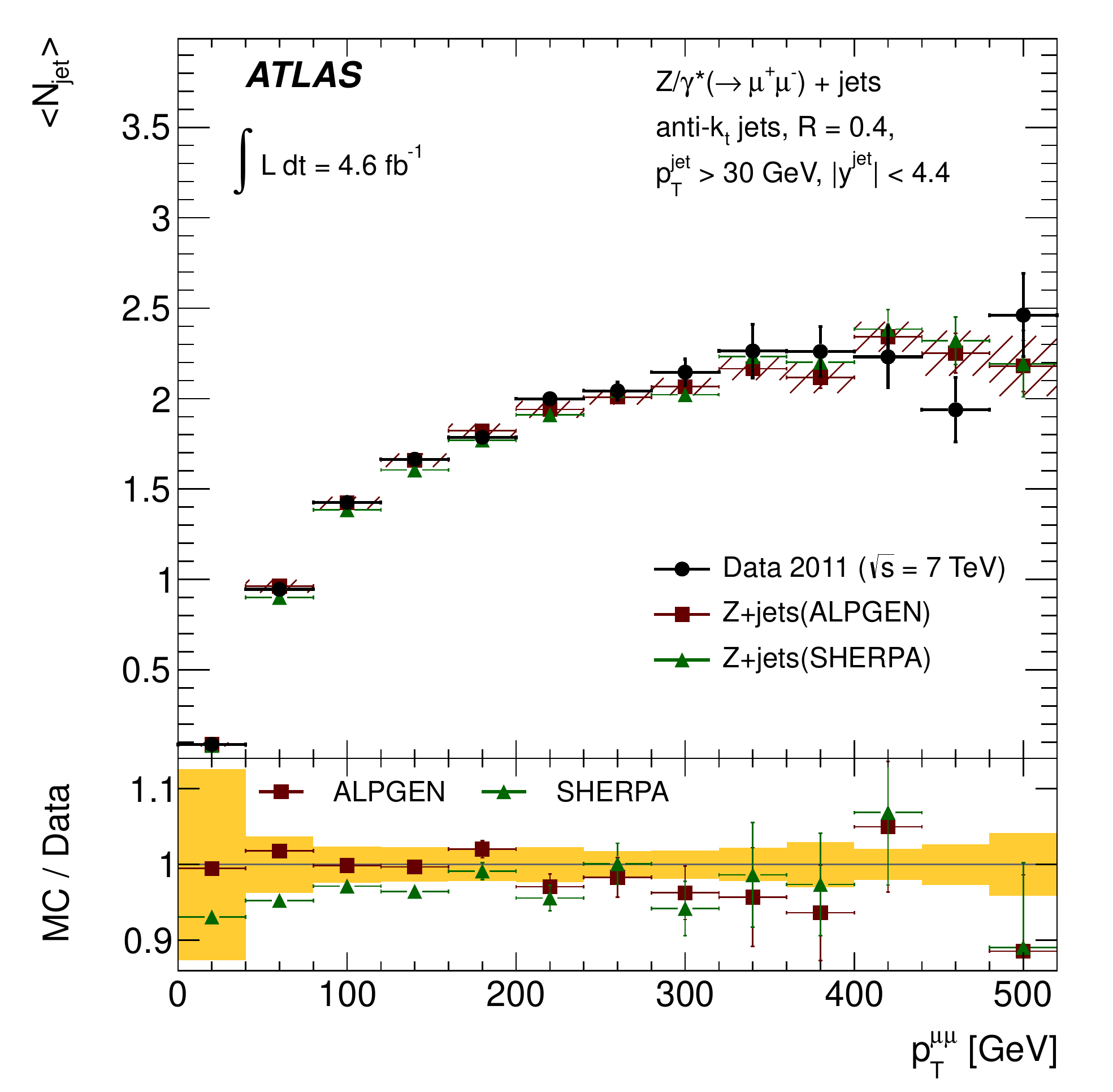}
  \label{fig:Subfigure2}
 }
\end{center}
\caption{(a) Average number of jets, ${\tt <}\nj {\tt >}$, in \Zeejets\ events as a function of the scalar \pt\ sum of the leptons and the jets,  \htj , and 
 (b) average number of jets in \Zmmjets\ events as a function of the transverse momentum of the \Zzero\ boson candidate, \ptll , 
 measured in data and predicted by the generators \alp\ and \she\  (see legend for details).
The data points indicate the measured distribution after subtraction of electroweak and multi-jet  background.
The hatched band corresponds to the combined  statistical and systematic uncertainty on the \Zjets\ prediction, modelled with \alp . 
The error bars on each data point show the combined statistical and systematic uncertainty on the data. The bottom panel shows the MC/data ratio. 
The shaded band corresponds to the total systematic  uncertainty and the 
 error bars to the statistical uncertainty on the MC/data ratio.
\label{fig:ElComb10}}
\end{figure}

Figures~\ref{fig:ElMuComb10a}(a) and~\ref{fig:ElMuComb10a}(b) replace the  fixed-order \bhs\  estimate for \htj\ and \ptll\ in 
figures~\ref{fig:ElMuComb9} and~\ref{fig:ElMuComb5b} with the `exclusive sum' 
of the cross sections for the first two jets: $(\Zzjex ) + (\Zzjj )$. The exclusive sum is consistent with the observed 
\htj\ and \ptll\ spectra  in  the phase space considered. These results support the  interpretation of the poor performance 
of the fixed-order calculation for inclusive quantities like \htj , \stj\ and \ptll\ as a sign of  missing higher jet multiplicities.
Agreement with the data can be restored by adding explicitly higher jet multiplicities via exclusive sums.

\begin{figure}[t]
\begin{center}

\subfigure[]{
 \includegraphics[width=0.47\textwidth]{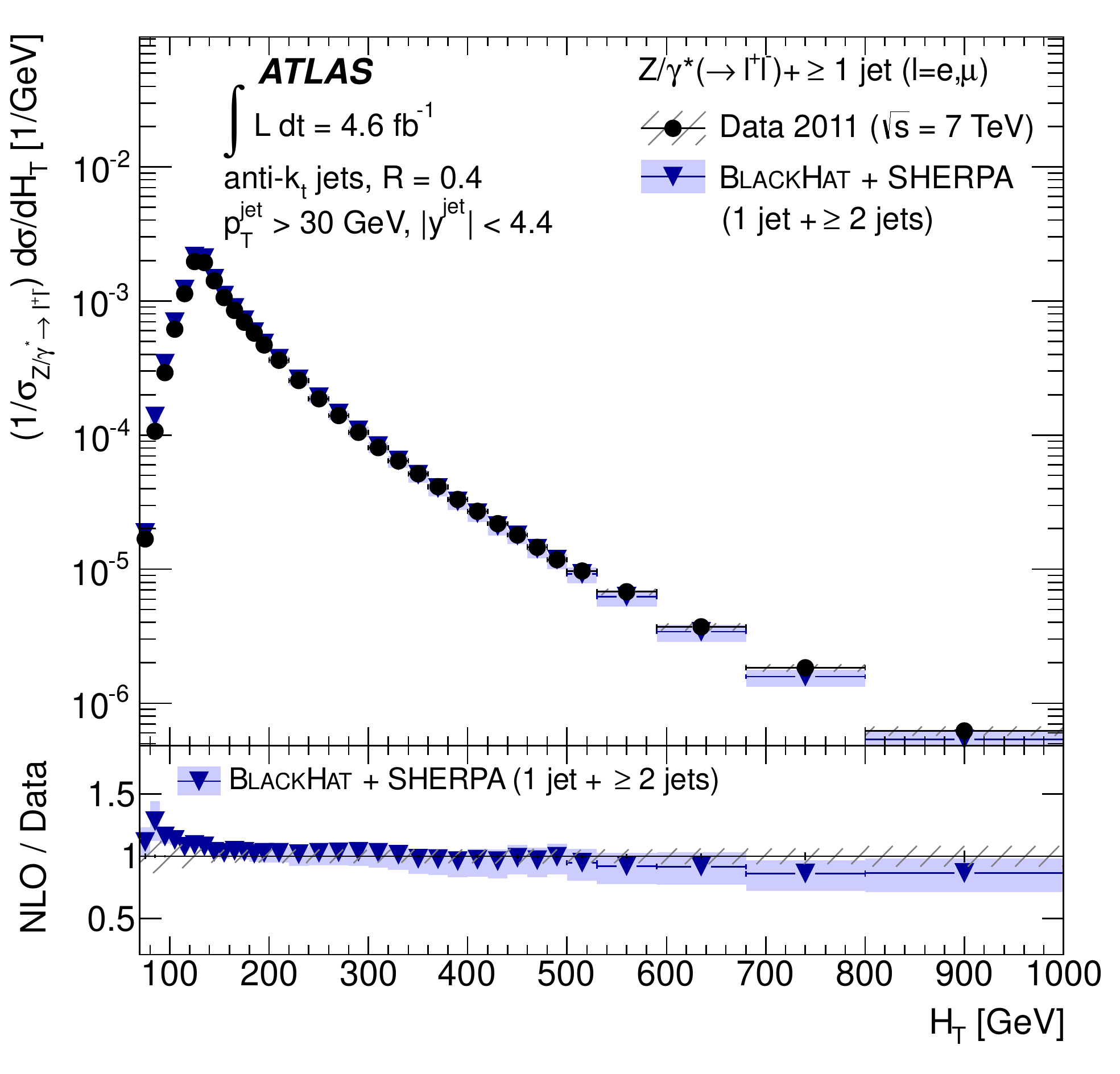}
  \label{fig:Subfigure1}
}
\subfigure[]{
\includegraphics[width= 0.47\textwidth]{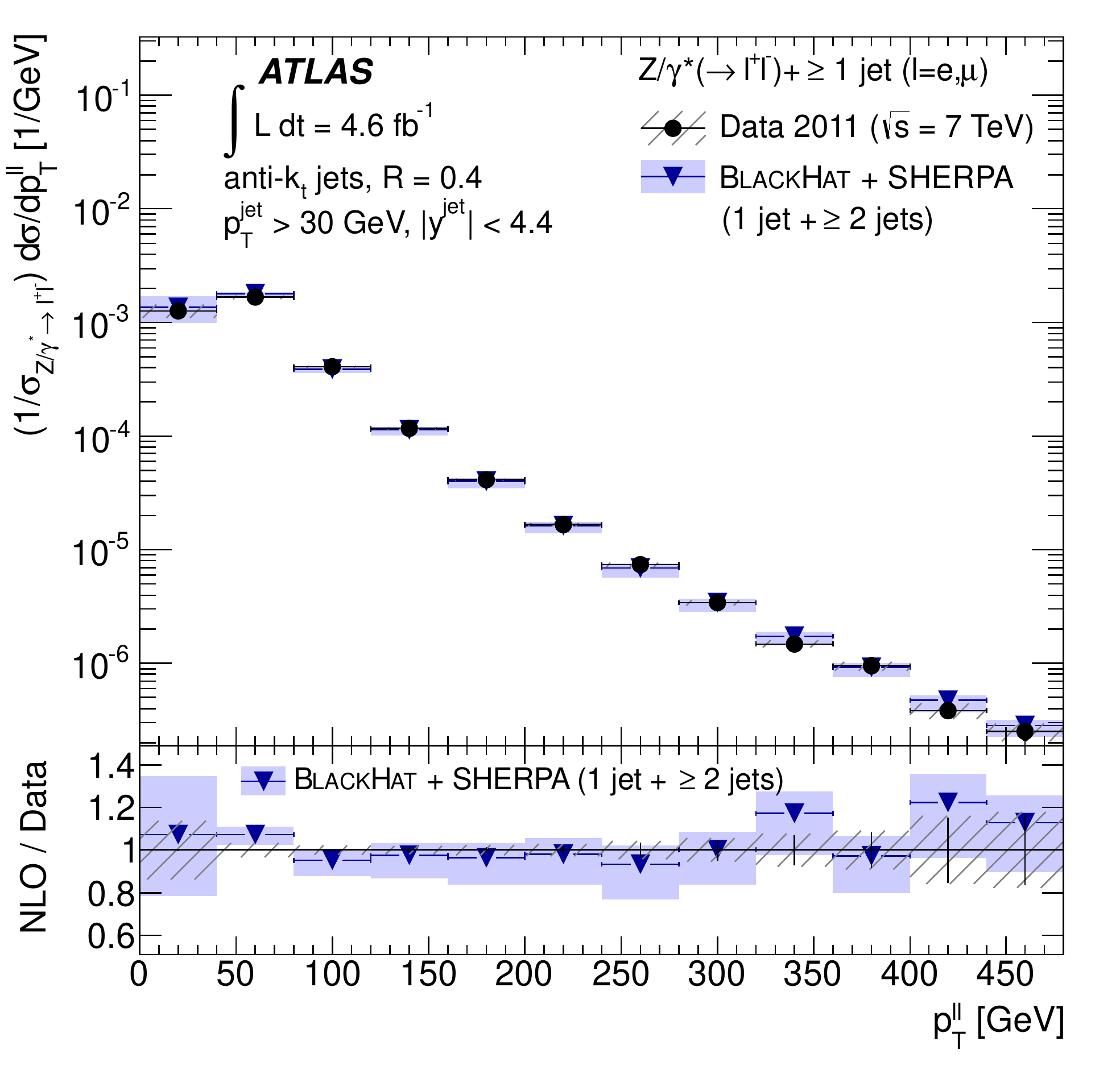}
  \label{fig:Subfigure2}
}
\end{center}
\caption{(a) Measured cross section for \Zlljets\ as a function of  the scalar \pt\ sum of the leptons and the jets, \htj , and 
(b) as a function of  the transverse momentum of the \Zzero\ candidate, \ptll ,   in  events with at least one jet
with $\ptj > 30 \gev$ and $\ayj<4.4$ in the final state.
The cross sections are normalized to the inclusive \Zgll\ cross section.
The unfolded data are compared to NLO pQCD predictions from \bhs , 
obtained by adding the exclusive \Zljex\ and the inclusive \Zljj\ calculations
and corrected to the particle level.
The error bars indicate the
statistical uncertainty on the data, and the hatched (shaded) bands the statistical and systematic
uncertainties on data (prediction) added in quadrature.
\label{fig:ElMuComb10a}}
\end{figure}

\section{Conclusions \label{Conclusions}}

Cross sections for  jets produced in
association with a \Zzero\  boson have been measured in proton--proton collisions at $\sqrt{s} =7 \tev$
with $4.6~\ifb$ of data observed with the ATLAS detector at the LHC, using
electron and muon decay modes of the \Zzero\  boson. The data have been unfolded to the particle
level and compared with predictions from the \she\ generator,  from \mca\ interfaced with \her , from  the  \alp\ generator, interfaced with \her , 
and with fixed-order calculations from \bhs .
The cross sections are quoted with respect to a phase-space region defined by \Zzero\ candidates constructed from opposite-sign leptons with $\pt  >20\gev$,   
$|\eta|<2.5$,  $\drll >0.2$ and $66\gev \leq \mll \leq 116\gev$ and for jets with $\ptj >  30\gev$, $ \ayj <4.4$ and $\drlj >0.5$. 

Cross sections as a function of the inclusive and exclusive jet multiplicities and their 
ratios have been compared, as well as differential cross sections as a function of
transverse momenta and rapidity of the jets, 
angular separation between the leading jets and the inclusive variables \htj\ and \stj . 
Compared with previous publications,  the sensitivity
has been extended  to regimes with larger jet
multiplicities and larger jet transverse momenta.  
In addition, the sample has been compared to theory in specific kinematic regions governed by large logarithmic corrections.

In general, the  predictions of the  matrix element plus parton shower generators  and the fixed-order
calculations are consistent with the measured values over a large 
kinematic range.  \mca\  fails to model not only higher jet multiplicities 
but also the  transverse momentum of the leading jet.
The transition from staircase to Poisson scaling
of the exclusive jet multiplicity ratio, expected from theory when introducing a large scale
difference,  is observed in the data.

 In events where  two jets have passed a VBF preselection,  the cross sections
for higher jet multiplicities  are overestimated by \alp .
This leads to a small underestimation of the probability
for \Zjets\ events to survive a veto on a soft third jet.

\alp\ predicts a too hard spectrum of the transverse momentum
of  the leading jet, of \ptll , \htj\ and \stj\ 
in a regime where large corrections from higher-order electroweak   and higher-order QCD processes
are expected.
The jet rapidity distribution is predicted to be too wide  in \bhs\  
and in \she . 
\bhs\ underestimates the cross section for large \ptll\
where more than one  jet can be resolved.
The  \htj\  or \stj\ spectra predicted by \bhs\ fixed order
 NLO calculations deviate  by several standard deviations from
the measured spectra in the hard \htj\ and \stj\  regime characterized
by  large  average jet multiplicities.
 The observed spectra of \htj\ and \ptll\ can be described by an  exclusive sum of \bhs\  
fixed-order calculations for \Zpex\  and \Zpp .


\section{Acknowledgements}

We thank CERN for the very successful operation of the LHC, as well as the
support staff from our institutions without whom ATLAS could not be
operated efficiently.

We acknowledge the support of ANPCyT, Argentina; YerPhI, Armenia; ARC,
Australia; BMWF and FWF, Austria; ANAS, Azerbaijan; SSTC, Belarus; CNPq and FAPESP,
Brazil; NSERC, NRC and CFI, Canada; CERN; CONICYT, Chile; CAS, MOST and NSFC,
China; COLCIENCIAS, Colombia; MSMT CR, MPO CR and VSC CR, Czech Republic;
DNRF, DNSRC and Lundbeck Foundation, Denmark; EPLANET, ERC and NSRF, European Union;
IN2P3-CNRS, CEA-DSM/IRFU, France; GNSF, Georgia; BMBF, DFG, HGF, MPG and AvH
Foundation, Germany; GSRT and NSRF, Greece; ISF, MINERVA, GIF, DIP and Benoziyo Center,
Israel; INFN, Italy; MEXT and JSPS, Japan; CNRST, Morocco; FOM and NWO,
Netherlands; BRF and RCN, Norway; MNiSW, Poland; GRICES and FCT, Portugal; MERYS
(MECTS), Romania; MES of Russia and ROSATOM, Russian Federation; JINR; MSTD,
Serbia; MSSR, Slovakia; ARRS and MIZ\v{S}, Slovenia; DST/NRF, South Africa;
MICINN, Spain; SRC and Wallenberg Foundation, Sweden; SER, SNSF and Cantons of
Bern and Geneva, Switzerland; NSC, Taiwan; TAEK, Turkey; STFC, the Royal
Society and Leverhulme Trust, United Kingdom; DOE and NSF, United States of
America.

The crucial computing support from all WLCG partners is acknowledged
gratefully, in particular from CERN and the ATLAS Tier-1 facilities at
TRIUMF (Canada), NDGF (Denmark, Norway, Sweden), CC-IN2P3 (France),
KIT/GridKA (Germany), INFN-CNAF (Italy), NL-T1 (Netherlands), PIC (Spain),
ASGC (Taiwan), RAL (UK) and BNL (USA) and in the Tier-2 facilities
worldwide.

\begin{table}[t]
\renewcommand{\arraystretch}{1.5}
{\footnotesize
\begin{tabular}{|c|c|c|c|}
\hline\hline
Incl.  jet multiplicity  & Data cross-section (pb) & $\rm \delta^{had}$ & $\rm \delta^{QED}$ \\
\hline\hline
$\geq$~1~jets   & [ 6.88$\pm$ 0.01 (stat)$\pm$ 0.52 (syst) $\pm$ 0.13 (lumi) ]$\times 10^{1}$ & 1.027$\pm$0.015 &  0.976$\pm$0.005\\
$\geq$~2~jets   & [ 1.51$\pm$ 0.01 (stat)$\pm$ 0.15 (syst) $\pm$ 0.03 (lumi) ]$\times 10^{1}$ &  1.036$\pm$0.017& 0.979$\pm$0.005\\
$\geq$~3~jets   &  3.09$\pm$ 0.03 (stat)$\pm$ 0.40 (syst) $\pm$ 0.06 (lumi)                             & 1.031$\pm$0.033 & 0.980$\pm$0.005\\
$\geq$4~jets   & [ 6.55$\pm$ 0.13 (stat)$\pm$ 1.06 (syst) $\pm$ 0.12 (lumi) ]$\times 10^{-1}$& 1.043$\pm$0.023 & 0.982$\pm$0.004\\
$\geq$~5~jets   & [ 1.35$\pm$ 0.06 (stat)$\pm$ 0.27 (syst) $\pm$ 0.02 (lumi) ]$\times 10^{-1}$&   &  \\
$\geq$~6~jets   & [ 2.53$\pm$ 0.27 (stat)$\pm$ 0.60 (syst) $\pm$ 0.05 (lumi) ]$\times 10^{-2}$&   &  \\
$\geq$~7~jets   & [ 6.23$\pm$ 1.46 (stat)$\pm$ 2.14 (syst) $\pm$ 0.11 (lumi) ]$\times 10^{-3}$&   &  \\
\hline\hline
Incl.  jet multiplicity ratio  & Data cross-section ratio & $\rm \delta^{had}$ & $\rm \delta^{QED}$ \\
\hline\hline
$\geq$~1~jets / $\geq$~0~jets   & [ 1.42$\pm$ 0.00 (stat)$\pm$ 0.11 (syst) ]$\times 10^{-1}$& 1.036$\pm$0.015 & 0.995$\pm$0.010\\ 
$\geq$~2~jets / $\geq$~1~jets   & [ 2.18$\pm$ 0.01 (stat)$\pm$ 0.07 (syst) ]$\times 10^{-1}$& 1.009$\pm$0.002 & 1.003$\pm$0.010\\
$\geq$~3~jets / $\geq$~2~jets   & [ 2.05$\pm$ 0.02 (stat)$\pm$ 0.07 (syst) ]$\times 10^{-1}$& 0.995$\pm$0.016 & 1.001$\pm$0.010\\
$\geq$~4~jets / $\geq$~3~jets   & [ 2.12$\pm$ 0.04 (stat)$\pm$ 0.08 (syst) ]$\times 10^{-1}$& 1.011$\pm$0.010 & 1.001$\pm$0.009\\
$\geq$~5~jets / $\geq$~4~jets   & [ 2.06$\pm$ 0.08 (stat)$\pm$ 0.10 (syst) ]$\times 10^{-1}$&  & \\
$\geq$~6~jets / $\geq$~5~jets   & [ 1.87$\pm$ 0.18 (stat)$\pm$ 0.13 (syst) ]$\times 10^{-1}$&  & \\
$\geq$~7~jets / $\geq$~6~jets   & [ 2.46$\pm$ 0.49 (stat)$\pm$ 0.39 (syst) ]$\times 10^{-1}$&  & \\
\hline\hline
Excl.  jet multiplicity ratio  & Data cross-section ratio &$\rm \delta^{had}$ & $\rm \delta^{QED}$  \\
\hline\hline
1~jet /  0~jet    & [ 1.29$\pm$ 0.00 (stat)$\pm$ 0.10 (syst) ]$\times 10^{-1}$ &1.032$\pm$0.013  & 0.994$\pm$0.010\\ 
2~jets / 1~jet    & [ 2.23$\pm$ 0.01 (stat)$\pm$ 0.08 (syst) ]$\times 10^{-1}$ &1.013$\pm$0.010  & 1.003$\pm$0.010\\
3~jets / 2~jets   & [ 2.03$\pm$ 0.02 (stat)$\pm$ 0.07 (syst) ]$\times 10^{-1}$& 0.990$\pm$0.032 & 1.001$\pm$0.010\\
4~jets / 3~jets   & [ 2.14$\pm$ 0.05 (stat)$\pm$ 0.08 (syst) ]$\times 10^{-1}$& 1.022$\pm$0.028 & 1.001$\pm$0.009\\
5~jets / 4~jets   & [ 2.11$\pm$ 0.11 (stat)$\pm$ 0.10 (syst) ]$\times 10^{-1}$&  & \\
6~jets / 5~jets   & [ 1.74$\pm$ 0.22 (stat)$\pm$ 0.10 (syst) ]$\times 10^{-1}$&  &\\
7~jets / 6~jets   & [ 2.60$\pm$ 0.79 (stat)$\pm$ 0.45 (syst) ]$\times
10^{-1}$&  &\\
\hline\hline
 Excl.  jet multiplicity ratio  & Data cross-section ratio &  $\rm \delta^{had}$ & $\rm \delta^{QED}$ \\
\ptj\ (1st jet) $>150\gev$ &  & & \\
\hline\hline
2~jets / 1~jet    &  1.04$\pm$ 0.03 (stat)$\pm$ 0.03 (syst)                              & 1.004$\pm$0.002 & 1.000$\pm$0.009\\ 
3~jets / 2~jets   & [ 4.82$\pm$ 0.13 (stat)$\pm$ 0.16 (syst) ]$\times 10^{-1}$& 0.989$\pm$0.037& 1.002$\pm$0.006\\
4~jets / 3~jets   & [ 3.71$\pm$ 0.17 (stat)$\pm$ 0.16 (syst) ]$\times 10^{-1}$& 1.025$\pm$0.040& 0.996$\pm$0.006\\
5~jets / 4~jets   & [ 2.85$\pm$ 0.21 (stat)$\pm$ 0.12 (syst) ]$\times 10^{-1}$&  &\\
6~jets / 5~jets   & [ 2.67$\pm$ 0.37 (stat)$\pm$ 0.21 (syst) ]$\times 10^{-1}$&  &\\
7~jets / 6~jets   & [ 2.57$\pm$ 0.78 (stat)$\pm$ 0.51 (syst) ]$\times
10^{-1}$&  &\\
\hline\hline
\end{tabular}}
\caption{Combined inclusive \Zlljets\ cross sections per lepton flavour and their ratios and exclusive cross-section ratios 
for various preselections measured in data together with the corresponding  non-perturbative  corrections  $\rm \delta^{had}$  and $\rm \delta^{QED}$.
The cross sections are quoted with respect to a phase-space region defined by \Zzero\ candidates constructed from opposite-sign leptons with $\pt  >20\gev$,   
$|\eta|<2.5$,  $\drll >0.2$ and $66\gev \leq \mll \leq 116\gev$ and for jets with $\ptj >  30\gev$, $ \ayj <4.4$ and $\drlj >0.5$. 
\label{t:DataTheoMulti}}
\end{table}

\onecolumn
\clearpage 
\begin{flushleft}
{\Large The ATLAS Collaboration}

\bigskip

G.~Aad$^{\rm 48}$,
T.~Abajyan$^{\rm 21}$,
B.~Abbott$^{\rm 111}$,
J.~Abdallah$^{\rm 12}$,
S.~Abdel~Khalek$^{\rm 115}$,
A.A.~Abdelalim$^{\rm 49}$,
O.~Abdinov$^{\rm 11}$,
R.~Aben$^{\rm 105}$,
B.~Abi$^{\rm 112}$,
M.~Abolins$^{\rm 88}$
% ...
\bigskip

% ...
$^{11}$ Institute of Physics, Azerbaijan Academy of Sciences, Baku, Azerbaijan\\
$^{12}$ Institut de F\'{i}sica d'Altes Energies and Departament de F\'{i}sica de la Universitat Aut\`{o}noma de Barcelona and ICREA, Barcelona, Spain\\
% ...
$^{21}$ Physikalisches Institut, University of Bonn, Bonn, Germany\\
% ...
$^{48}$ Fakult\"{a}t f\"{u}r Mathematik und Physik, Albert-Ludwigs-Universit\"{a}t, Freiburg, Germany\\
$^{49}$ Section de Physique, Universit\'{e} de Gen\`{e}ve, Geneva, Switzerland\\
% ...
$^{88}$ Department of Physics and Astronomy, Michigan State University, East Lansing MI, United States of America\\
% ...
$^{105}$ Nikhef National Institute for Subatomic Physics and University of Amsterdam, Amsterdam, Netherlands\\
% ...
$^{111}$ Homer L. Dodge Department of Physics and Astronomy, University of Oklahoma, Norman OK, United States of America\\
$^{112}$ Department of Physics, Oklahoma State University, Stillwater OK, United States of America\\
% ...
$^{115}$ LAL, Universit\'{e} Paris-Sud and CNRS/IN2P3, Orsay, France\\
% ...
\end{flushleft}


\end{document}